\documentclass[twocolumn,twocolappendix]{aastex63}
\usepackage[caption=false]{subfig}

\usepackage{savesym}
\usepackage{natbib}
\savesymbol{tablenum}
\usepackage{siunitx}
\restoresymbol{SIX}{tablenum}
\submitjournal{ApJ}
\accepted{October 19, 2019}
\shortauthors{N. Trueba et al.}
\shorttitle{The Disk Wind in 4U 1630$-$472}

\begin{document}

\title{A Comprehensive Chandra Study of the Disk Wind in the Black
  Hole Candidate 4U 1630$-$472}
\correspondingauthor{Nicolas Trueba}
\email{ntrueba@umich.edu}

\author{Nicolas Trueba}
\affiliation{Department of Astronomy, University of Michigan, Ann Arbor, MI}
\author{J.~M.~Miller}
\affiliation{Department of Astronomy, University of Michigan, Ann Arbor, MI}
\author{J.~Kaastra}
\affiliation{SRON, Netherlands Institute for Space Research, Sorbonnelaan 2, 3584 CA Utrecht, The Netherlands}
\author{A.~Zoghbi}
\affiliation{Department of Astronomy, University of Michigan, Ann Arbor, MI}
\author{A.~C.~Fabian}
\affiliation{Institute of Astronomy, University of Cambridge, Madingley Road, Cambridge CB3 OHA, UK}
\author{T.~Kallman}
\affiliation{NASA Goddard Space Flight Center, Code 662, Greedbelt, MD 20771, USA}
\author{D.~Proga}
\affiliation{Department of Physics, University of Nevada, Las Vegas, Las Vegas, NV 89154, USA}
\author{J.~Raymond}
\affiliation{Harvard-Smithsonian Center for Astrophysics, 60 Garden Street, Cambridge, MA 02138, USA}

\begin{abstract}

The mechanisms that drive disk winds are a window into the physical processes that underlie the disk. 
Stellar-mass black holes are an ideal
setting in which to explore these mechanisms, in part because their
outbursts span a broad range in mass accretion rate.  
We performed a spectral analysis of the disk wind found in six Chandra/HETG 
observations of the black hole candidate 4U~1630$-$472, covering a range of luminosities over two distinct spectral states. 
We modeled both wind absorption and extended wind re-emission components using PION, a self-consistent photoionized absorption model.
In all but one case, two photoionization zones were required in order to obtain acceptable fits. 
Two independent constraints on launching radii, obtained via
the ionization parameter formalism and the dynamical broadening of the re-emission,
helped characterize the geometry of the wind.
The innermost wind components ($r \simeq {10}^{2-3}$ $GM/{c}^{2}$) tend towards small volume filling factors, high ionization, densities up to $n \simeq {10}^{15-16} {\text{cm}}^{-3}$, and outflow velocities of $\sim 0.003c$. 
These small launching radii and large densities require magnetic driving, as they are inconsistent with numerical and analytical treatments of thermally driven winds.
Outer wind components ($r \simeq {10}^{5}$ $GM/{c}^{2}$) are significantly less ionized and have filling factors near unity. 
Their larger launching radii, lower densities ($n \simeq {10}^{12} {\text{cm}}^{-3}$), and outflow velocities ($\sim 0.0007c$) are nominally consistent with thermally driven winds. 
The overall wind structure suggests that these components may also be part of a broader MHD outflow and perhaps better described as magneto-thermal hybrid winds.

\end{abstract}

\section{Introduction}\label{sec:intro}
A detailed analysis of the disk winds from low-mass X-ray binaries
(LMXBs) is critical to understanding the accretion flow in these
systems and, more generally, in forming a complete picture of
accretion onto compact objects.  Indeed, winds are a sizable component
in terms of mass transfer in the disk-- estimates of wind mass-loss
rates range from a fraction of the accretion rate to, in some cases,
drastically exceeding the mass inflow rate.  A highly non-conservative
accretion flow of this kind would impact several aspects of cusrrent
LMXB evolution models.  Moreover, the scale of these outflows suggests 
that disk winds may play a fundamental role in accreting systems.

Insights gained from studying stellar-mass black hole winds may 
further our understanding of outflows spanning the black hole mass scale.  
Analyses of AGN winds through Chandra/HETG deep exposures \citep[e.g.,][]{Young2005} have revealed highly ionized X-ray
wind components associated with the broad line region (BLR),
similar to the wind found in some stellar-mass black holes
\citep[e.g.,][]{Miller2006a}.  These similarities in column density,
ionization, and outflow velocity (and consequently kinetic power and
launching radii) suggest that LMXB winds may probe similar physics as
these inner disk AGN winds. Unlike supermassive black holes, however,
analyses of stellar-mass black hole winds are unimpeded by complex
SEDs (including wind absorption components not associated with the
inner disk) and can be performed at higher sensitivities.

Most notably, understanding the mechanisms driving disk winds may bring insights into the physical processes mediating angular momentum 
and mass transport within the disk.
Evidence of magnetically driven winds has been uncovered in several black hole LMBXs, including 
GRO J1655-40 \citep{Miller2006a,Miller2008,Miller2015,Fukumura2017,NeilsenHoman,Kallman2009}, GRS 1915+105 \citep{Miller2015,Miller2016},
IGR J17091-3624 \citep{King2012}, V404 Cyg \citep{King2015}, and H1743-322 \citep{Miller2015}.
These tentative results suggest that magnetic processes may not only drive disk winds, but mediate mass transfer within the disk itself.
Simulations of magnetically viscous disks, wherein turbulence arises due to the magnetorotational instability \citep[MRI;][]{BalbusHawley}, 
predict the presence of disk winds driven via the resulting magnetohydrodynamic (MHD) pressure.
Alternatively, magnetocentrifugal acceleration \citep{BlandfordPayne} can drive winds 
that transport angular momentum as they are accelerated outwards along magnetic field lines.
These outflows would be compact analogs to those seen in some FU Ori and T Tauri systems \citep{Calvet1993}.
With the prevailing view of accretion as a fundamentally magnetic process, 
disk winds are a key observational counterpart to theoretical work.  

There are other mechanisms apart from magnetic forces that can drive
a disk wind, namely radiative and thermal driving. The dominant absorption components in black hole binaries are often too
ionized to be driven via radiation pressure: Ions in the absorbing gas have
been stripped of most electrons involved in the UV transitions where
cross-section spikes occur. 
Alternatively, Compton heating of the disk can effectively drive a thermal wind
from outside the Compton radius (or, ${R}_{C}$), though this limit 
may extend down to 0.1${R}_{C}$ \citep[see][]{Begelman1983,Woods1996}. 
However, Compton heating cannot account for winds launched nearest to the black hole.
Robust estimates of wind launching radii are therefore the primary means in identifying and differentiating magnetic winds.

Assuming the bulk of the absorbing gas column density is located
at or near its launch point, wind launching radii, kinetic power, and outflow rates can be estimated through
photoionized absorption (or, PIA) modeling. The ionization parameter,
$\xi=L/nr^{2}$, links the degree of ionization of the gas to its density, source luminosity,
and distance to the photoionizing source.  Although this is possible in a few sources \citep[e.g. GRO J1655-40 and MAXI J1305-704,
where Fe XXII line ratios suggest a density of n ${\simeq {10}^{14} {\text{cm}}^{-3}}$, see][]{Miller2008,Miller2014b},
gas densities cannot be directly measured in most cases. Instead, only an upper limit can be set using the column density via $r\leq L/N\xi$, the limit at which the wind has a filling factor of unity.

Previous efforts have also included re-emission from	
the same absorbing gas and obtained independent launching radii 
estimates based on dynamical broadening, assuming winds rotate at the local 
Keplerian velocity. Wind re-emission is visually apparent in a handful of LMXB spectra, such as strong P-Cygni profiles
corresponding to He-like Fe XXV found in some observations of GRS 1915+105 \citep{Miller2015,Miller2016}.
Evidence of Fe K band P-Cygni profiles has been uncovered in the NuSTAR and XMM Newton spectra of some AGN, 
including PDS 456 \citep{Reeves2018}, PG1211+143 \citep{PoundsReeves}, and Cygnus A \citep{Reynolds2015}.
Photoionization modeling of these features is complicated by overlapping Fe K reflection, while high outflow velocities
(ranging from 0.08-0.25c) means that a large portion of the broadening is not Keplerian. 
The Chandra HETGS spectrum of NGC 7469 shows clear P-Cygni profiles (e.g. Ne X Ly$\alpha$), 
yet again the emission line broadening is dominated by the wind outflow velocity rather than the orbital motion of the gas \citep{Mehdipour2018}.
Obtaining geometric information from these sources will require more sensitive spectra and at higher resolution.
In LMXBs, the lack of obvious emission features would point towards 
highly broadened emission and/or a small wind covering factor, depending on the assumed geometry. 
Alternatively, models that neglect re-emission lack full self-consistency
and often yield worse statistical fits, as line ratios can be significantly affected by emission lines (see Section \ref{sec:soft}). 

Modeling of X-ray winds through single and multiple absorption zones has improved in recent years; 
the physical self-consistency of this approach outweighs the simplicity of
line-by-line fitting through Gaussian functions, given statistically acceptable fits. 
Despite the success of leading ionization codes such as \texttt{Cloudy} \citep{Ferland2017} and \texttt{XSTAR} \citep{KallmanBautista}, 
photoionization grid models still lack self-consistency: 
the initial estimate of the illuminating unabsorbed continuum used to calculate the grid rarely coincides with 
the resulting continuum after fitting, and iterating this process until convergence is inefficient in more complex problems.
This issue is compounded when using multiple absorption zones, as electron scattering from inner absorbing zones can
significantly change the ionizing continuum incident on each successive zone.
If the absorbing wind column is large, then using the same grid model for each zone would be less physical than calculating separate grids for each zone. 
However, trying to converge individual continua for each grid is inefficient.

In this work, we performed photoionization analysis on all Chandra/HETG observations of 4U 1630$-$472 for which a wind can be confidently detected. We utilized the spectral analysis package \texttt{SPEX} \citep{Kaastra1996}
and modeled the wind absorption with \texttt{PION}, a self-consistent photoionized absorption model. Instead of assuming an input SED, \texttt{PION} calculates a new ionization balance as the source spectrum changes, allowing for simultaneous fitting of the absorber and the intrinsic source continuum. When using multiple \texttt{PION} components, each successive layer is illuminated by a different, successively more obscured ionizing continuum. This is an improvement over the pre-calculated photoionization grid models discussed earlier, as accounting for the reprocessing of the continuum incident on each layer yields more robust constraints on wind geometry. This level of self-consistency is required when testing alternative forms of wind driving:
If winds launched below the Compton radius are not magnetic, but instead are the pressure confined outer layer of a highly ionized thermal wind,
these two components should be nearly co-spatial yet illuminated by very different continua.
We modeled the wind with two absorption zones and included wind re-emission from the same absorbing gas layer.
Throughout this work we refer to these wind absorption plus emission zones as ``photoionization zones".

The black hole candidate 4U 1630-472 lies close to the Galactic center, at an estimated distance of 10 kpc \citep{Augusteijn2001}. 
This line of sight carries a very high ISM column density \citep[{} ${N}_{H,ISM}$ $\simeq$ $10^{23}$ ${\text{cm}}^{-2}$; e.g.][]{King2014}. 
Previous efforts have been unable constrain its mass given the difficulties in identifying an optical or IR counterpart. 
In this work, we assumed a mass of 10 ${M}_{\odot}$ based on work by \citet{Seifina2014}. 4U 1630-472 is likely viewed at a high inclination \citep[{$\theta$} $\sim$ $\ang{70}$; e.g.][]{Tomsick1998,Seifina2014}, in line with disk winds being largely equatorial \citep{Miller2006b,King2012,Ponti2012}.

The recurring disk wind in 4U 1630-472 has been detected in absorption numerous times. Photoionization analyses have been performed on NuSTAR and XMM-Newton spectra 
\citep[see ][]{King2014,Diaz2014,WangMendez}, yet a robust analysis of these winds require the superior energy resolution and absolute calibration of Chandra/HETGS. Of the six Chandra/HETG observations of 4U 1630$-$472 with strong evidence of wind absorption, ObsID 13715 was been the subject of detailed photoionization analysis. \citet{Miller2015} identified two distinct wind components in this spectrum with different outflow velocities, ionization, and column density; this was modeled with \texttt{XSTAR} photoionization grids and included wind re-emission.
Concurrent with this work, \citet{Gatuzz2019} analyzed these six observations using the photoionized absorption model "warmabs", an analytic implementation of the \texttt{XSTAR} code.
Their work only used one absorption zone for each observation and did not include wind re-emission.

%----------------------------------------------------------------------
\section{Observations and Data Reduction}\label{sec:data}
%-----------------------------------------------------------------------------------------------------------
\begin{table*}
\caption{Observation Details}
\vspace{-1.0\baselineskip}
\begin{footnotesize}
\begin{center}
\begin{tabular*}{\textwidth}{@{\extracolsep{\fill}}lccccccl}
\tableline
\\ [-3.0ex]
\tableline
ObsID& Obs Label & Duration & Count Rate & Start Date & Data Mode & Wind & Comments \\
  & & (${10}^{3}$ s) & (Avg.)& (YYYY/MM/DD) & &Absorption\\
\\ [-2.5ex]
\tableline
\\ [-2.5ex]

$4568 $ &I1& $49.99$ & 77.64 &$2004/08/04$ & CC & Yes & \scriptsize Absorption line at 7 keV\\
$13714$&S1 & $28.92$ & 66.84 &$2012/01/17$ & TE & Yes & \scriptsize Strong Fe XXV and Fe XXVI absorption\\
$13715$ &S2& $29.28$ & 65.82 &$2012/01/20$ & TE & Yes & --    \\
$13716$ &S3& $29.28$ & 62.71&$2012/01/26$ & TE & Yes & --\\
$13717$ & S4&$29.44$ & 69.56&$2012/01/30$ & TE & Yes & --\\
$19904$ & I2&$30.93$ & 68.95 &$2016/10/21$ & CC & Yes & --\\
\\ [-2.5ex]
\tableline 
\\ [-2.5ex]
$14441 $ & ... &$19.00$ & ... & 2012/06/03 & CC & No & \scriptsize Dips in HETG effective area, no real lines \\
$15511$ &... &$49.39$ & ... & 2013/04/25 & TE & No & --\\

\\ [-2.5ex]
\tableline 
\\ [-2.5ex]

$15524$ &... &$48.91$ & ... & 2013/05/27 & TE & No & Source in quiescence. \\

\\ [-2.5ex]
\tableline
\\ [-3.5ex]
\tableline

\end{tabular*}
\vspace*{-1.0\baselineskip}~\\ \end{center} 
\tablecomments{\footnotesize{Basic parameters for all nine Chandra/HETG observations of 4U 1630-472 are listed above. 
The columns list Chandra ObsID, exposure time, start date, ACIS data mode, whether there is significant evidence of wind absorption, and additional comments. See text for details.}}
\vspace{-1.0\baselineskip}
\end{footnotesize}
\vspace{1.0\baselineskip}
\label{tab:par}
\end{table*}
%------------------------------------------------------------------------------------

4U 1630-472 has been observed in outburst\footnote[1]{An additional observation (ObsID 15524) took place while 4U 1630-472 was in quiescence.} by {\it Chandra} with the High Energy Transmission Grating (or, HETG) on eight occasions, five of which show clear evidence of wind absorption (ObsID 13714, 13715, 13716, 13717, and 19904). A single feature near 7 keV, 
perhaps a weak Fe XXVI absorption line, 
is present in each remaining observations (ObsID 4568, 14441, and 15511), possibly indicating weak wind absorption. We found that the features in ObsID 14441 and 15511 are likely instrumental, and therefore we did not include these observations in our analysis (see Section \ref{sec:analysis}).

The data for all eight archival HETG observations of 4U~1630$-$472
considered in this work (including the two observations without
evidence of line absorption) were reduced using CIAO version 4.9 and
CALDB version 4.7.6.  While the bulk of data analysis was performed on
first-order HEG spectra, we also extracted third-order HEG spectra due
to its higher spectral resolution, at the cost of significantly lower
sensitivity.  Due to the lower resolution and collecting area in the Fe K band of
the MEG, this work makes use of the HEG exclusively.

Spectral files for HEG first and third orders were extracted from
level-2 event files using the routines \texttt{tg\_finzo},
\texttt{tg\_create\_mask}, \texttt{tg\_resolve\_events}, and
\texttt{tgextract}.  In order to reduce the contamination of dispersed
MEG photons overlapping with the HEG Fe K band, the
\texttt{tg\_create\_mask} parameter \texttt{width\_factor\_hetg}
 was set to a value of 10 (significantly lower that the default of 35),
resulting in a narrower extraction region for the HEG and thus
  allowing for better sensitivity at higher energies. The
  corresponding RMF and ARF response files for each separate order
  were created using the CIAO routines \texttt{mkgrmf} and
    \texttt{fullgarf}.

In order to increase the sensitivity of the HEG first and
third-order spectra, the {\texttt{combine\_grating\_spectra} routine is used to combine the plus and minus components for each order,
  therefore combining the HEG first-order spectra into a single
  spectrum, and the HEG third-order spectra into an independent single
  spectrum.  Combined RMF and ARF response files were also created
  using the \texttt{combine\_grating\_spectra} routine.

Of the eight observations considered in this work, three were made
with the ACIS-S array in ``continuous clocking'' (or, CC) mode
(ObsID 4568, 14441, and 19904), while the remaining five were made
using ``timed exposure'' (or, TE) mode (ObsID 13714, 13715, 13716,
13717, and 15511).  With the exception of ObsID 15511, a ``grey''
filter was applied to the zeroth-order of all observations, creating a
window (100 by 100 in TE-mode, 1024 by 100 in CC-mode) around the
zeroth-order order where only one in 10 or one in 20 events is
recorded. This prevents frames in the ACIS S3 chip from being dropped
in the telemetry stream if the zeroth order is too bright.  The photon
flux for this observation, however, is on par with the other
observations we are considering, and thus the zeroth-order suffers
from significant pile-up and frames were likely dropped from the telemetry
stream.  This results in a lower effective exposure for ObsID 15511.

\section{Analysis \& Results}\label{sec:analysis}

Spectral analysis for all observations was performed using SPEX version 3.03.00 and SPEXACT (SPEX Atomic Code and Tables) version 3.03.00. 
Fitting of wind parameters was primarily done through MCMC analysis, using \texttt{emcee} \citep{Foreman2013}, a python Markov chain Monte Carlo (MCMC) package. Fits of the underlying source continuum were obtained using the internal fitting routines in SPEX (see Section \ref{sec:rad_mdot}). This includes fitting the continuum normalization at each step of a chain during the MCMC phase, where the normalization is treated as a nuisance parameter (see Section \ref{sec:rad_mdot}). We used the $\chi^{2}$ fit statistic exclusively throughout our analysis, with the standard (data) weighting. The data did not require any additional binning. All errors reported are at the 1$\sigma$ level.

Unlike XSPEC, the plasma routines in SPEX require the use of physical dimensions, rather than ratios, when defining parameters such as normalization, meaning that a distance must be specified before spectral fitting. A distance of 10 kpc was assumed for all spectra based on work by \citet{Augusteijn2001}, and also for the purpose of convenient scaling

%------------------------------------------------------------------------------------------------------------------------------------------------------------------------------------------------------------
%------------------------------------------------------------------------------------------------------------------------------------------------------------------------------------------------------------
\begin{table*}
\caption{Continuum Parameters}
\vspace{-1.0\baselineskip}
\begin{footnotesize}
\begin{center}
\begin{tabular*}{\textwidth}{l  @{\extracolsep{\fill}}  cc  ccccc}
\tableline
\\ [-3.0ex]
\tableline
ObsID & Obs. Label & ${T}_{\text{i, Dbb}}$  & ${T}_{\text{max, Dbb}}$& ${T}_{\text{e, Comt}}$& ${T}_{\text{seed, Comt}}$ & ${\tau}_{\text{plasma}}$ & ${L}_{\text{Comt}}/{L}_{\text{Dbb}}$ \\
&&(keV)&(keV)&(keV)&(keV)&& (13.6 eV - 13.6 keV)\\
\\ [-3.0ex]
\tableline 
\\ [-3.0ex]
13714 & S1 &  3.03 $\pm$ 0.02 &  1.52 $\pm$ 0.01 &  ... & ... & ... & ...   \\
13715 & S2 & 2.96 $\pm$ 0.02  & 1.48 $\pm$ 0.01 &  ... & ... & ... & ...  \\
13716 & S3 & 2.92 $\pm$ 0.02 & 1.46 $\pm$ 0.01 &  ... & ... & ... & ...   \\
13717 & S4 & 3.13 $\pm$ 0.02 & 1.57 $\pm$ 0.01  &  ... & ... & ... & ...   \\
\\ [-3.0ex]
\tableline 
\\ [-3.0ex]
4568 & I1 & 2.26 $\pm$ 0.06 & 1.13 $\pm$ 0.03  & 50.0 (fixed) & $=$  ${T}_{max, disk}$   & 0.3 (fixed) & 0.25 \\
19904 & I2 & 2.35 $\pm$ 0.06  & 1.17 $\pm$ 0.03  & 50.0 (fixed) & $=$  ${T}_{max, disk}$   & 0.3 (fixed) & 0.09 \\

\\ [-3.0ex]
\tableline
\\ [-3.0ex]
\tableline
\end{tabular*}
\vspace*{-1.0\baselineskip}~\\ \end{center} 
\tablecomments{\footnotesize The parameters of the best-fit continuum model for each observation of 4U 1630$-$472.  All errors are $1\sigma$.
Observations are modeled either as a thermal disk only (``Dbb'' in SPEX), and as a disk blackbody plus Comptonization (``Comt'' in SPEX), as required. 
Interstellar absorption was modeled with ``Absm'', with ${N}_{H}$ fixed at $9.17 \times {10}^{22} {cm}^{-2}$. 
We report both the nominal disk temperature (${T}_{i}$, free parameter in ``Dbb'' model) 
as well as the maximum temperature of the disk, ${T}_{max}$, which is more representative of the average disk photon energy.
Due to difficulties in constraining the parameters of the thermal Comptonization component, 
I1 and I2 were modeled with a fixed plasma optical depth and electron temperature at assumed values of $\tau$ = 0.3 and ${T}_{\text{e, Comt}} = $ 50 keV.
Please see the text for additional details.}
\end{footnotesize}
\label{tab:con}
\end{table*}

%------------------------------------------------------------------------------------------------------------------------------------------------------------------------------------------------------------
%------------------------------------------------------------------------------------------------------------------------------------------------------------------------------------------------------------

Our initial efforts at modeling the wind absorption via the built-in fitting routines in SPEX revealed major issues with this approach. 
First was the complexity of the eight-dimensional parameter space: Broadband fits (3-10 keV range) were generally unsatisfactory (particularly in the Fe K band) and yielded poorly constrained parameters.
In contrast, fits over the 5-10 keV range generally resulted in over-predicting of the dominant Fe XXV and XXVI lines while failing to capture lines at other energies (Fe XXV and XXVI $\beta$ lines at higher energies, and Ar XVIII and Ca XX lines at lower energies). 
These issues were compounded by long computation times, as implementing the full plasma physics and re-emission in SPEX comes at a considerable computational cost.

Markov chain Monte Carlo either addressed or eliminated most of these issues.
In addition to sampling the parameter space more efficiently and allowing for dynamic parameter ranges, implementing MCMC allowed us to treat the continuum normalization as a nuisance parameter: 
at each step of a chain, a best-fit normalization value was obtained (via the built-in fitting routines) before implementing the full SPEX plasma physics needed to fit the line absorption separately. 
This treatment vastly reduced the computation time of analysis by reducing the number of free parameters. 
It also allowed for the use of separate specialized fitting ranges for the broadband continuum and wind absorption. 
For more detail see Section \ref{sec:cont}.
A fitting range of 3-10 keV minus small portions corresponding to the strongest absorption lines was used when fitting the continuum. 
When fitting the wind absorption, a segmented range consisting of 6.5 to 7.2 keV (Fe XXV and XXVI), 7.7 to 8.7 keV (Ni and Fe $\beta$), plus 4.08 to 4.13 keV (Ca XX), was used instead.
This choice of fitting range ensures that the ${\chi}^{2}$ used when fitting wind parameters is mostly determined by how well it models the line profiles relative to the continuum, rather than the quality of the continuum fit. 
Again, this is only possible because, at each point of parameter space, a best fit continuum is obtained before calculating a ${\chi}^{2}$ for the absorption lines. 

The superior effective area of the MEG near 1 keV could allow us to fit additional absorption lines (such as Fe XXIV) at energies where the HEG spectrum becomes too noisy. 
We found that, between high galactic absorbing column combined with the loss of sensitivity of the ACIS detector at lower energies, the MEG contains no useful flux near 1 keV.
We limited our analysis the HEG first-order. 

Of the eight Chandra/HETG spectra of 4U 1630-472 in outburst, only six displayed blueshifted absorption lines at a confidence above the $3\sigma$ level.
In the case of both ObsID 14441 and 15511, possible absorption features coincide with large and narrow drops in the HETG/ACIS effective area, and were ruled out as non-detections.
The feature in ObsID 4568 can be detected above the $3\sigma$ level while avoiding any narrow dips in effective area by 0.1$-$0.2 keV. 

Unsurprisingly, the detection of a wind in an observation matches its location in the MAXI hardness-intensity diagram (Figure \ref{fig:HR}):
observations with the strongest wind absorption (e.g. ObsID 13716 and 13717, blue and cyan) are in a high-soft state.
Despite the comparatively weaker absorption lines and the presence of an additional non-disk component in the continuum of ObsID 19904 (magenta),
the system appears to be in a comparable accretion state as the high-soft state (when the strongest winds are detected).
The lack of winds in ObsID 15511 (yellow) and 14441 (red) is consistent with the disappearance of winds during spectrally hard states, 
the former occurring as the system transitioned to low-hard state (as in ObsID 15524 $\sim32$ days later, in green), while the latter as 
the system transitioned from a luminous hard state to a low-hard state. For a more detailed discussion, see \cite{Neilsen}.

The six spectra considered in this work is divided into two distinct groups. 
The first comprised of the four consecutive observations that occurred during a relatively flat phase of the same outburst (ObsID. 13714, 13715, 13716, and 13716). 
These spectra display disk-dominated continua and the strongest absorption lines. 
The spectra in second group (ObsID. 4568 and 19904) display significantly weaker absorption lines and their continua cannot be described by a disk blackbody alone. For simplicity, we refer to these groups as either soft-state or intermediate-state observations, while using S1-S4 (ObsID. 13714, 13715, 13716, and 13716) and I1-I2 (ObsID. 4568 and 19904) when referring to individual observations (see Table \ref{tab:par}).

\subsection{Continuum Fits}\label{sec:cont}

%------------------------------------------------------------------------------------------------------------------------------------------------------------------------------------------------------------
%------------------------------------------------------------------------------------------------------------------------------------------------------------------------------------------------------------
\begin{figure}
\centering
\includegraphics[width=0.48\textwidth]{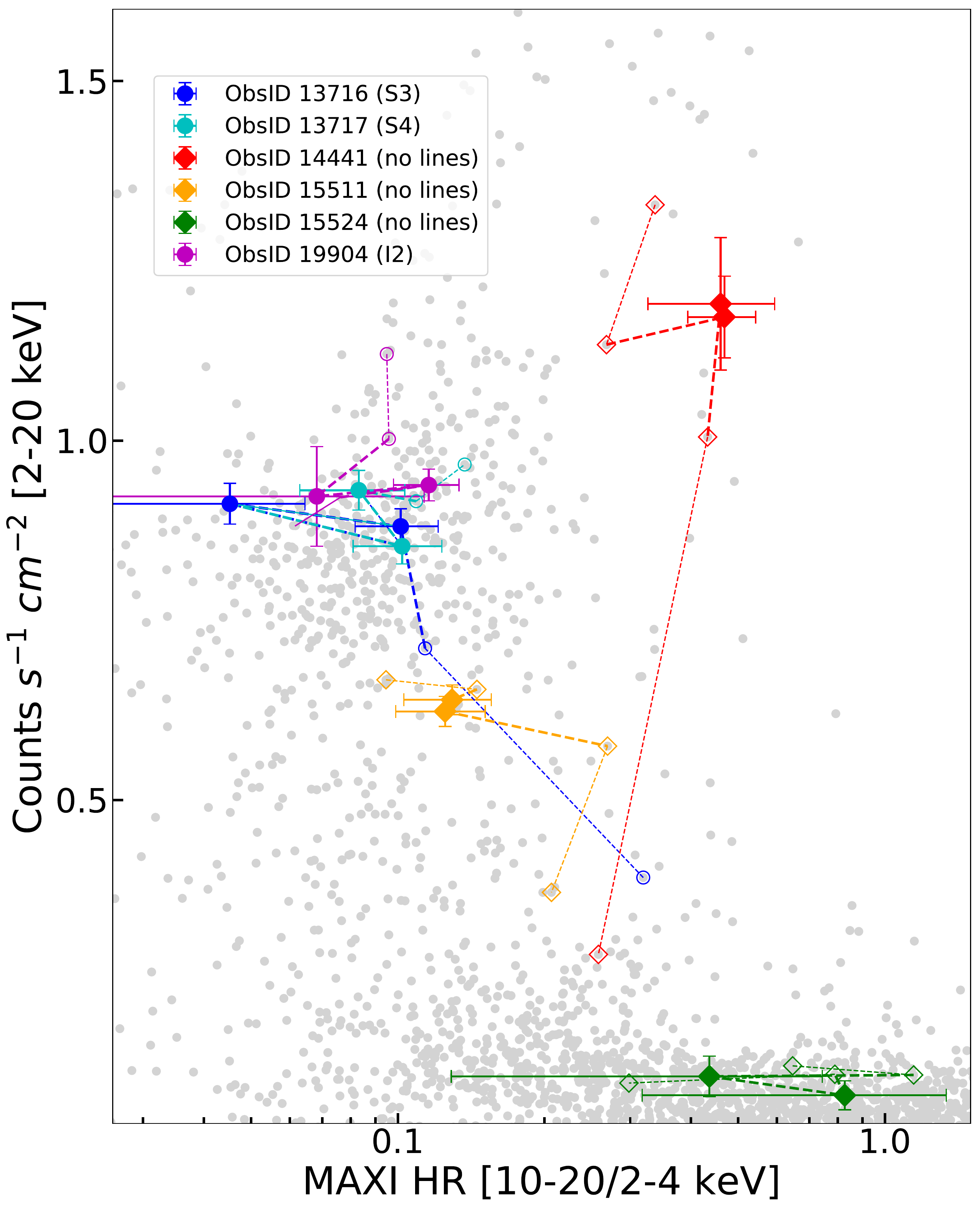}
  \figcaption[t]{\footnotesize MAXI hardness-intensity diagram for 4U 1630$-$472. 
  Filled markers represent the two MAXI data points closest to the time at which each Chandra observation occurred, each color coded by ObsID. 
  Empty markers connected by dashed lines plot additional MAXI points contemporaneous with each Chandra observation, and are meant to roughly contextualize the spectral state at the time of observation. 
    \label{fig:HR}}
\end{figure}

%------------------------------------------------------------------------------------------------------------------------------------------------------------------------------------------------------------

In addition to strong absorption lines, photoionized gases produce continuum absorption through various processes. 
At lower ionizations, the opacity is dominated by bound-free transitions and the attenuation of the continuum is stronger at lower photon energies.
Fitting the shape of the underlying continuum in a source obscured by a gas of lower ionization would depend strongly on the ionization and absorbing column of the gas.
At higher ionizations (above log~$\xi \sim 3.0$), electron scattering becomes the primary source of opacity and the attenuation is largely independent of photon energy. 
We did not find any of the strong lines that would suggest significant absorption of a gas below log~$\xi \sim 3.0$ in our spectra of 4U 1630-472, 
meaning that the majority of the observed absorption lines originate in wind layers with log~$\xi > 3.0$.
During the fitting process, we assumed that any attenuation of the continuum by the absorbing wind is be due to electron scattering (and therefore 
mostly act as a grey absorber). 
This allowed us to fit the underlying continuum shape before introducing wind absorption. 
Once wind absorption is implemented, the additive continuum components would then only require a shift in normalization to compensate for the attenuation, reducing the number of free parameters. 

All HEG first-order spectra were modeled in SPEX with a phenomenological 
multi-temperature disk blackbody model (``Dbb'') plus, in the case of I1 and I2, an optically thin thermal Comptonization model (``Comt''), modified by interstellar absorption (``Absm'').
We found that replacing ``Comt'' with an unbroken power-law model resulted in complete ionization of the absorbing gas due to additional heating via free-free absorption of low energy photons. 
Although SPEX defines the ionization parameter using a 1-1000 Rydberg flux range, the entire SED is utilized when calculating the ionization balance (including heating processes). 
On the other hand, the ionization balance is largely insensitive to X-ray photons above 13.6 keV.
Given the relatively low luminosity of the powerlaw component, there was no noticeable change in ${\chi}^{2}$ regardless of whether a high energy cutoff is present. 
This insensitivity to hard X-ray photons is also true when using  ``Comt'': there with no significant change in ${\chi}^{2}$ regardless of the specific model parameters provided that the continuum in the Chandra energy band is fit properly. This is not to say that hard X-rays are irrelevant in this scenario, as they are known to affect both the thermal stability and Compton temperature of the gas \citep{Chakravorty2013,HigginbottomProga,Bianchi2017}. However, the lack of simultaneous observations with facilities such as NuSTAR makes it difficult to properly explore these effects.

In principle, the flux responsible for setting the photoionization balance is almost entirely encapsulated between 13.6 eV and 13.6 keV (as per the definition of the ionization parameter in SPEX).
Although the contribution by photons below 13.6 eV may ultimately be important, an unbroken powerlaw is likely a poor description of the flux at these energies. 
Given that I2 occupies the same space in the hardness-intensity diagram as the soft-state observations (Figure \ref{fig:HR}), we chose to model this component with a physically motivated Comptonization model, with some parameters fixed at canonical values \citep[see][]{Tomsick2005}, as it is likely a more realistic description of the flux below the energy range available to us.

Table \ref{tab:con} lists the best-fit continuum parameters.
Given the agreement in column density of the neutral absorber when each spectrum was fit separately, the fits listed in Table \ref{tab:con} were obtained with ${N}_{H}$ fixed at the weighted average of $9.17 \times {10}^{22}$ ${\text{cm}}^{-2}$. 
The continua in S1 to S4 (occurring within 13 days of the same outburst) are well described solely with disk blackbody model with well constrained parameters. 
SPEX's ``Dbb'' includes the torque-free condition at the inner boundary of the disk, where ${T}_{i}$ is the fitting parameter (${T}_{max} \simeq 0.49 {T}_{i}$ roughly corresponds to the temperature in ``diskbb'').

Continuum parameters in I1 and I2, primarily those in ``Comt'', were much harder to constrain in the limited energy range of Chandra and (at low energies) given the high ISM column. 
After coupling the seed photon energy (${T}_{0}$) to ${T}_{i}$ in ``Dbb'' by a factor of 0.49, 
electron temperature (${T}_{1}$) and optical depth ($\tau$), in particular, remained highly degenerate.
We adopted fixed values of ${T}_{1}$ =  50 keV and $\tau$ = 0.3 \citep[resulting in a photon index of $\Gamma\sim 2.5$, see][]{Titarchuk1994} and fit the continua with ``Dbb'' and ``Comt'' normalizations, as well as coupled disk and plasma seed temperatures, as free parameters.
In section \ref{sec:PIA}, ``Comt" normalization is coupled to disk normalization by the same relative factor in the best-fit model.

``Absm'' models the transmission of the neutral gas in the ISM with fixed \citet{MorrisonMcCammon} abundances. This model has drawbacks-- most importantly, fixed abundances and imperfect location of absorption edges. 
The SPEX user manual recommends using the collisional ionization equilibrium slab model ``Hot'' (fixed at a low temperature) if more higher precision is required when modeling these features. This resulted in a noticeable shift in the location of some absorption edges, but with negligible change in $\chi^{2}$ and similar best-fit ${N}_{H,ISM}$. 
We opted for ``Absm'' given the small impact of absorption edges in our analysis. 

\subsection{Photoionization analysis}\label{sec:PIA}
\label{sec:phot}

Compared to line-by-line fitting, photoionized absorption grid models (including XSTAR and Cloudy) 
are a vastly superior tool for characterizing the physical properties of an absorbing gas, 
but do not achieve full self-consistency.
In these models, the ionization balance of the absorbing gas is pre-calculated by assuming the shape and luminosity of photoionizing continuum (i.e. the naked source continuum) before spectral fitting of the combined absorption and continuum models. 
After importing and fitting this pre-calculated absorber, the resulting best-fit continuum may diverge significantly from the assumed continuum (initially used to create the grid model) if the optical depth is high enough. 
This mismatch can become problematic when using multiple absorbers, where the ionizing flux from the central engine is reprocessed repeatedly as it passes through each successive absorption layer.  
The ionization balance in a particular photoionization zone is therefore dictated by this new incident flux, 
reprocessed by the absorbers located between the central engine and the zone in question, and not the naked source continuum.
As the optical depth increases, using the same pre-calculated grid to model multiple absorption layers would lack some self-consistency.
In order to address these concerns, we modeled the photoionized wind absorption in our spectra with PION, a self-consistent PIA model within SPEX.

SPEX requires the user to define a geometry, where a source continuum is first chosen from standard additive components and, most importantly, where the order in which multiplicative components reprocess the flux of the additive components is specified. When PION is included as an absorber, this same reprocessed flux is what SPEX utilizes to calculate the ionization balance of the absorber. For a given geometry, SPEX can fit the continuum and PIA simultaneously using its internal plasma routines. PION also calculates re-emission from the same plasma, where an emission covering factor (as well as the fraction of backwards/forwards emission) can be specified. This acts as an additional additive component.

For each PION component in our analysis, we set fixed values of hydrogen number density at ${n}_{H}$ = ${10}^{14}$ ${\text{cm}}^{-3}$ and turbulent velocity of ${v}_{turb}$ = 400 km/s \citep{Miller2015}. It is important to note that, given the parameter regime, energy range, and resolution of the data, changing ${n}_{H}$ by several orders of magnitudes in either direction has no observable effect on the model and produces no change in $\chi^{2}$. The ${n}_{H}$ values derived in Section \ref{sec:rad_mdot} were not obtained through fitting. 
The emission covering factor ($\Omega/4\pi$) determines the normalization of the re-emission component, which is calculated internally. 
Given our limited understanding of wind geometry, we assumed a fixed value of $\Omega/4\pi$ = 0.5 \citep{Miller2015}.
We performed an additional test fit of S3 with a significantly lower $\Omega/4\pi$ = 0.2 to test the validity of this assumption.
The mix parameter in PION allows you to specify the geometry of the emitter. A value of mix = 1 would result in only forwards emission (a lamp-post geometry where your X-ray source would be behind a slab),
while mix = 0 would result in all backwards emission (where the slab is behind the X-ray source).
We assumed that we observe roughly equal amounts of re-emission from forwards and backwards portions of an axially symmetric wind, and set a fixed value of mix = 0.5. 

In the analysis by \citet{Miller2015}, the complexity and asymmetry of Fe XXV and Fe XXVI lines in first-order HETG spectra of S2 strongly suggested separate wind components with different outflow velocities and ionization, in agreement with individual lines found when examining higher resolution third-order HETG spectra. Using Gaussians, we find a similar trend in outflow velocity and relative ionization between photoionization zones in S1-S4 and I2 as Miller et al. (2015) did in S2, requiring the use of at least two distinct photoionization zones. However, we still performed single-zone fits to select observations for comparison to our two-zone models (see Section \ref{sec:soft}). 

%------------------------------------------------------------------------------------------------------------------------------------------------------------------------------------------------------------
%------------------------------------------------------------------------------------------------------------------------------------------------------------------------------------------------------------
\begin{table*}
\caption{Parameters for Best Fit Wind Model}
\vspace{-1.0\baselineskip}
\begin{footnotesize}
\begin{center}
\begin{tabular*}{\textwidth}{l c c @{\extracolsep{\fill}}  c cc c c c }
\tableline
\\ [-3.0ex]
\tableline

Obs. Label  && Zone & ${N}_{H}$  & log ($\xi$) & ${v}_{abs}$  & ${\sigma}_{emis}$  &  ${L}_{illum}$  &  $\chi^{2}/\nu$ \\  
&&&  (${10}^{22} {\text{cm}}^{-2}$) && (km/s) &(km/s) & (${10}^{38}$ erg/s) \\

\\ [-3.0ex]
\tableline 
\\ [-3.0ex]
 
S1 & & 1 &$54^{+{15}^{\dagger}}_{-17}$ & $5.25^{+0.12}_{-0.16}$ & $-1000^{+190}_{-180}$ & $15000^{*}_{-3500}$ & $2.93^{+0.30}_{-0.22}$& 159/155 = 1.03 \\
& &2 & $14.4^{+3.8}_{-3.0}$ & $4.02 \pm 0.10$ & $-160 \pm 80 $ & $1000 \pm 200$ & $1.89^{+0.05}_{-0.06}$&  \\

\\ [-3.0ex]
\tableline 
\\ [-3.0ex]

S2 &  &1 & $43.5^{+14}_{-17}$ & $5.40^{+0.11}_{-0.19}$ & $-1010^{+280}_{-320}$ & $15000^{*}_{-4200}$ & $2.68^{+0.29}_{-0.21}$& 139/155 = 0.90\\
  	  &  &2 &$17.6^{+3.6}_{-3.4}$ & $3.90^{+0.07}_{-0.07}$ & $-210^{+80}_{-80}$ & $900^{+200}_{-100}$ & $1.91^{+0.06}_{-0.06}$\\
	  
\\ [-3.0ex]
\tableline 
\\ [-3.0ex]

S3  & &1 & $38.4^{+11}_{-10}$ & $4.95 \pm 0.10 $ & $-600^{+120}_{-90}$ & $15000^{*}_{-3100}$ & $2.35^{+0.19}_{-0.16}$ &  156.07/155 = 1.01\\
 	& &2 & $12.4^{+1.7}_{-1.4}$ & $3.39^{+0.09}_{-0.07}$ & $-120^{+{50}^{\dagger}}_{-70}$ & $1000^{+300}_{-200}$ & $1.77^{+0.02}_{-0.02}$ \\

\\ [-3.0ex]
\tableline 
\\ [-3.0ex]
 
 S4  &  & 1 &$ 56^{+{16}^{\dagger}}_{-17}$ & $5.21^{+0.14}_{-0.12}$ & $-890^{+180}_{-140}$ & $15000^{*}_{-3100}$ & $3.19^{+0.39}_{-0.16}$&  165.4/155 = 1.07 \\
  & &2 & $21.3 \pm 3.7$ & $3.86 \pm 0.06 $ & $-240 \pm 70 $ & $ 900 \pm 100 $ & $1.92^{+0.05}_{-0.07}$& \\
  
\\ [-3.0ex]
\tableline 
\\ [-3.0ex]
\tableline 
\\ [-3.0ex]

I1 &&1 & $23.7^{+15.4}_{-6.0}$ & $5.41^{+0.18}_{-0.27}$ & $-420^{+210}_{-280}$ & $13200^{+{1800}^{\dagger}}_{-3400}$ & $3.11^{+0.25}_{-0.11}$ & 129.27/130 = 0.99 \\
\\ [-3.0ex]
\tableline 
\\ [-3.0ex]

I2 & &1& $3.5^{+2.4}_{-1.8}$ & $4.51^{+0.17}_{-0.07}$ & $-1260^{+450}_{-430}$ & $15000^{*}_{-4300}$ & $2.92^{+0.05}_{-0.02}$& 167.4/155 = 1.08 \\
  && 2 & $7.7^{+2.2}_{-2.4}$ & $4.38(6) $ & $-240^{+130}_{-20}$ & $7100 \pm 3500 $ & $2.87^{+0.03}_{-0.05}$&  \\

\\ [-2.0ex]
\tableline 
\\ [-3.0ex]
\tableline

\end{tabular*}
\vspace*{-1.0\baselineskip}~\\ \end{center} 
\tablecomments{\footnotesize{The table above lists the best-fit wind model for each observation, grouped by spectral state. 
Quoted errors are at the $1\sigma$ level and were obtained empirically
from MCMC analysis.
Parameter values are listed for both inner and outer photoionization zones, listed as Zone 1 and 2, respectively. 
Errors marked with * indicate that a parameter is unconstrained within $1\sigma$ in that direction 
(as is the case for the upper limit of ${\sigma}_{\text{emis}}$ in Zone 1).
Errors annotated with $\dagger$ indicate a parameter with unconstrained behavior at more than $1\sigma$ away, 
as in the case of ${N}_{H}$ in Zone 1 of S1 and S4 (see Figures~\ref{fig:mc1}$~\&~$\ref{fig:mc2}).
}}
\end{footnotesize}
\label{tab:fit}
\end{table*}
%------------------------------------------------------------------------------------------------------------------------------------------------------------------------------------------------------------
%------------------------------------------------------------------------------------------------------------------------------------------------------------------------------------------------------------

For each observation, our model was constructed in the following manner: 
The additive components of the naked continuum are first reprocessed by two successive photoionization zones, 
such that the flux incident on the outer photoionization zone (Zone 2) is the absorbed source continuum after reprocessing by the inner zone (Zone 1). 
Wind re-emission (an additive component in PION) for both zones were each modified with ``Vgau'', a Gaussian velocity-broadening model, 
to model the dynamical broadening due to Keplerian motion of the orbiting gas.  
Finally the entire model was then modified by ``Absm'', with ${N}_{H}$ fixed at $9.17 \times {10}^{22} {\text{cm}}^{-2}$. 
With only one absorption feature in Observation 1, the model was constructed with a single photoionization zone.
This velocity broadening is applied only to the re-emission component: 
the pencil-beam geometry of the absorber relative to the emitting region of the inner disk means that the 
orbital motion of the absorber is almost entirely perpendicular to our line-of-sight.
Instead, the broadening of the absorber could arise due to turbulent motion in the gas, or perhaps as a result of
velocity shearing due to large changes in orbital velocity within a single gas layer. 
We account for these effects using the ${v}_{turb}$ parameter.

For each wind zone, we fit four free parameters: The equivalent neutral hydrogen column density (${N}_{H}$), the ionization parameter ($\xi$), the radial velocity (${v}_{\text{abs}}$), and 
the velocity broadening (${\sigma}_{\text{emis}}$). The first three parameters dictate the gas properties of both absorption and re-emission for that zone. 
The continuum normalizations (either just ${K}_{disk}$ or the coupled ${K}_{\text{disk}}$ + ${K}_{\text{Comt}}$) are free, 
but are treated as nuisance parameters in our analysis (see Appendix \ref{sec:MCMC}).

In principle, the systematic radial velocity of re-emission in an axially symmetric wind should be zero. 
Currently, PION does not allow for separate absorption and emission velocities, requiring two components in order to model each wind zone.
After several experiments, we found that fits with separate velocities (e.g. ${\chi}^{2}/\nu$ = 161/155) yielded nearly identical results to fits with a single velocity (e.g. ${\chi}^{2}/\nu$ = 159/155).
It is important to note that the data still require re-emission, as the observed ratio of Fe K$\alpha$ and K$\beta$ lines cannot be achieved with absorption lines alone.
While the model is still sensitive to the degree of broadening of re-emission, in this particular case it is largely insensitive to the systematic velocity of the emitter given the absence of strong P-Cygni profiles. In soft-state observations, the systematic velocity of the absorber was either too small compared to the absorption line width (${v}_{\text{abs}} \sim 150$ km/s and ${v}_{turb}$ $\sim 400$ km/s, Zone 2) or too small compared to the broadening of the re-emission (${v}_{\text{abs}} \sim 1000$ km/s and ${\sigma}_{\text{emis}}$ $\sim 15000$ km/s, Zone 1). 
For Zone 2 in particular, the outflow velocity is small enough that the combined emission-plus-absorption line profile is largely unaffected
regardless of whether the emission line is centered at ${v} = 0$ (6.700 keV) or at ${v}={v}_{\text{abs}} \sim -150 km/s$ (6.704 keV),
but is sensitive to how much flux from the broad emission line lies within the core of the absorption line, which is primarily controlled by the dynamical broadening of the re-emission.

%------------------------------------------------------------------------------------------------------------------------------------------------------------------------------------------------------------
%------------------------------------------------------------------------------------------------------------------------------------------------------------------------------------------------------------
\begin{figure*}
\centering
\subfloat{\includegraphics[width=0.49\textwidth]{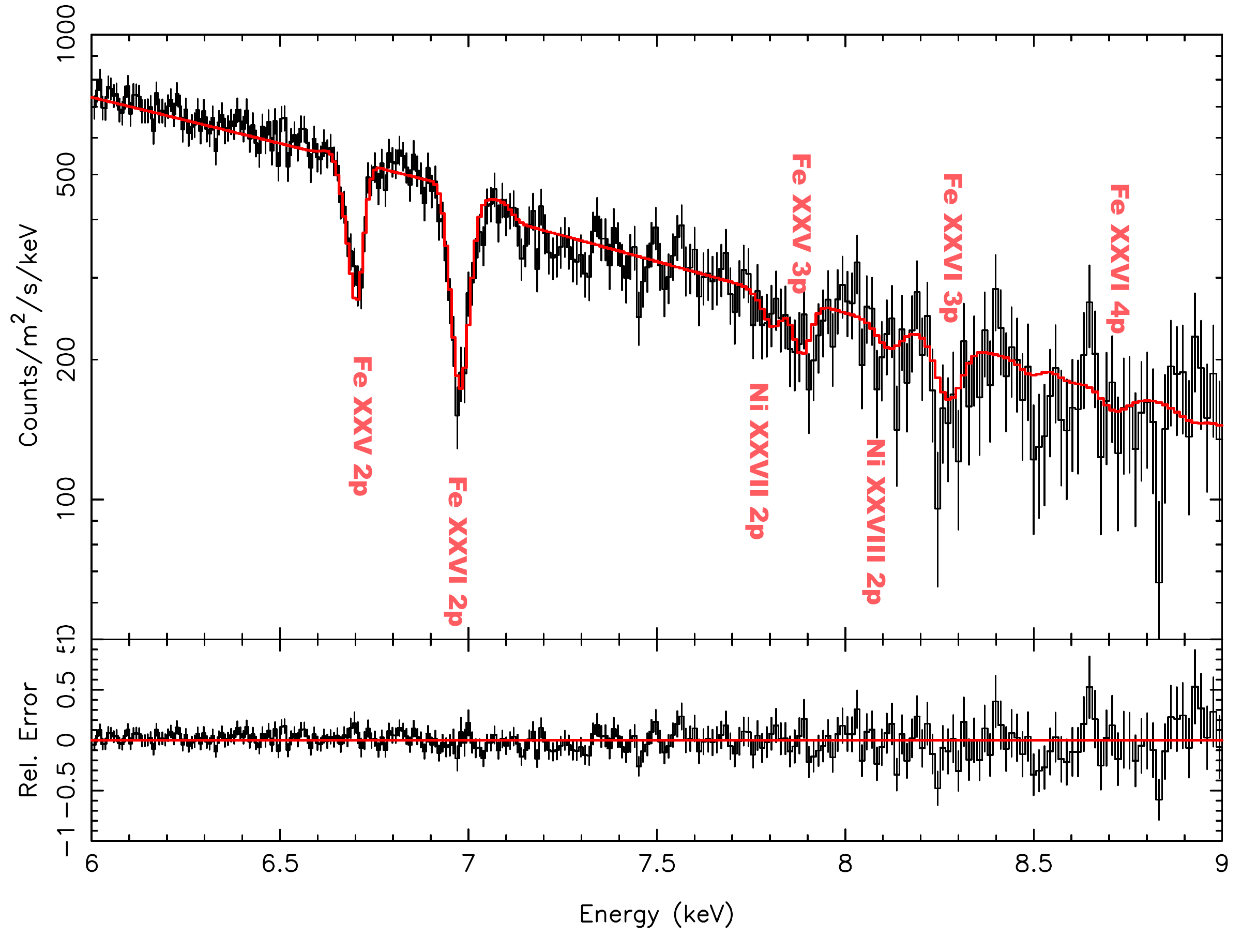}}
\subfloat{\includegraphics[width=0.49\textwidth]{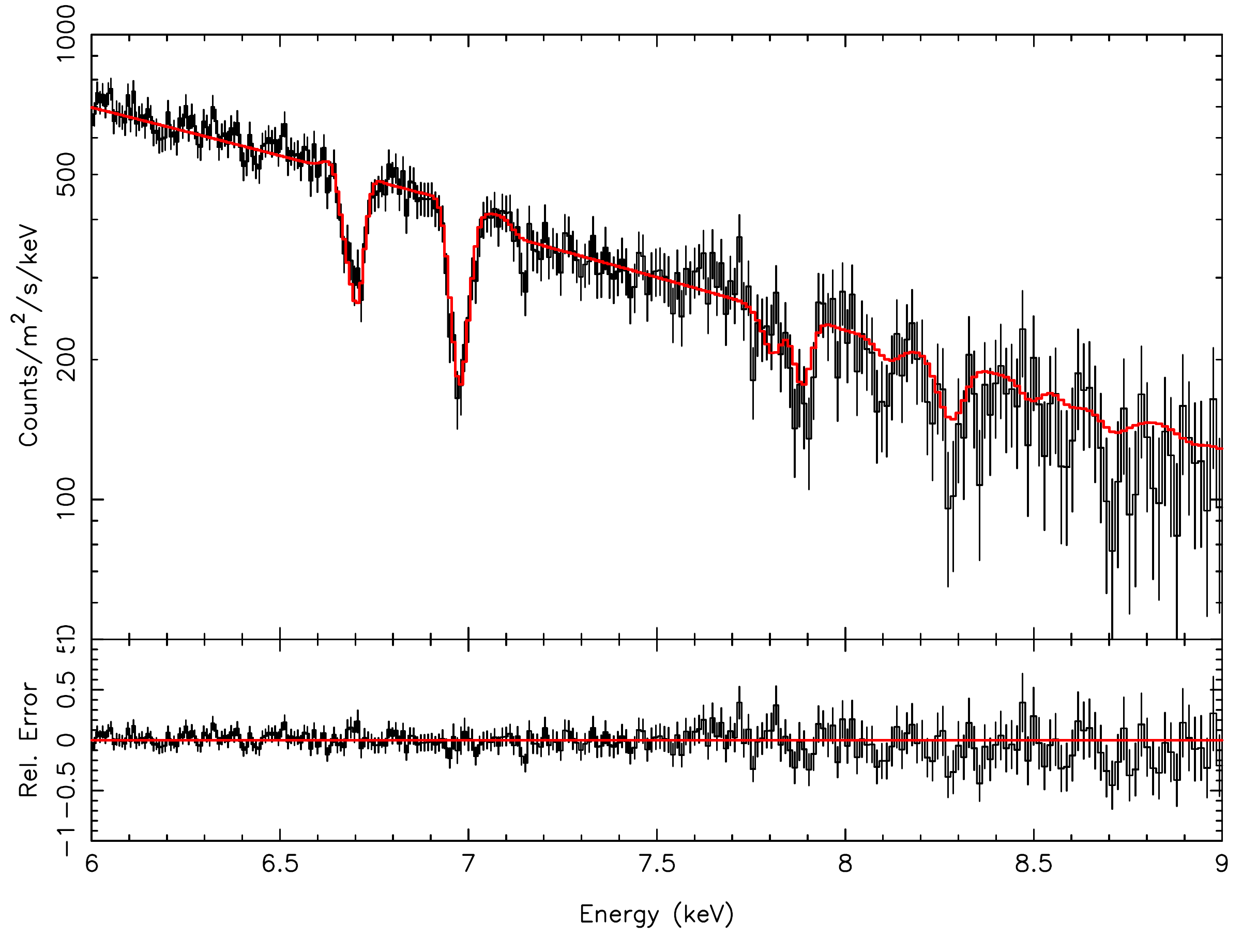}}\\
\subfloat{\includegraphics[width=0.49\textwidth]{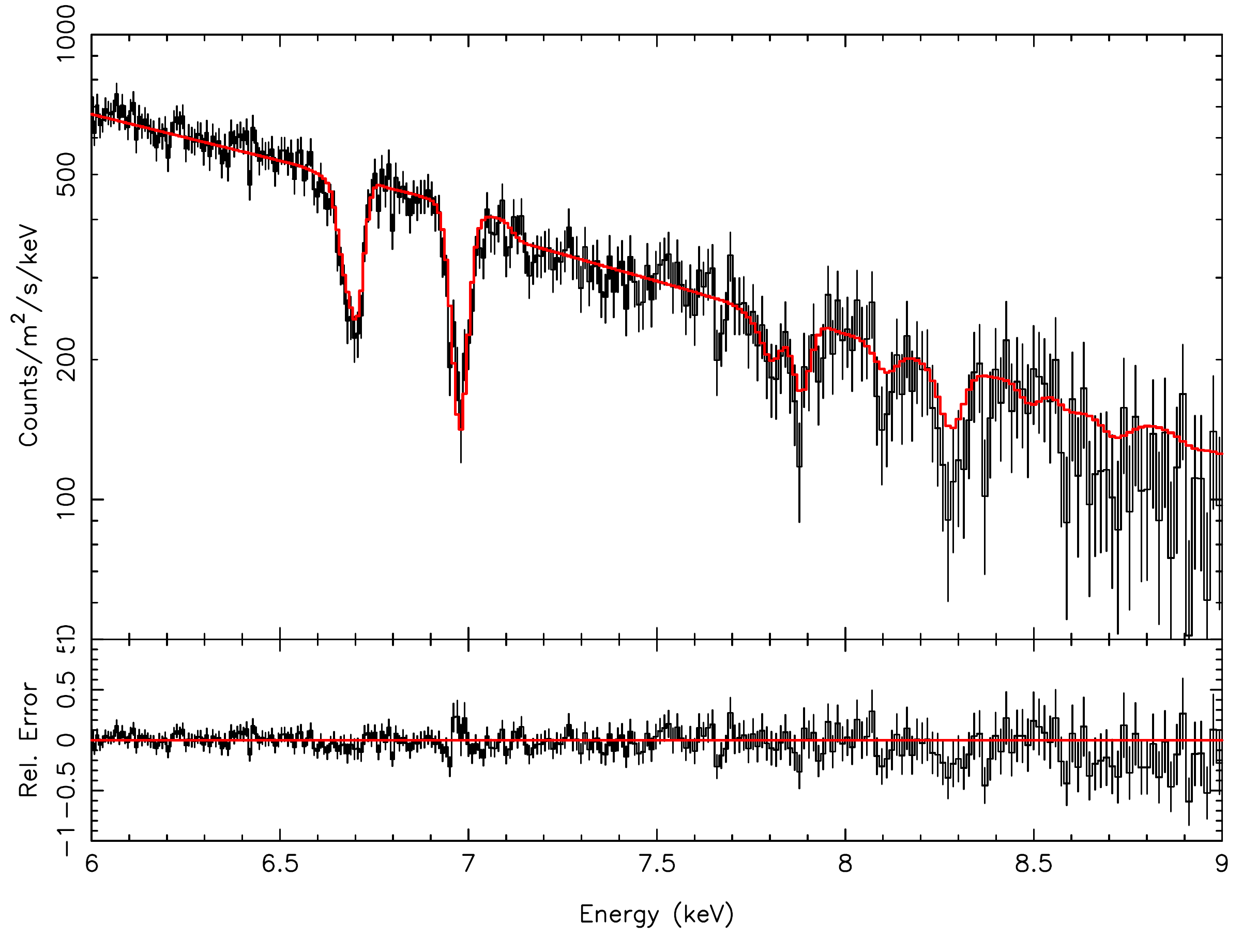}}
\subfloat{\includegraphics[width=0.49\textwidth]{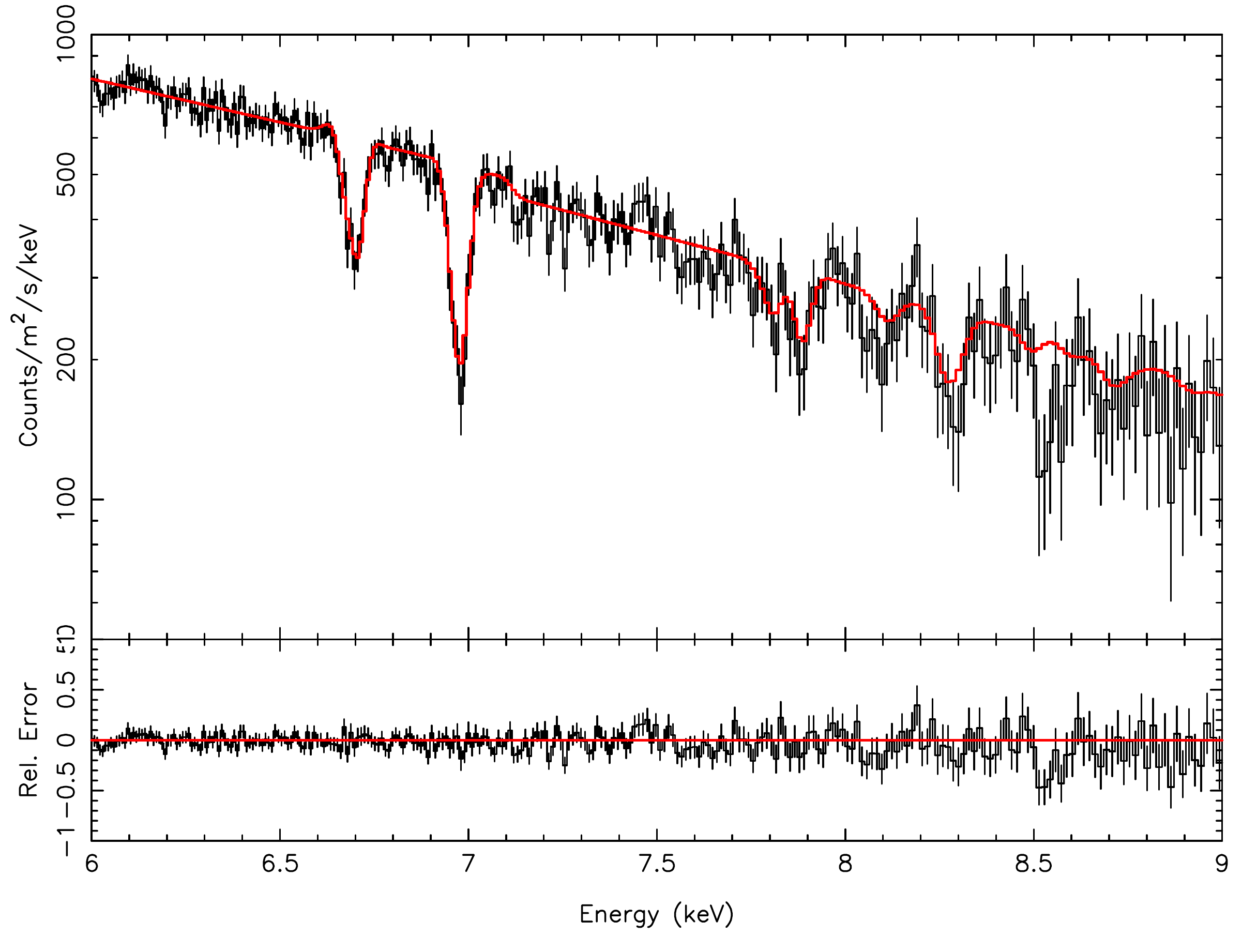}}
  \figcaption[t]{\footnotesize The first-order HEG spectrum of observations S1 (ObsID. 13714, top-left),  S2 (ObsID. 13715, top-right), S3 (ObsID. 13716, bottom-left), and S4 (ObsID. 13717, bottom-right)
  of 4U~1630$-$472, fit
    with the two photoionization zone models listed in Table \ref{tab:fit}. 
    The data require a dynamically broadened emission component for each corresponding photoionization zone
    in order to achieve the $\alpha$/$\beta$ line ratio for Fe XXV and Fe XXVI.
    Please see the text and Table \ref{tab:fit} for additional details.
    There is feature in the spectrum of S4 (bottom-right) at $\sim 8.5$ keV that coincides with a discrete drop in the HEG effective area. 
    This feature has a significance of less than 2$\sigma$ (via Gaussian fitting) and therefore may be instrumental.
    \label{fig:14}}

\end{figure*}
%------------------------------------------------------------------------------------------------------------------------------------------------------------------------------------------------------------
%------------------------------------------------------------------------------------------------------------------------------------------------------------------------------------------------------------

We initially constrained the wind equivalent hydrogen column density to ${10}^{22}\leq{N}_{H}\leq{10}^{24}$ in ${\text{cm}}^{-2}$, 
the upper bound corresponding to Compton-thick winds. The ionization parameter was restricted to $3.0\leq$ log~$\xi \leq 6.0$, 
although these bounds were tightened as minima where found. Winds require a net outflow velocity, so we constrained ${v}_{\text{abs}}\leq 0$ km/s.
Finally, the velocity broadening was constrained between $500\leq$ ${\sigma}_{\text{emis}}$ $\leq 15000$ in km/s, 
therefore constraining orbital radii to ${R}_{\text{orbital}} \geq 400~GM/{c}^{2}$.

\subsection{Fits}\label{sec:fits}

Results of our MCMC analysis of the photoionized absorption in both soft- and intermediate-state observations, including best-fit parameter values with 1$\sigma$ errors and $\chi^{2}$ values, are detailed in Table \ref{tab:fit}.
The luminosities listed correspond to the illuminating luminosity incident on a specific wind layer, which in the case of Zone 1 corresponds to the projection of the intrinsic luminosity of the disk
at our viewing angle (more detail in Section \ref{sec:rad_mdot}). The best-fit models for each observation are shown in Figures \ref{fig:14}
and \ref{fig:19s}, while corner plots of parameter posterior distributions are show in Figures~\ref{fig:mc1}$-$\ref{fig:mc2}.
Figure~\ref{fig:py_spec} shows the contribution of the dynamically broadened re-emission to the line depths of Fe XXV and XXVI.

\subsubsection{Soft State Observations}\label{sec:soft}

Modeling of S1-S4 resulted very good statistical fits: $\chi^{2}/\nu$ values range from 139/155 = 0.90 to 165.4/155 = 1.07. 
The models also do a good job fitting the lines in the 7.5 to 9 keV region, as can be seen in Figure \ref{fig:14}.
Figure \ref{fig:cal} shows that our best-fit models also do a good job describing the lines in the 3 to 5 keV range, despite including only a few bins of this range during fitting.
This demonstrates the strength of our specialized fitting range: 
Although satisfactory $\chi^{2}/\nu$ values can be obtained by fitting over the 6-10 keV range, 
$\chi^{2}$ is dominated by the prominent Fe XXV and XXVI $\alpha$ lines. 
The resulting fits failed to capture the 7.5-9 keV (Fe K$\beta$ and Ni XVIII) and the 3-5 keV energy bands. 
Our approach of anchoring the fit to a small portion of the low energy spectrum achieved the right balance between 
the Fe K and the 3-5 keV energy bands. 
By not allowing either region to dominate, we obtain good fits to the Fe K band while still capturing the 7.5-9 keV and the 3-5 keV energy bands.

%------------------------------------------------------------------------------------------------------------------------------------------------------------------------------------------------------------
%------------------------------------------------------------------------------------------------------------------------------------------------------------------------------------------------------------
\begin{figure}

\centering
\subfloat{\includegraphics[scale=0.34,angle=0]{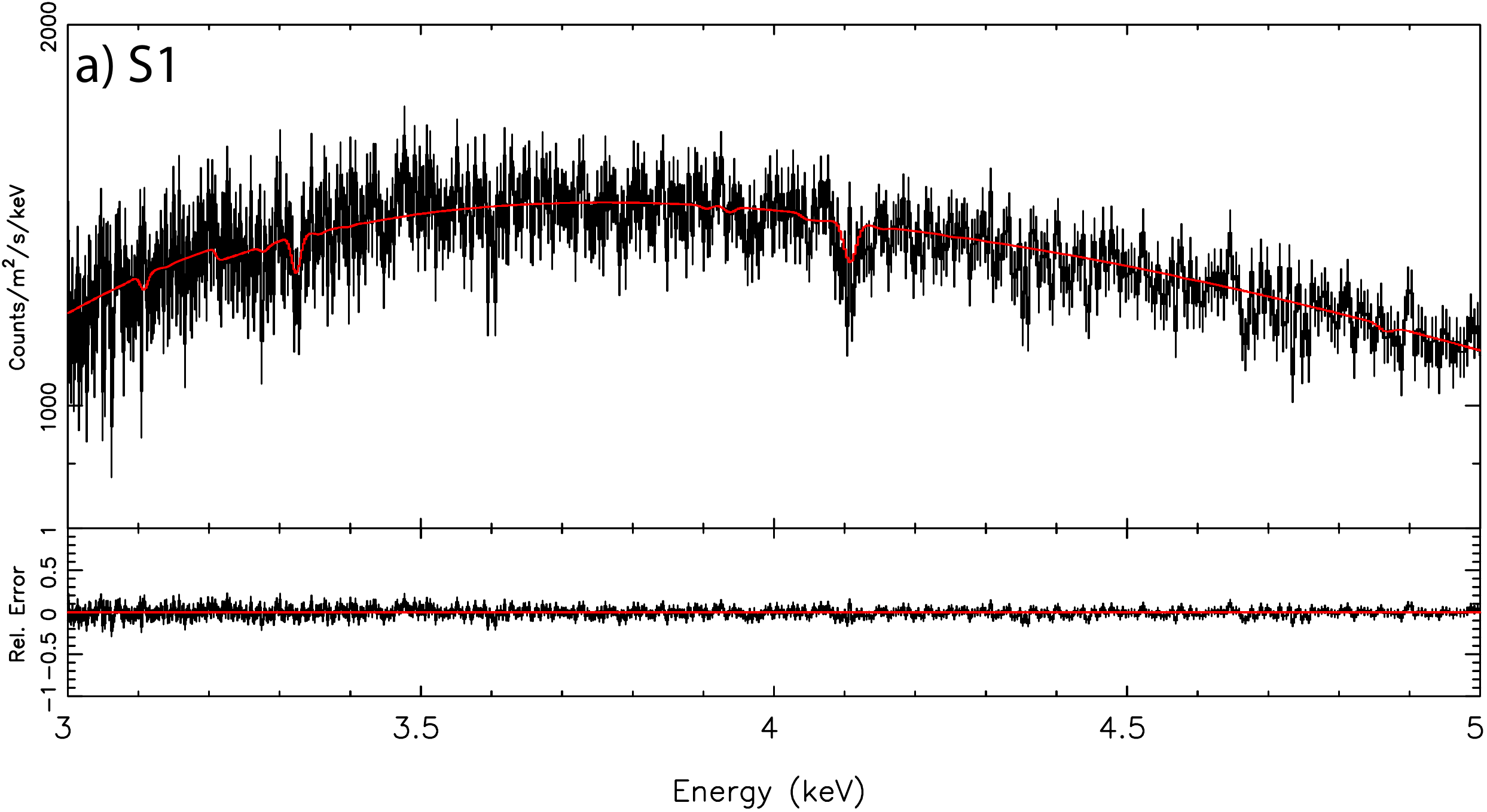}}\\
\vspace{-0.1in}
\subfloat{\includegraphics[scale=0.34,angle=0]{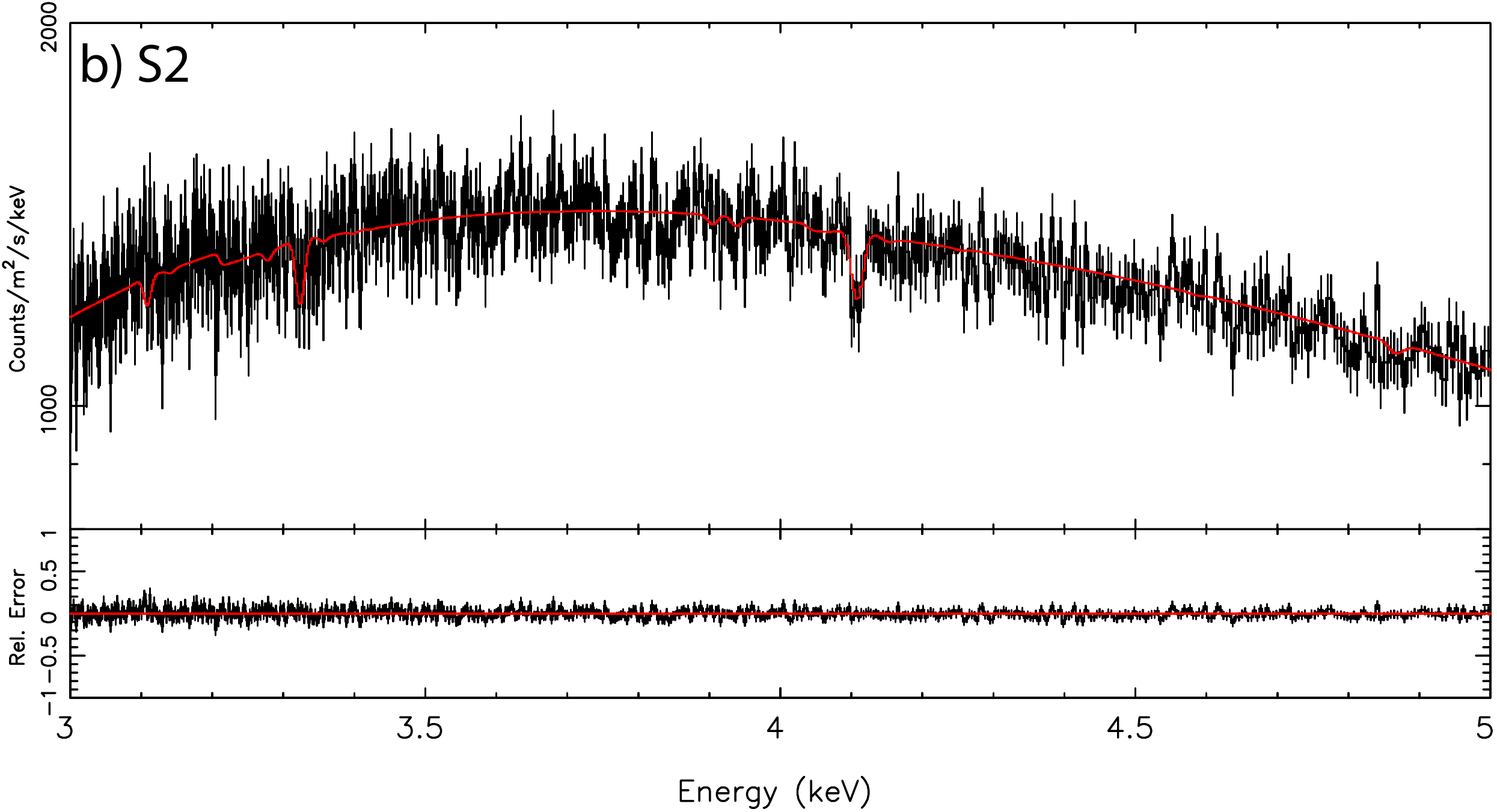}}\\
\vspace{-0.1in}
\subfloat{\includegraphics[scale=0.34,angle=0]{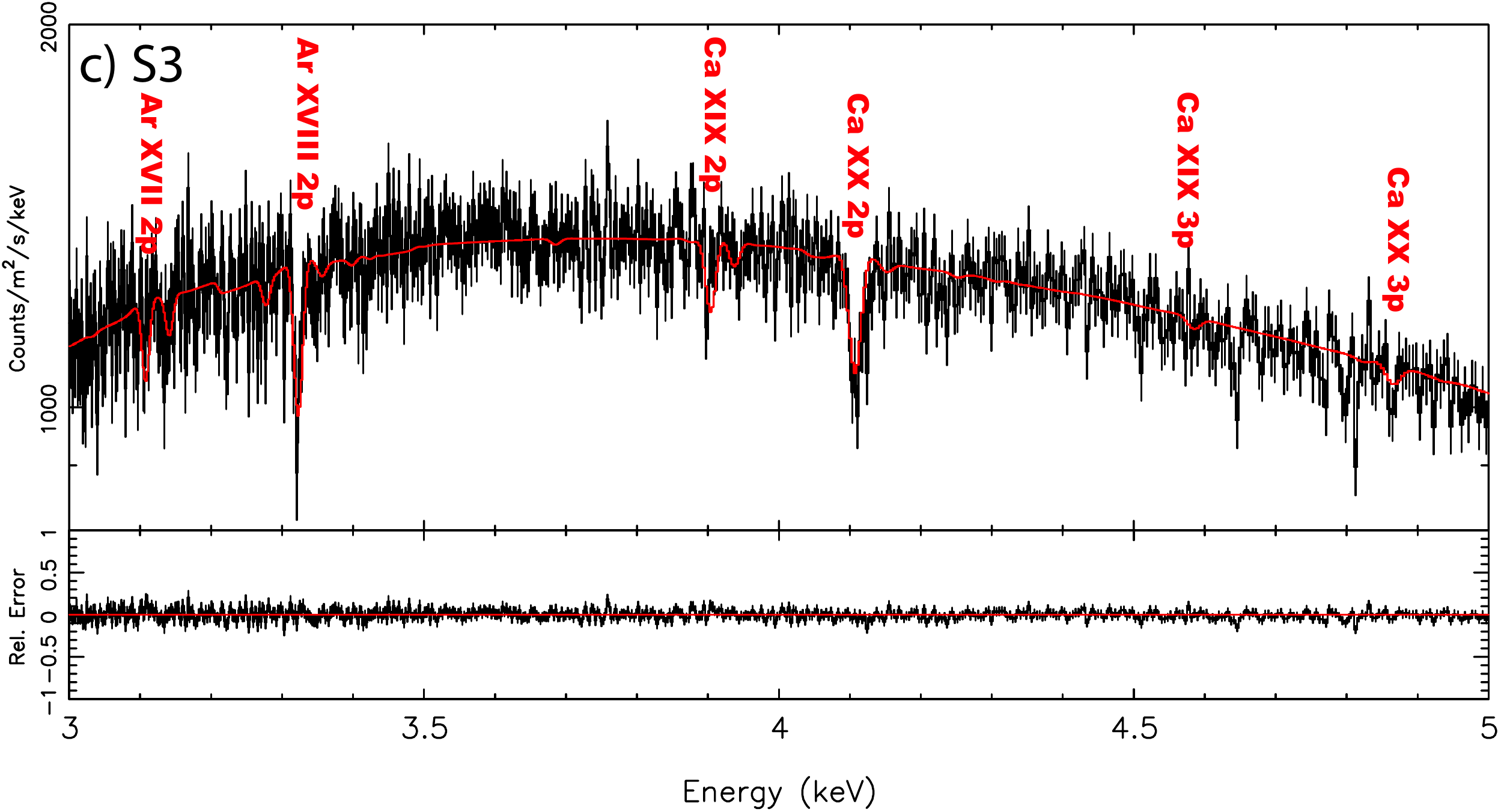}}\\
\vspace{-0.1in}
\subfloat{\includegraphics[scale=0.34,angle=0]{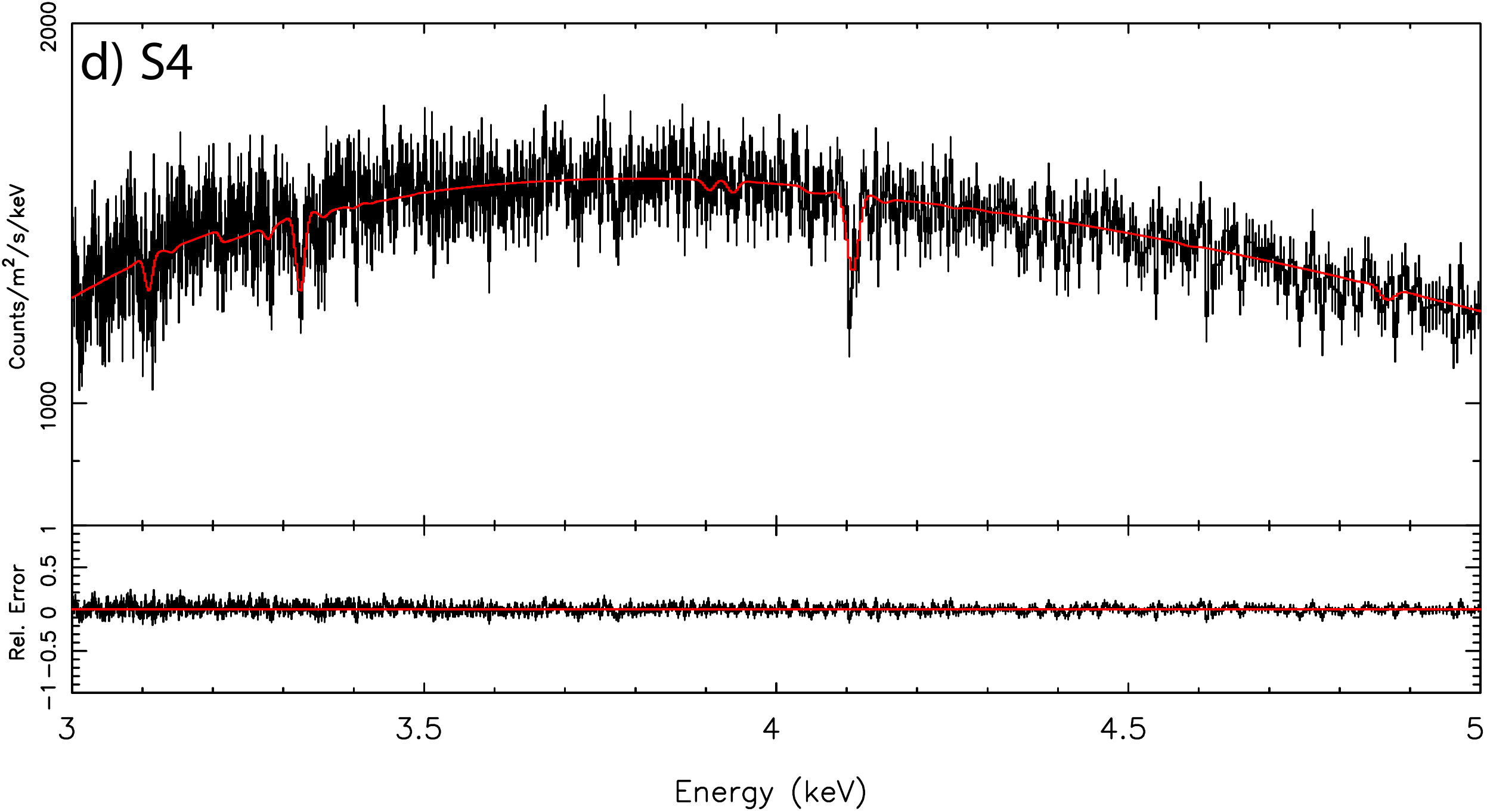}}

  \caption{\footnotesize 
    The 3$-$5 keV HEG spectrum of all soft-state observations of 4U~1630$-$472, 
    fit with the two photoionization zone models listed in Table \ref{tab:fit}. 
    Our fits adequately reproduce most of the absorption lines at lower energies
    despite the fact that only a few bins from this part of the spectrum 
    (4.08 to 4.13 keV) were included during spectral fitting.
    Please see the text and Table \ref{tab:fit} for additional details.\label{fig:cal}}
\end{figure}
%------------------------------------------------------------------------------------------------------------------------------------------------------------------------------------------------------------
%------------------------------------------------------------------------------------------------------------------------------------------------------------------------------------------------------------
\begin{figure*}

\centering

\includegraphics[width=0.76\textwidth,angle=0]{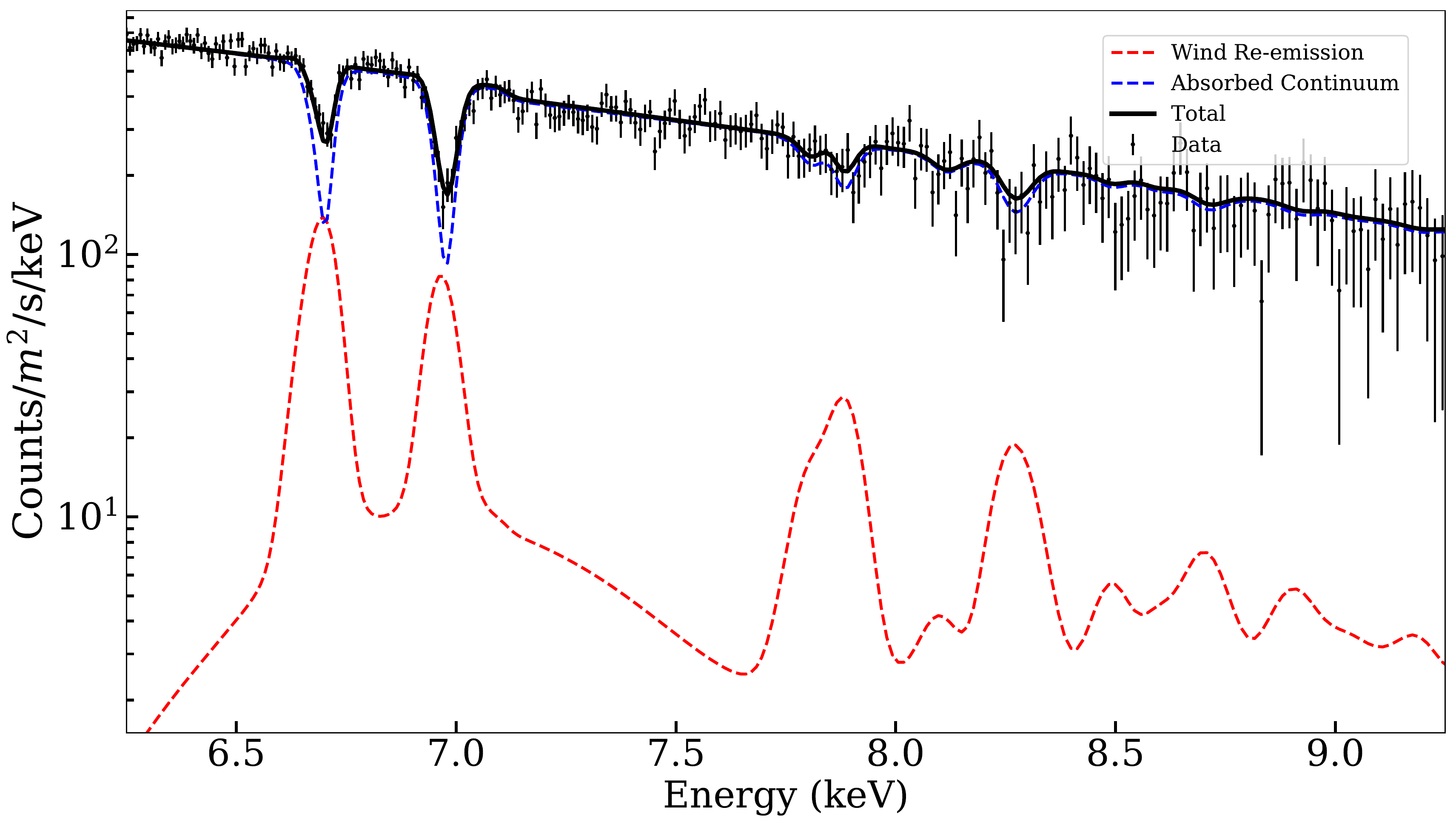}
  \figcaption[t]{\footnotesize Dynamically broadened re-emission (red, both zones) contributes significantly to the line depths of Fe XXV and XXVI near 6.7 and 6.97 keV. 
Compared to the absorption only model (blue, both zones), re-emission has a much weaker effect on the corresponding Fe K$\beta$ lines. 
This is vital for achieving the observed Fe K$\alpha/\beta$ line ratios. 
 \label{fig:py_spec} }

\end{figure*}

%------------------------------------------------------------------------------------------------------------------------------------------------------------------------------------------------------------
%------------------------------------------------------------------------------------------------------------------------------------------------------------------------------------------------------------

In our models, most of the observed Fe XXV absorption originates in a gas with lower ionization relative to those found by \citet{Miller2015}:
Fe XXV intercombination lines become more prominent at these ionizations and,
at HETG resolution, blend with the primary resonance line into a single, highly asymmetric line profile.
This is consistent with Fe XXV line shape seen in all soft-state observations, as well as the abundance of lower energy lines.
This lower ionization gas makes up only a portion of the Fe XXVI line; the rest originates in a highly ionized 
gas with large absorbing columns. Outflow velocities for this component are about half those found by \citet{Miller2015}.

For comparison, the analysis by \citet{Gatuzz2019} was the result of fitting a single absorption zone and therefore
represents a rough weighted average of the wind properties of the system.
For the soft-state observations, they obtain outflow velocities of about -600 km/s (as opposed to two separate zones at -200 km/s and -1000 km/s) and ionizations closer, but systematically higher to our outer wind zones ($\Delta$ log~$\xi \sim 0.1$).
These ionizations are needed in order to achieve the Fe XXV/Fe XXVI line ratios, while the larger outflow velocity is required to fit Fe XXVI at line center.
However, the large outflow velocities are inconsistent with lower energy lines such as Ca XX, which is why their model does a comparatively poor job at fitting most prominent lines below the Fe K band. 
In our single-zone fit to S3, we obtained a considerably worse $\chi^{2}/\nu$ = 198/159 = 1.25 compared to the two-zone model ($\chi^{2}/\nu$ = 156/155 = 1.01). 
In this case, the best-fit ionization (driven by the Fe XXV/Fe XXVI line ratio) results in almost no Ca XX absorption, 
while the best-fit velocity (-200 km/s) fails to capture a significant blue wing in Fe XXVI. 
Even in S1, the soft-state observation with the weakest Ca XX absorption, the single-zone model yielded worse fits ($\chi^{2}/\nu$ = 1.18) compared to the two-zone model ($\chi^{2}/\nu$ = 1.03).

\citet{Gatuzz2019} did not implement wind re-emission (which is necessary in order to achieve the Fe K$\alpha$/$\beta$ ratio, see Figure \ref{fig:py_spec}) 
and instead obtained approximate fits to the Fe K$\beta$ line complex by relaxing ${v}_{turb}$ when fitting the ``warmabs" model.
At log~$\xi \sim 4.0$, the Fe XXVI $\alpha$ line transitions to the flat portion of curve of growth at ${N}_{H} \sim {10}^{23}$ ${\text{cm}}^{-2}$, 
where its equivalent width (EW) becomes sensitive to the turbulent velocity broadening. 
Due to their lower oscillator strengths, Fe XXVI $\beta$ lines are still in the linear regime and their EWs depend only on ${N}_{H}$.
Their best-fit $\alpha$/$\beta$ ratios require turbulent velocities of $\sim 150-200$ km/s, 
which are substantially lower than those typical of LMXB winds \citep[300-500 km/s,][]{Miller2008, Miller2015,Lee2002}.
If the observed velocity broadening is dominated velocity shearing between wind layers, then 
${\sigma}_{v} \sim \Delta r \times (dv/dr) = 0.5 (\Delta r/r) \times {v}_{\text{orbital}}$ \citep{Fukumura2010,Fukumura2017}. 
Their launching radius estimates at a filling factor of unity would correspond to a velocity broadening of 420 km/s.

As an additional test, we fit Voigt profiles to select lines that almost entirely originate in Zone 2 and therefore do not appear broadened 
due to blending with Zone 1 lines.
Two Voigt profiles were used to model the doublets for each H-like line profile with their normalizations coupled using their laboratory measured ratios. 
With their Lorentzian $\gamma$ frozen at laboratory values, we coupled the velocity shift of all the lines in question within a single observation and then coupled their velocity broadening 
across all four soft-state observations. By simultaneously fitting the velocity broadening in these four observations, we obtained a $1\sigma$ confidence interval on 
the turbulent velocity of Zone 2 (after accounting for thermal motions) ranging from 340 to 560 km/s, consistent with our assumed ${v}_{turb} =$ 400 km/s.
Although relaxing the turbulent velocity parameter may help achieve the observed line ratios when modeling a large set of lines with a single absorber, 
this closer examination of line profiles suggests that the turbulent velocities in Zone 2 are considerably higher than those obtained in \citet{Gatuzz2019},
and therefore the data likely require some re-emission.

Our implementation of wind re-emission, however, is dependent on the geometry of the wind. 
Notably, we assumed a wind emission covering factor of $\Omega/4\pi = 0.5$, 
when (in a simplified geometry) this value could be lower. 
Roughly, the lower-limit on $\Omega$ would be the inclination angle at which the system is observed relative to the disk surface (as it is the minimum vertical extent of the wind), 
integrated over the entire azimuthal angle. 
As a simple test of our assumed geometry, we performed the same fitting procedure to S3 using a two-zone model with $\Omega$=0.2, 
corresponding to the minimum vertical extent of this wind given an inclination between 20 and 30 degrees relative to the disk surface.
The resulting best-fit parameter values do not change significantly from those obtained with $\Omega = 0.5$, yet the fit is noticeably worse ($\chi^{2}/\nu$ = 186/155 = 1.2 compared to 156/155 = 1.01).
Ultimately, our incomplete understanding of the wind geometry is a weakness of this type of analysis. 
However, our choice of $\Omega$ did not qualitatively affect our results, while $\Omega$=0.5 yielded better statistical fits.

Although we found that the turbulent velocities in Zone 2 are likely higher than those required to model the data without re-emission \citep[as in][]{Gatuzz2019},
we performed two alternative fits to S3 assuming ${v}_{turb}$ = 200 km/s and either $\Omega = 0.25$ or $\Omega = 0.5$ in order to explore the degeneracy between these parameters.
For Zone 2, we found in the first case ($\Omega = 0.25$) that lowering the emission covering factor results in essentially the same fit as the results listed in Table \ref{tab:fit} ($\Omega = 0.5$ and ${v}_{turb} = 400$ km/s),
with ${N}_{H}= 11 \pm 1 \times {10}^{22} {cm}^{-2}$ (vs. ${N}_{H} \sim 12.4 \pm 1.5$), log~$\xi = 3.35 \pm 0.05$ (vs. log~$\xi = 3.39 \pm 0.08$), and ${\sigma}_{\text{emis}} = {1200}^{+1000}_{-400}$ km/s (vs. ${\sigma}_{\text{emis}} = {1000}^{+300}_{-200}$ km/s).
This suggests a positive correlation between ${v}_{turb}$ and $\Omega$ that does not appear to add significant scatter to the best-fit parameters when comparing extreme values for either.
In the second alternative scenario, however, a lower turbulent velocity combined with a high emission covering factor resulted in a higher discrepancy among best fit parameters, 
with ${N}_{H}= 9 \pm 1 \times {10}^{22} {cm}^{-2}$, log~$\xi = 3.20 \pm 0.08$, and ${\sigma}_{\text{emis}} = {2100}^{+1000}_{-700}$ km/s. 
This fit, however, likely lies in an unphysical region in parameter space: 
With $\Omega=0.5$ and log~$\xi \sim$ 3.20, the re-emission is prominent and therefore the model is sensitive to ${\sigma}_{\text{emis}}$. 
Physically, the increased dynamical broadening of the re-emission, lower ionization, and lower absorbing column would result filling factors of 
$f={0.024}^{+0.029}$. This high degree of clumpiness would likely result in variability that is not observed in the lightcurves of any of our observations,
as in the case of the highly clumpy stellar winds in high-mass X-ray binaries such as Cygnus X-1 \citep{Hanke2009,Grinberg2015,Miskovicova2016} and Vela X-1 \citep{Grinberg2017}.
Although we cannot ultimately rule out the possibility of very clumpy yet homogeneous structure, as could be the case in the cold and partially neutral gas in the BLR and/or tori of AGN, 
it is unlikely for such a small filling factor to occur in either a hot thermal wind  (T $>$ 1 keV) or highly ionized magnetic wind without some additional instability to drive this highly specific type of clumpiness. 

Given that (a) fitting Voigt profiles suggest larger ${v}_{turb}$ values consistent with our assumed 400 km/s, 
(b) small covering factors yield poor fits given ${v}_{turb}$ = 400 km/s, 
(c) a small ${v}_{turb}$ (200 km/s) with a small covering factor yields nearly identical fits to our original fits,
and (d) a small ${v}_{turb}$ (200 km/s) with a large covering factor yields questionably small filling factors,
it is likely that the best fit models listed in Table \ref{tab:fit} (with ${v}_{turb}$ = 400 km/s and $\Omega = 0.5$) provide a better description 
of the winds in this system. The remainder of this work focuses exclusively on the results listed in Table \ref{tab:fit}.

Our results also demonstrate the benefits of PION's self-consistency. 
The ${L}_{Illum}$ column in Table \ref{tab:fit} lists the effective luminosity each PION layer ``sees'' when calculating its ionization balance.
In soft-state observations, after the naked source continuum has been reprocessed by Zone 1, the effective luminosities incident on Zone 2 are between 25-40\% lower than those incident on Zone 1.
The ionization parameter is defined as $\xi = L/{r}^{2}n$, which means that any densities (or, upper limits on the launching radii, $r < L/{N}_{H}\xi$)
derived without this correction may be overestimated by as much as 40\%.
In addition, this has a significant effect on the re-emission component: using the exact same model parameters, 
the re-emission in Zone 2 is 60\% more luminous if it is instead illuminated by the naked source continuum.
By switching the order in which each layer absorbs the continuum, our fits worsened from ${\chi}^{2}/\nu = 1.03$ to 1.56.

In this particular case, the attenuation of the continuum is mostly due to electron scattering and has little overall effect on the shape of the continuum, 
meaning that the ionization balance in Zone 2 is not affected by a change in the shape in the ionizing flux.
Winds with lower ionizations have been observed in other accreting black holes (e.g. GRO J1655-40), 
in which case a change in the shape of the ionizing flux may also have a noticeable effect. 

%------------------------------------------------------------------------------------------------------------------------------------------------------------------------------------------------------------
%------------------------------------------------------------------------------------------------------------------------------------------------------------------------------------------------------------
\begin{figure}
\centering
\includegraphics[width=0.45\textwidth,angle=0]{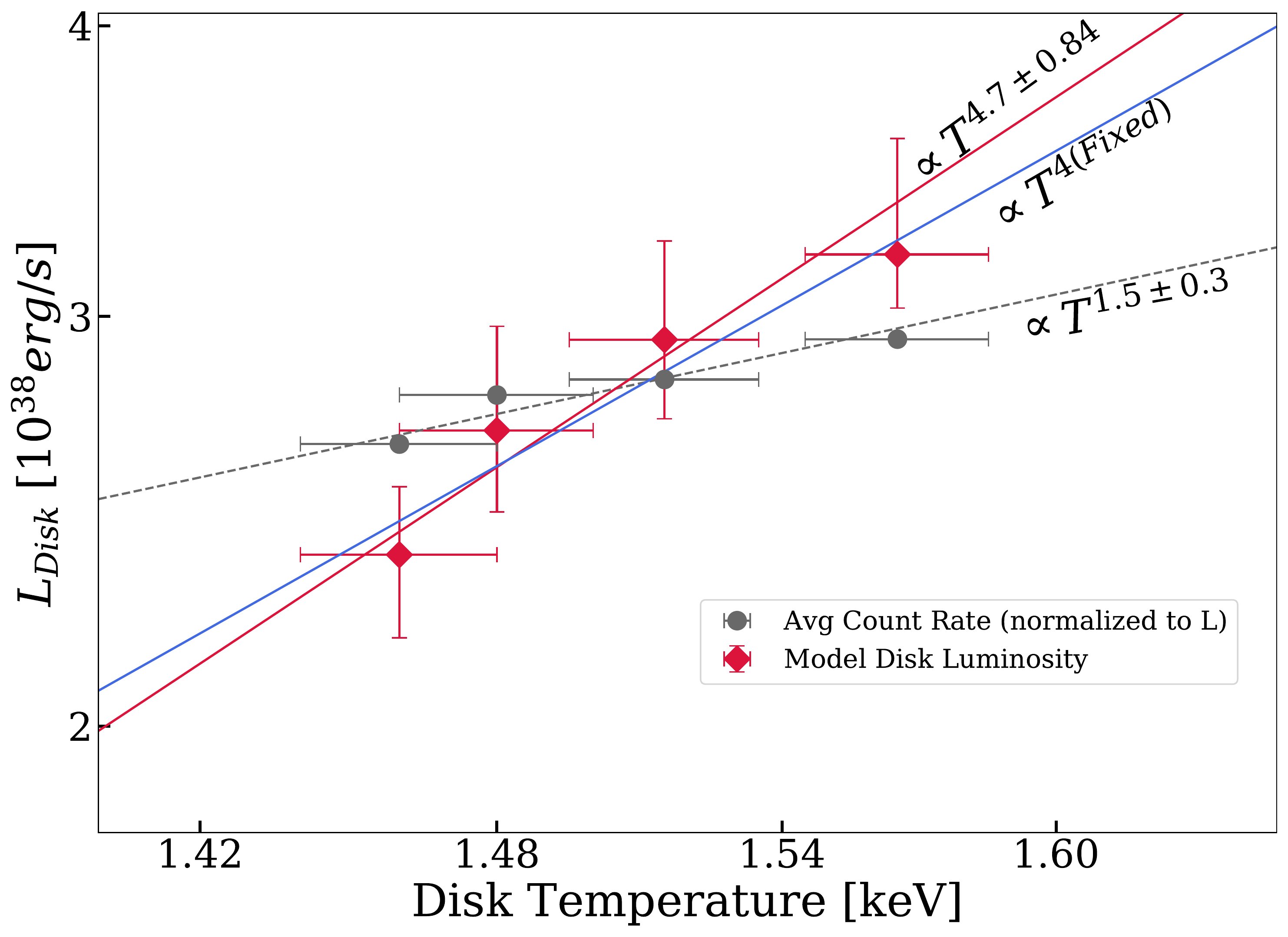}
  \figcaption[t]{\footnotesize Intrinsic disk luminosity in soft-state observations, versus disk color temperature.
  We also plot the luminosities implied by the average detector count-rate in grey which, for visual comparison, 
  were normalized to the average model luminosity. 
  Our model luminosities reflect the with the ${T}^{4}$ scaling expected in disk-dominated states and are a strong indication that ${L} < {L}_{Edd}$.   
  This seems to support the high ${N}_{H}$ values we found in Zone 1 (and large changes in ${N}_{H}$ between observations), 
  over models with small ${N}_{H}$ values.\label{fig:temp}
  }
\end{figure}

%------------------------------------------------------------------------------------------------------------------------------------------------------------------------------------------------------------
%------------------------------------------------------------------------------------------------------------------------------------------------------------------------------------------------------------

The spectra for the four soft-state observations display well-behaved disk-dominated continua, absorption lines of similar depth, and very little change in measured flux between them. 
This stability is reflected in the flatness of the MAXI light curve during this 13 day period, meaning that best-fit wind models for these observations must be broadly consistent with each other. 
This was very helpful at discarding local minima: 
We do not expect, for example, in the four days separating S3 and S4, the system to evolve from being highly obscured and luminous to a low luminosity state obscured by low ${N}_{H}$ winds, particularly when the measured flux, line depths, and model temperatures trend in the opposite direction. 

Our best-fit wind models for the soft state observations achieve this consistency. 
The general picture is that of two distinct photoionization zones: 
An inner and outer absorption layer (Zone 1 and Zone 2) with high/low values for wind ionization, column density, outflow velocities, and dynamical broadening, respectively.
Zone 2 values for ${N}_{H}$, ${v}_{\text{abs}}$, log~$\xi$, and ${\sigma}_{\text{emis}}$ are well-constrained and generally display modest variation between observations. 
The dip in ionization seen in S3 is consistent with its spectra containing the highest number of low-ionization lines and, according to the best-fit model, the lowest Zone 2 incident flux.

As in Zone 2, we observe consistent trends across observations in Zone 1. 
Outflow velocities (${v}_{\text{abs}}$) are well constrained and roughly five times greater than those in Zone 2.
Values of ${\sigma}_{\text{emis}}$ trend towards the upper bound of 15000 km/s (or 0.05 c) for all observations.
For ${N}_{H}$ (which displays some degeneracy with $\xi$), the trend is towards high values ($38 \times {10}^{22}$ ${\text{cm}}^{-2}$ $\leq{N}_{H}$) and, 
as with ${\sigma}_{\text{emis}}$, values approach the upper boundary for some observations.
The behavior of ${\sigma}_{\text{emis}}$ in Zone 1 is not due to the priors described in Section \ref{sec:PIA}:
Despite high ${N}_{H,1}$ and ${\xi}_{1}$ values, $GM/{\sigma}_{\text{emis}}^{2}$ is still three to ten times smaller than ${L}/{{N}_{H}{\xi}}$.

The presence of unconstrained parameters requires further examination.
Implementing wind re-emission is not only crucial in order to achieve the observed Fe $\alpha/\beta$ line ratios, but
as evidenced by our fits of Zone 2, it is possible to constrain the velocity broadening even in the absence of strong P-Cygni profiles.
At very high velocities, emission lines become so broad that the model becomes insensitive to ${\sigma}_{\text{emis}}$. 
We chose a limit of ${\sigma}_{\text{emis}} \leq 0.05 c$, or $r \geq$ 400 $GM/{c}^{2}$, 
allowing us to extract velocity information from gas orbiting at small radii without 
imposing an arbitrarily large cutoff radius, affecting the quality of the fit, or giving meaninglessly small radius values.

The partially unconstrained behavior of ${N}_{H}$ in Zone 1 of some observations is due to only one prominent line (a portion of Fe XXVI $\alpha$) originating from this zone.
At the spectral resolution of the HEG first-order, this means that ${N}_{H}$ and $\xi$ can become degenerate and explode towards higher values. 
Once a wind reaches ${N}_{H} \sim {10}^{24} {\text{cm}}^{-2}$, it becomes Compton-thick. 
These winds are clumpy and result in highly variable light curves \citep{King2015}, neither of which we observe in our spectra. 
Assuming ${N}_{H}$ would converge well below this point, we allowed an initial fitting range up to $\sim {10}^{24} {\text{cm}}^{-2}$.
If instead ${N}_{H}$ became unconstrained and started approaching Compton limit, we tightened this limit down to $6 \times {10}^{23} {\text{cm}}^{-2}$ and 
reported where it becomes unconstrained.
The value of the lower error bar would then be set to the lower bound on the top 68$\%$ of posterior distribution.
For all observations, we found that ${N}_{H}$ for Zone 1 is constrained within 1$\sigma$ from the peak of the posterior distribution, and only becomes unconstrained beyond 2$\sigma$ above the peak in observations S1 and S4 (see Figures \ref{fig:mc1} \& \ref{fig:mc2}).

%------------------------------------------------------------------------------------------------------------------------------------------------------------------------------------------------------------
%------------------------------------------------------------------------------------------------------------------------------------------------------------------------------------------------------------
\begin{figure*}
\centering
\subfloat{\includegraphics[width=0.49\textwidth]{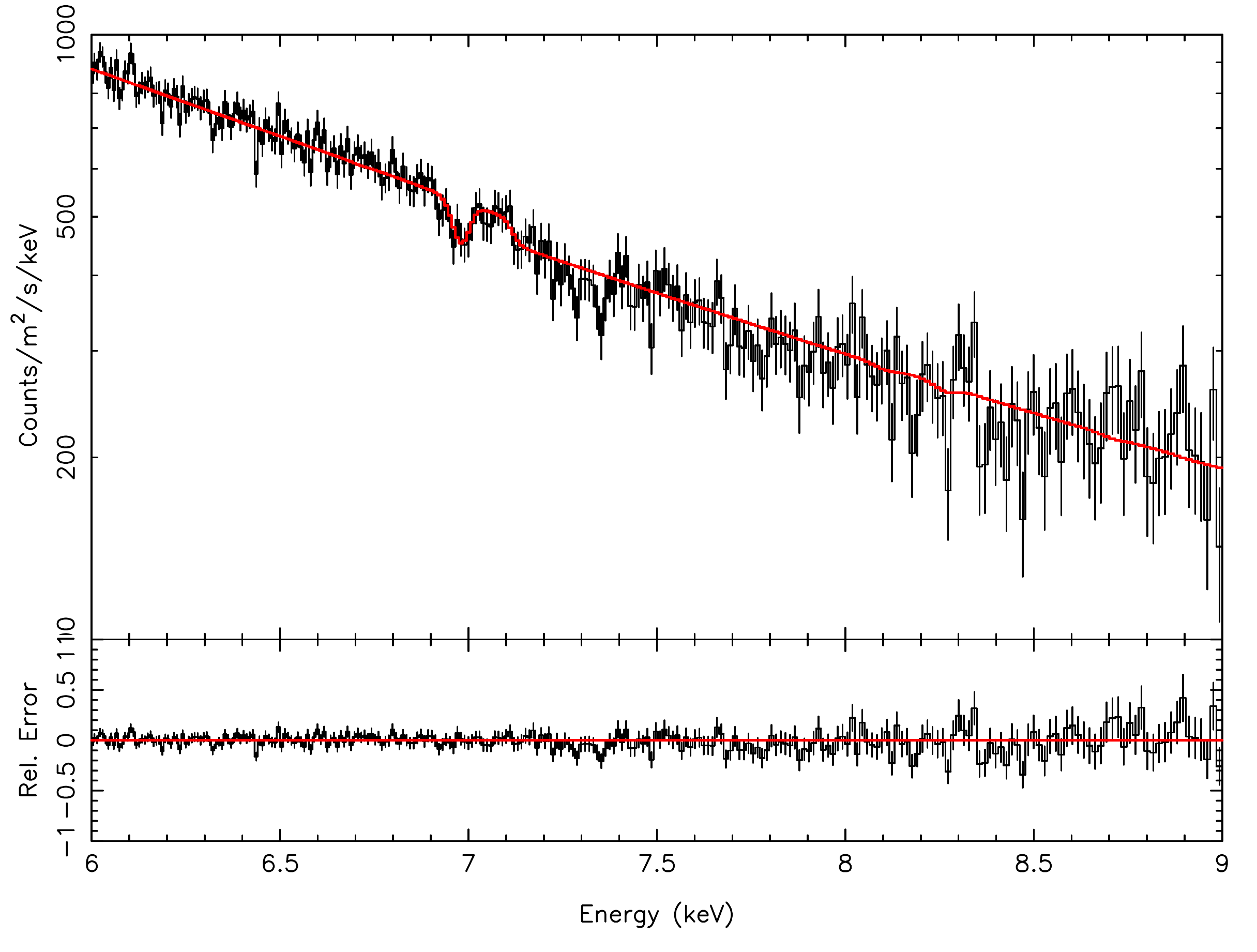}}
\subfloat{\includegraphics[width=0.49\textwidth]{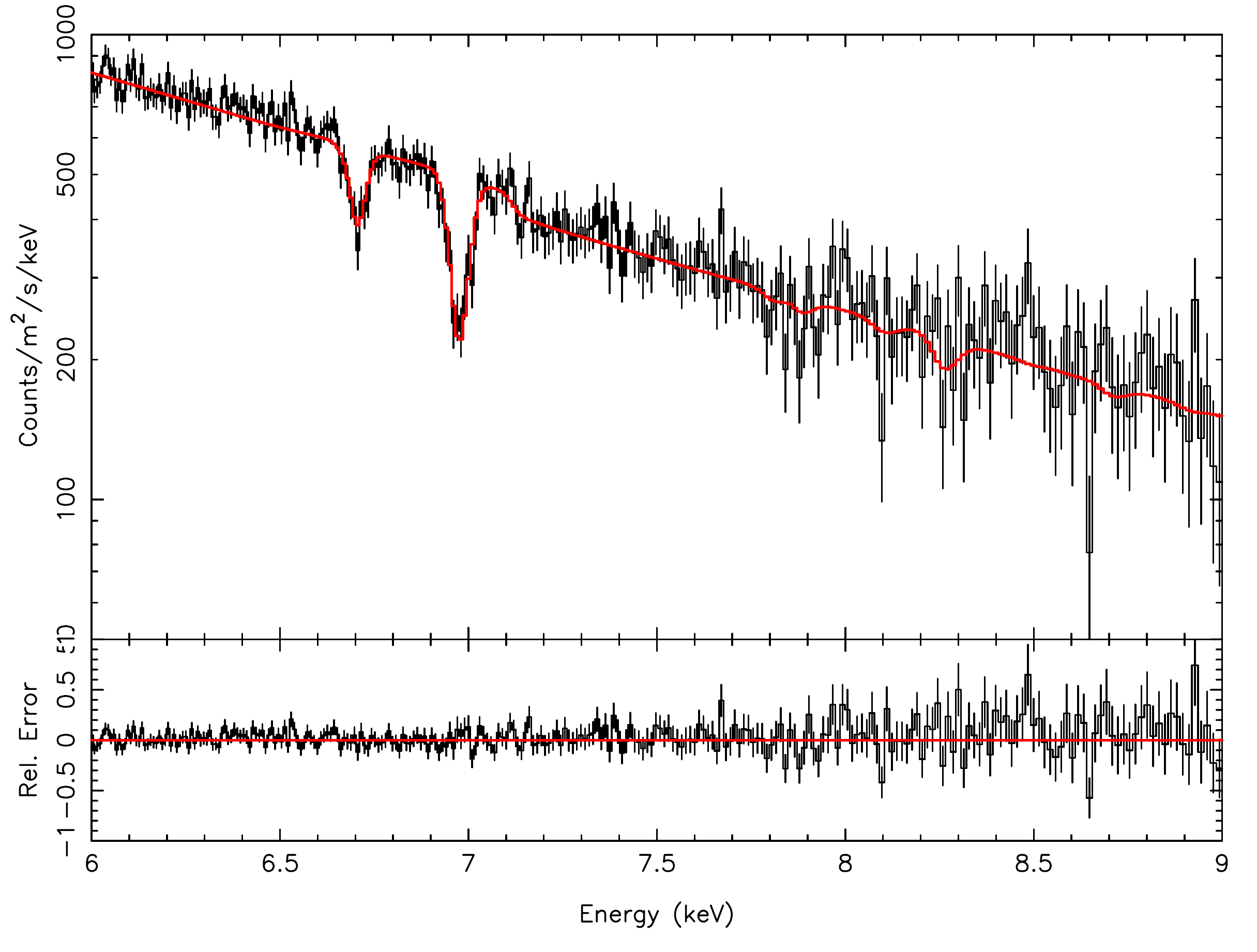}}
  \figcaption[t]{\footnotesize Left: the first-order HEG spectrum of observation I1 (ObsID. 4568) of 4U~1630$-$472, fit
    with the single photoionization zone model listed in Table \ref{tab:fit}. 
    In addition, the data requires a dynamically broadened emission component for each corresponding photoionization zone.
    Right: the first-order HEG spectrum of observation I2 (ObsID. 19904) of 4U~1630$-$472, fit
    with the two photoionization zone model listed in Table \ref{tab:fit}. 
    The data requires a dynamically broadened emission component for each corresponding photoionization zone
    in order to achieve the $\alpha$/$\beta$ line ratio for Fe XXV and Fe XXVI.
    Please see the text and Table \ref{tab:fit} for additional details.\label{fig:19s}}
\end{figure*}
%------------------------------------------------------------------------------------------------------------------------------------------------------------------------------------------------------------
%------------------------------------------------------------------------------------------------------------------------------------------------------------------------------------------------------------

The model unabsorbed source luminosities for these observations are not flat, as suggested by the average detector count rate, but instead are rank-correlated with disk temperature. 
As can be seen in Figure~\ref{fig:temp}, our model luminosities are consistent with the ${T}^{4}$ scaling expected in disk-dominated states and are a strong indication that ${L} < {L}_{Edd}$ 
\citep{Kubota2001, KubotaMakishima,GierlinskiDone,Abe2005,McClintock2009, McClintock2014}.
This trend seems to favor our models which have large ${N}_{H}$ values in Zone 1 (and large changes in ${N}_{H}$ between observations), 
over models with small ${N}_{H}$ values (which have a much flatter ${T}^{1.5}$ scaling).

\subsubsection{Intermediate State Observations}\label{sec:inter}

Fits to I1 and I2 still resulted in good statistical fits, with $\chi^{2}/\nu$ values of 129/130 = 0.99 and 167/155 = 1.08, respectively. 
The relative lack of strong absorption features in these spectra, however, made it considerably harder to constrain wind parameters.

Our best-fit model for I2 consists of two photoionization zones of ``moderate" ionization (log~$\xi$ = 4.4 to 4.6) with low absorbing columns (${N}_{H} \sim 4 \times {10}^{22} {\text{cm}}^{-2}$ in Zone 1, a near order of magnitude decrease when compared with S1-S4).
As with the soft-state spectra, the same trend of inner winds having higher ionizations, outflow velocities, and velocity broadening is observed. 
Velocity broadening in Zone 2 is well constrained but at higher velocities, corresponding to an increase in ionization at smaller radii, 
while it trends towards the upper bound of 15000 km/s for Zone 1. 

The bounds on $\xi$ for Zone 1 were also tightened in order to only probe the minima in which Zone 1 contributed Fe XXV line absorption.
We found that models in which Zone 1 becomes ionized to the point where it only contributes to Fe XXVI absorption resulted in worse statistical fits.
The combined Fe XXV profile is too broad for a single Zone at ``moderate" ionizations, while the total lack of low energy lines and symmetry of the lines rules out a ``low" ionization gas. 

At first glance, strong Fe XXV and XXVI $\alpha$ absorption lines in I2 seem to indicate winds similar to those in S1-S4. 
The stark differences between the best-fit model for I2 and the high ${N}_{H}$ winds in soft-state spectra are consistent with discrepancies outside these two lines.
First, although individual low-count bins coincide with the location of Ni and Fe lines above 7.5 keV, the spectrum is far too noisy for any of these lines to be significant. 
Moreover, the complete lack of low energy lines indicates that differences between I2 and soft-state spectra are greater than the Fe XXV and XXVI line profiles would suggest.
Physically, our best-fit model is consistent with winds being correlated with disk activity:
the connection between the presence of the additional continuum component and weaker absorption lines is mainly due to a decrease in the measured absorbing column of the wind, not over-ionization from powerlaw photons. In particular, the presence of Fe XXV absorption lines constrains the fit away from the much higher ionizations that, in turn, would require higher columns.
Especially in the case of I2, our fits strongly indicate that the observed absorption lines originate in absorbers with low column densities.
This picture is also consistent with I2 appearing in the same location as soft-state observations in Figure \ref{fig:HR}, and further reinforces the notion that most of the flux responsible for dictating ionization balance is within the Chandra energy band. 
However, this does not rule out the possibility that additional over-ionized and optically-thin absorbers with high columns may be present, as these are inherently difficult to detect. 
This is further complicated by the lack of simultaneous observations with other facilities that would allow us to constrain the broadband continuum.
Although the disappearance of winds in spectrally-hard states may indicate an anti-correlation between disk winds and jets \citep{Miller2012},
the disappearance of winds at different spectral states may instead signal a change in disk geometry \citep{Ueda2010}, or a combination of lower columns and increased ionization \citep{Diaz2014}.

%------------------------------------------------------------------------------------------------------------------------------------------------------------------------------------------------------------
%------------------------------------------------------------------------------------------------------------------------------------------------------------------------------------------------------------
\begin{figure*}

\centering

\includegraphics[scale=1.0,angle=0]{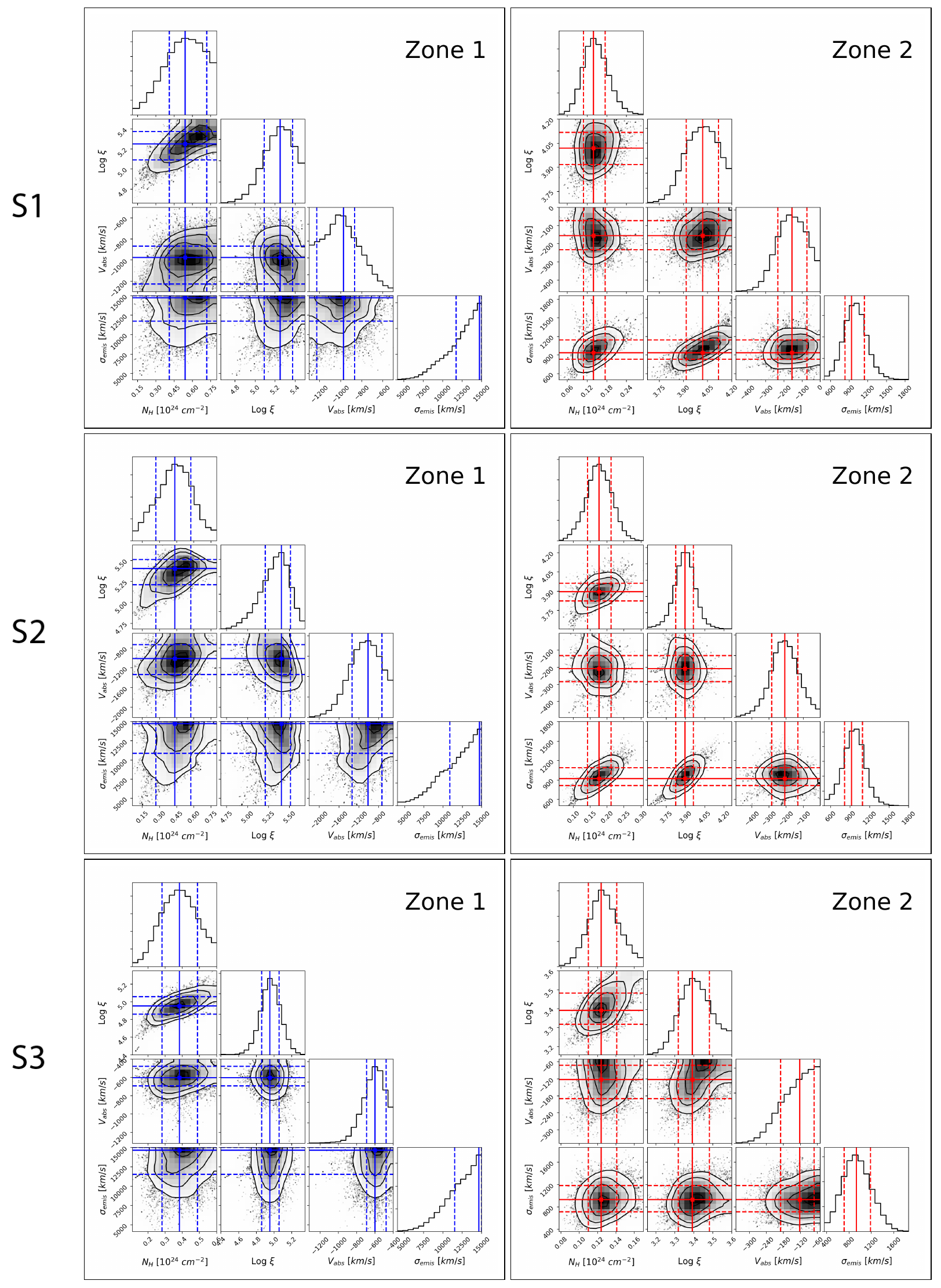}

  \caption{\footnotesize Posteriors of wind absorbing column, ionization parameter, outflow velocity, and emission velocity broadening for observations S1 (top), S2 (middle), and S3 (bottom). 
 See text for details.\label{fig:mc1}}
\end{figure*}
%------------------------------------------------------------------------------------------------------------------------------------------------------------------------------------------------------------
\begin{figure*}

\centering
\includegraphics[scale=1.0,angle=0]{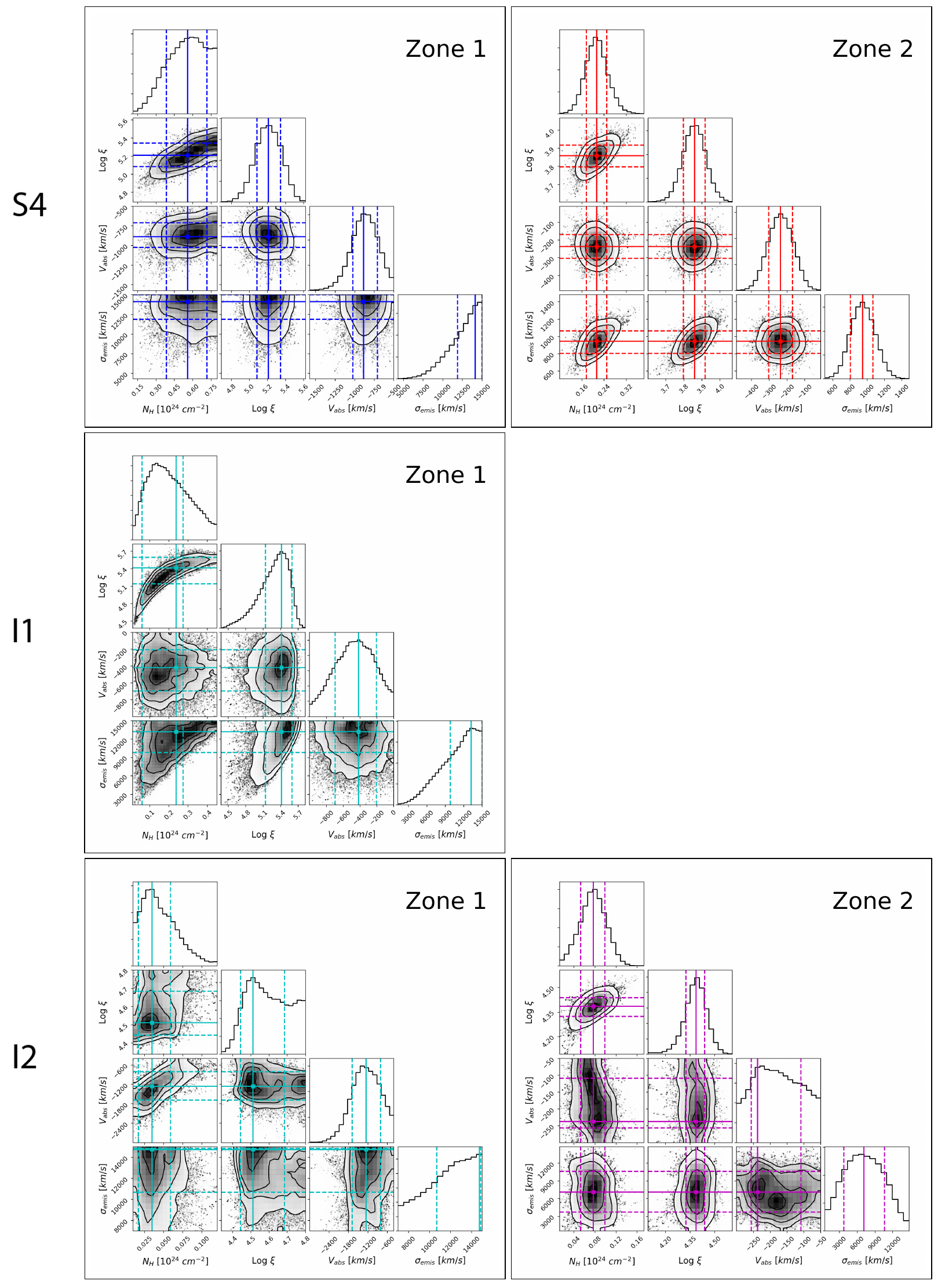}
  \caption{\footnotesize Posteriors of wind absorbing column, ionization parameter, outflow velocity, and emission velocity broadening for observations S4 (top), I1 (middle), and I2 (bottom). See text for details.\label{fig:mc2}}%
\end{figure*}

%------------------------------------------------------------------------------------------------------------------------------------------------------------------------------------------------------------
%------------------------------------------------------------------------------------------------------------------------------------------------------------------------------------------------------------

%------------------------------------------------------------------------------------------------------------------------------------------------------------------------------------------------------------
%------------------------------------------------------------------------------------------------------------------------------------------------------------------------------------------------------------
\begin{table*}
\caption{Wind Launching Radii}
\begin{footnotesize}
\begin{center}
\begin{tabular*}{\textwidth}{l c @{\extracolsep{\fill}} c c c c c c c}
\tableline
\\ [-3.0ex]
\tableline 
Obs. Label & Zone & ${R}_{phot, n = {10}^{14}}$ & ${R}_{\text{upper}}$ & ${R}_{\text{upper}}$/${{R}_{c}}$ & ${R}_{\text{orbital}}$ & ${R}_{\text{orbital}}/{{R}_{c}}$   &log ${n}$ & $f$ \\
 & & (GM/${c}^{2}$) & (GM/${c}^{2}$) & & (GM/${c}^{2}$) &&$({\text{cm}}^{-3})$\\

\\ [-3.0ex]
\tableline 
\\ [-3.0ex]

S1 & 1 & $2800^{+300}_{-400}$ & $2100^{+600}_{-{1100}^{\dagger}}$ &0.005& $400^{+200}_{*}$ & 0.001& $15.4^{+0.2}_{-0.3}$ & $0.3^{+0.1}_{-0.2}$\\
	 & 2 & $9100^{+1200}_{-900}$ & $80000^{+26000}_{-27000}$ & 0.2 & $77000^{+39000}_{-1700}$ & 0.20 & $12.0^{+0.2}_{-0.2}$ & $1.0^{*}_{-0.215}$\\

\\ [-3.0ex]
\tableline 
\\ [-3.0ex]

S2 & 1 &$2300^{+300}_{-400}$ & $1300^{+1400}_{-{200}^{\dagger}}$ & 0.003 & $400^{+400}_{*}$ & 0.001& $15.2^{+0.2}_{-{0.4}^{\dagger}}$ & $0.35^{+0.1}_{-0.2}$\\

	 & 2 &$11000^{+700}_{-800}$ & $93000^{+1600}_{-34000}$ & 0.24 & $97000^{+22000}_{-32000}$ & 0.25 & $12.0^{+0.3}_{-0.2}$ & $1.0(1)$\\

\\ [-3.0ex]
\tableline 
\\ [-3.0ex]

S3 & 1 &$3400(300)$ & $4100(1700)$   						&0.01 & $400^{+300}_{*}$  & 0.001& $15.6^{+0.2}_{-0.3}$ & $0.10^{+0.08}_{-0.04}$\\	 
	 & 2 & $18000^{+1800}_{-1500}$ & $395000^{+54000}_{-140000}$ & 0.99 & $104000^{+36000}_{-58000}$ & 0.26 & $12.5^{+0.3}_{-{0.4}^{\dagger}}$ & $0.3^{+0.1}_{-{0.2}^{\dagger}}$ \\

\\ [-3.0ex]
\tableline 
\\ [-3.0ex]

S4 & 1 & $2900^{+400}_{-300}$ & $2400^{+700}_{-{1300^{\dagger}}}$ &0.006 & $400^{+200}_{*}$ & 0.001& $15.6^{+0.1}_{-0.3}$ & $0.13^{+0.21}_{-0.04}$\\
	& 2 & $11000(600)$ & $78000^{+24000}_{-18000}$ & 0.21 & $102000^{+20000}_{-34000}$ & 0.28 & $12.1^{+0.1}_{-0.2}$ & $1.3^{+0.1}_{-0.3}$\\
\\ [-3.0ex]
\tableline 
\\ [-3.0ex]
\tableline 
\\ [-3.0ex]

I1 & 1 & $2400^{+800}_{-400}$ & $3700^{+7700}_{-{300^{\dagger}}}$ &0.007 & $500^{+500}_{-{100}^{\dagger}}$ & 0.001& $15.1^{+0.2}_{-{0.4^{\dagger}}}$ & $0.16^{+0.18}_{-0.05}$ \\

\\ [-3.0ex]
\tableline 
\\ [-3.0ex]
I2 & 1 & $6500^{+400}_{-1300}$ & $120000^{+60000}_{-75000}$ &0.20 & $400^{+400}_{*}$ & 0.001& $15.9^{+0.2}_{-{0.4}^{\dagger}}$ & $0.004^{+0.004}_{-0.003}$\\
	& 2 & $7400^{+400}_{-500}$ & $89000^{+44000}_{-30000}$ & 0.18 & $1800^{+1300}_{-{1300^{\dagger}}}$ & 0.004 & $15.2^{+0.5}_{-{0.4}^{\dagger}}$ & $0.017^{+0.013}_{-{0.015^{\dagger}}}$\\

\\ [-3.0ex]
\tableline 
\\ [-3.0ex]
\tableline

\end{tabular*}
\vspace*{-1.0\baselineskip}~\\ \end{center} 
\tablecomments{\footnotesize{The table above lists estimates of wind launching radii, wind density, and volume filling factors derived from the best-fit models listed in Table \ref{tab:fit}.
Quoted errors are at the $1\sigma$ level. Errors marked with * indicate that a parameter is unconstrained within $1\sigma$ in that direction, 
while those annotated with $\dagger$ indicate a parameter with unconstrained behavior outside the $1\sigma$ confidence interval.
For comparison, we included a column for the photoionization radius, ${R}_{phot,~n = {10}^{14}}$, assuming a density of $n = {10}^{14}$. 
Launching radii were derived as ${R}_{upper} = {L}/{{N}_{H}{\xi}}$ and ${R}_{\text{orbital}}$ = $({{c}^{2}}/{{\sigma}^{2}_{\text{emis}}})$ GM/${c}^{2}$,
where ${R}_{\text{upper}}$ is the upper limit on the photoionization radius. We also list the ratio of these radii over the corresponding Compton radius (or, ${R}_{c}$)
given the disk color temperature at the time of the observation (see section 3.4).  
Filling factor and density estimates derived using $f = \Delta r/r = {{r}}/{{R}_{\text{upper}}}$ and ${n} = {{L}/{{r}^{2}{\xi}}}$, with ${r} = {R}_{\text{orbital}}$.}
}

\end{footnotesize}
\label{tab:rad}
\end{table*}

%------------------------------------------------------------------------------------------------------------------------------------------------------------------------------------------------------------
%------------------------------------------------------------------------------------------------------------------------------------------------------------------------------------------------------------

With a single absorption feature, it is difficult to justify the use of two photoionization zones when modeling I1 given the spectral resolution of the HEG first-order.
Our best-fit single-zone model for I1 details a highly ionized (log~$\xi$ $\sim 5.4$) wind with a well constrained outflow velocity (${570}^{+350}_{-130}$ km/s), launched from small radii (${\sigma}_{\text{emis}} = {15000}^{*}_{-4000}$ km/s). 
The largest difficulty in fitting this model was the degeneracy between $\xi$ and ${N}_{H}$. 
As shown in Figure~\ref{fig:mc2}, the posterior distribution is not Gaussian, and although there is clearly a preferred minimum in the 2-D histogram between these parameters, it neither corresponds to the median or peak of the 1-D distributions of either parameter. 
We take the point of highest 2-D probability as the best-fit value for both parameters.

\subsection{Wind Launching Radii and Outflow Properties}\label{sec:rad_mdot}

Estimates for wind launching radii derived from our best-fit models, as well as estimates on wind density and filling factor, are listed in Table \ref{tab:rad}.
Errors for launching radii and other wind properties were determined empirically:
For complex dependencies on observed parameters, as is the case for most wind properties, propagating errors analytically can result in either greatly over or underestimating the propagated error. 
Chains constructed during spectral fitting contain important information about parameter correlations that more accurately represent the uncertainty on a model parameter. This is particularly useful when dealing with unconstrained parameters, where unbounded behavior can be cancelled out by a reciprocal correlation.

%------------------------------------------------------------------------------------------------------------------------------------------------------------------------------------------------------------
%------------------------------------------------------------------------------------------------------------------------------------------------------------------------------------------------------------
\begin{figure*}
\centering
\hspace{0.2 in}
\includegraphics[scale=0.38,angle=0]{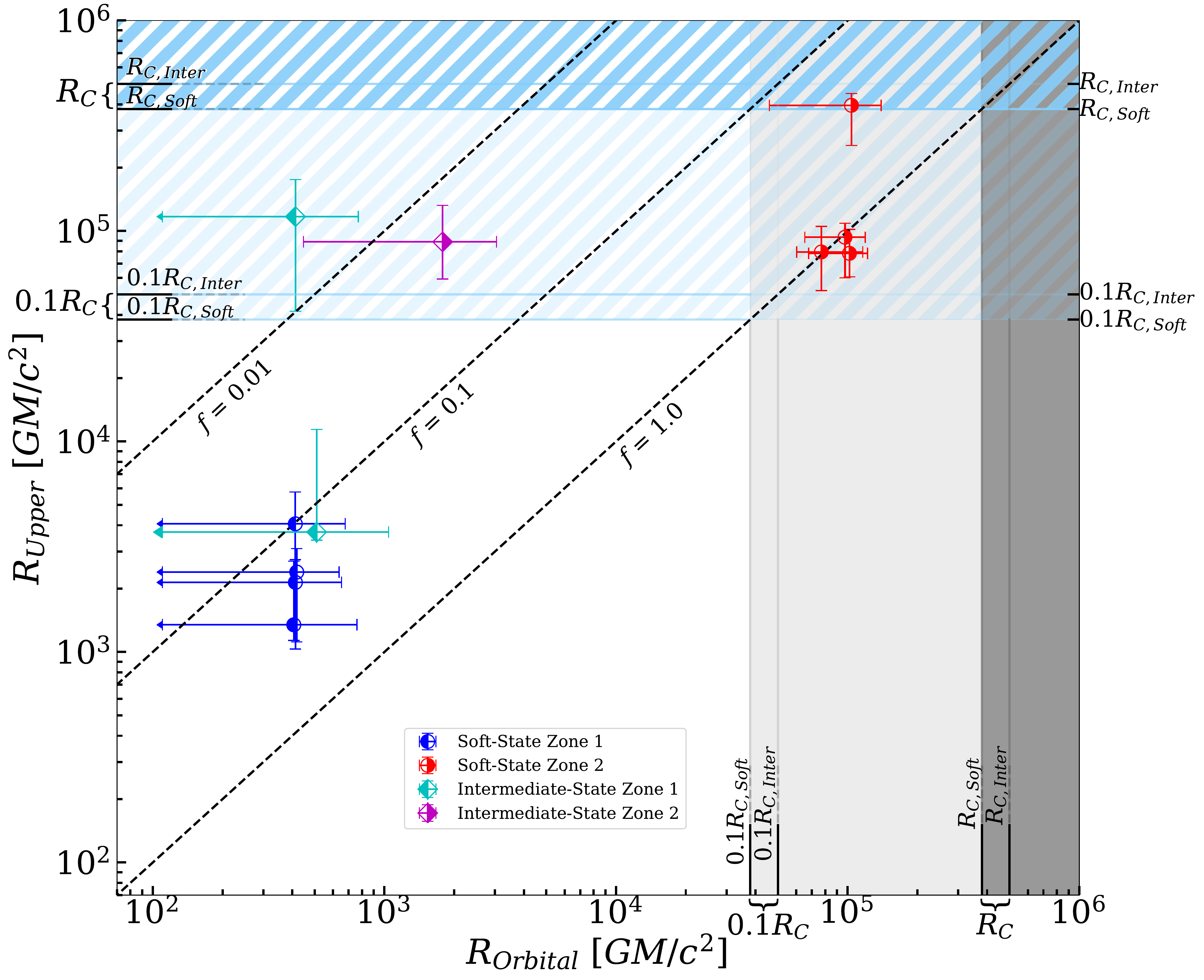}
  \figcaption[t]{\footnotesize Wind launching radii for all observations plotted in radius vs. radius space, 
  with ${R}_{orbital}$ and ${R}_{upper}$ corresponding to the x and y coordinates, respectively. 
  Dark-grey shaded and dark-blue hatched regions correspond to radii above 1.0${R}_{C}$ (and above 0.1${R}_{C}$ in light-grey and light-blue)
  for ${R}_{orbital}$ and ${R}_{upper}$, respectively.
  Points in the white region lie below this limit in both coordinates and therefore rule out thermal driving.
  An arrow at the end of an error bar indicates that a parameter is unconstrained in that direction.
  Lines of constant volume filling factor are plotted as dashed diagonal lines.
  By design, no points should have filling factors significantly above unity (see Appendix \ref{sec:MCMC}).
  \label{fig:py_rad}  }
\end{figure*}
%------------------------------------------------------------------------------------------------------------------------------------------------------------------------------------------------------------
%------------------------------------------------------------------------------------------------------------------------------------------------------------------------------------------------------------

In cases where the gas density can be measured directly, the wind absorption radius (or, photoionization radius) is given by ${R}_{phot} = \sqrt{{L}/{{n}_{H}{\xi}}}$.
If gas density is not known, an upper limit on this radius can be obtained directly from observables, $r \leq {R}_{\text{upper}} \equiv {L}/{{N}_{H}{\xi}} $. 
Finally, the velocity broadening of the re-emission give us a measure of the local Keplerian velocity in an axially-symmetric disk-wind, 
and therefore we can obtain an independent launching radius estimate of ${R}_{\text{orbital}}$ = ${GM}/{{\sigma}^{2}_{\text{emis}}}$.
We provide estimates for wind launching radii via these three metrics, including ${R}_{phot}$ assuming a fiducial density of log~${n}_{H} = 14$ 
\citep[based on Fe XXII line ratios of other LMXB winds,][]{Miller2008}, as a point of comparison. 
Filling factor and density estimates listed in Table \ref{tab:rad} were derived by assuming ${r} = {R}_{\text{orbital}}$, where ${n} = {{L}/{{r}^{2}{\xi}}}$ and $f = \Delta r/r = ({N}_{H}/n)/r $ is 
arithmetically equivalent to $f = ({N}_{H}{\xi}/L)\times {r}^{2}/r = r/{R}_{\text{upper}}$.

Measurements and estimates of BH and NS disk wind densities span several orders of magnitude, 
an uncertainty that is often not reflected in many published radius estimates that rely on assumed densities.
For ${R}_{phot}$, assuming a density is equivalent to assuming radius. 
For the remainder of our analysis, we relied exclusively on ${R}_{\text{upper}}$ and ${R}_{\text{orbital}}$, which are mutually independent and derived strictly from observables.
Despite their individual limitations, the combined information from ${R}_{\text{upper}}$ and ${R}_{\text{orbital}}$ is far better representation of the wind launching radii, their uncertainty, and their limits, than an assumed ${R}_{phot}$. This also allows us to obtain density estimates, as well as density-dependent wind parameters.

Figure \ref{fig:py_rad} shows a plot of wind launching radii estimates for all six observations and photoionization zones, in radius vs radius space. 
The x- and y-coordinate values for each point correspond to their ${R}_{\text{orbital}}$ and ${R}_{\text{upper}}$ values, respectively. 
Since ${R}_{upper}$ is simply ${R}_{phot}$ with a filling factor of unity, y-axis values should be interpreted as an upper limit with errors on its value. 
Points with arrows indicate that the parameter is unconstrained in that direction:
For example, cases in which ${\sigma}_{\text{emis}}$ trends towards values above the upper bound $0.05 c$, ${R}_{\text{orbital}}$ would then be unconstrained towards very small radii.
Soft-state outer and inner zones are plotted in red and blue, respectively, while intermediate state inner and outer zones are plotted in cyan and magenta, respectively.
Because we use $f = {{R}_{\text{orbital}}}/{{R}_{upper}}$, we can plot lines of constant filling factor in this space. 
Because of the priors set while fitting, no points should lie significantly below the $f = 1.0$ line (see Appendix \ref{sec:MCMC}).

Radiation pressure can drive winds via line interactions and/or electron scattering. 
Line-driven winds are gases of relatively low ionization: Although the force multiplier at log~$\xi \sim 3$ is non-zero, the line force becomes negligible above log~$\xi \geq 2$ \citep{Proga2000a, Proga2000b}. 
Electron scattering is significantly weaker than line-driving, requiring near-Eddington luminosities in order to efficiently drive a wind (Proga et al. 2003a).
Given that $L \leq 0.25 {L}_{Edd}$ and log~$\xi \geq 3.4$ in all photoionization zones, we can rule out radiation pressure as a driving mechanism.
This leaves thermal and magnetic driving as the remaining possibilities.

Compton heated winds can be driven ballistically from ${R}_{C}$ \citep{Begelman1983}, 
though \citet{Woods1996} suggests that this limit may extend down 0.1-0.2${R}_{C}$. 
The precise nature and location of this boundary between a gravitationally bound corona and a free thermal wind is likely sensitive to many disk parameters and the subject of much debate. Radiation pressure enhancement at luminosities near Eddington and pressure confinement of outer layers via completely ionized winds have been suggested as plausible scenarios in which these outflows may still be thermal in nature \citep{ProgaKallman,Done2018}.
For a more detailed discussion on how our results compare to these alternative scenarios, see Section \ref{sec:discussions}. 

In this section, we will discuss our results relative to ${R}_{C}$ and 0.1${R}_{C}$ as described by \citep{Begelman1983}, 
where ${R}_{C}$ is a function of the temperature at the surface of disk and the mass of the black hole. 
The disk surface is assumed to be in thermal equilibrium as it is Compton heated by flux from the inner disk, 
so we can approximate this temperature as being equal to the disk color temperature. 
Our Compton radius is then ${R}_{C} = {10}^{10} \times ({M}_{BH}/{M}_{\odot})/{T}_{C8}$ cm or ${R}_{C} = (5.82/{kT}_{\text{keV}}) \times {10}^{5}$ $GM/{c}^{2}$. In the latter definition, both ${R}_{C}$ and ${R}_{G} = GM/{c}^{2}$ 
are proportional to the mass of the accretor, and therefore the value of ${R}_{C}$ is independent of mass when measured in gravitational radii.
For soft state observations, these range from ${R}_{C,Soft} \simeq 5.5$ to $5.9 \times {10}^{11}$ cm, or 
${R}_{C,S} \simeq 3.9 \times {10}^{5}$ $GM/{c}^{2}$. For intermediate state observations, ${R}_{C,Inter} \simeq 5.1 \times {10}^{5}$ $GM/{c}^{2}$.
The light and dark shaded regions in Figure \ref{fig:py_rad} correspond to values of ${R}_{\text{orbital}}$ that lie above 0.1${R}_{C}$ and ${R}_{C}$, respectively.
For ${R}_{upper}$, these values are plotted as light and dark blue dashed regions

%------------------------------------------------------------------------------------------------------------------------------------------------------------------------------------------------------------
%------------------------------------------------------------------------------------------------------------------------------------------------------------------------------------------------------------
\begin{figure}
\centering
   \subfloat{\includegraphics[width=0.47\textwidth,angle=0]{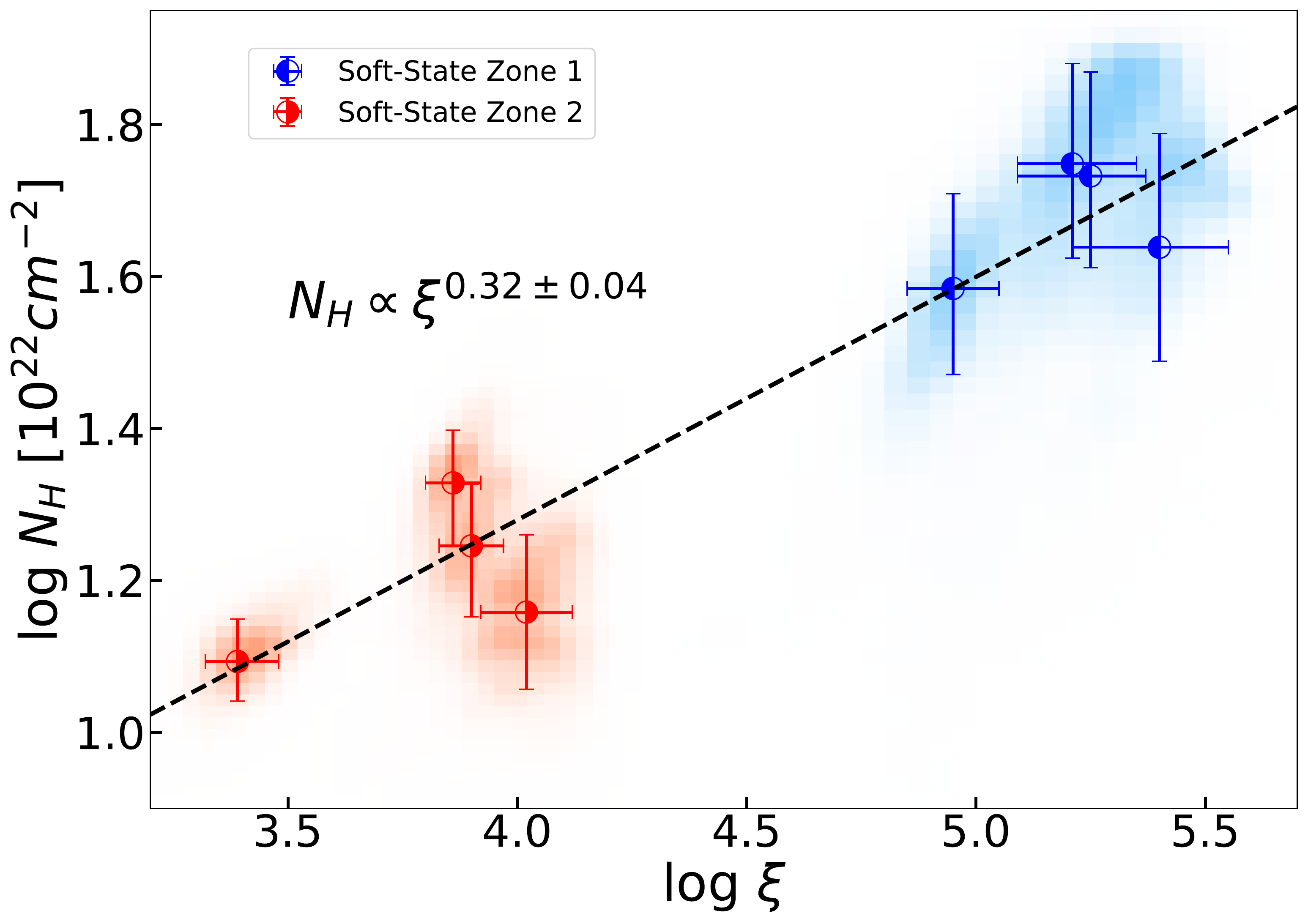}}\\
   \vspace{-0.1in}
     \subfloat{\includegraphics[width=0.47\textwidth,angle=0]{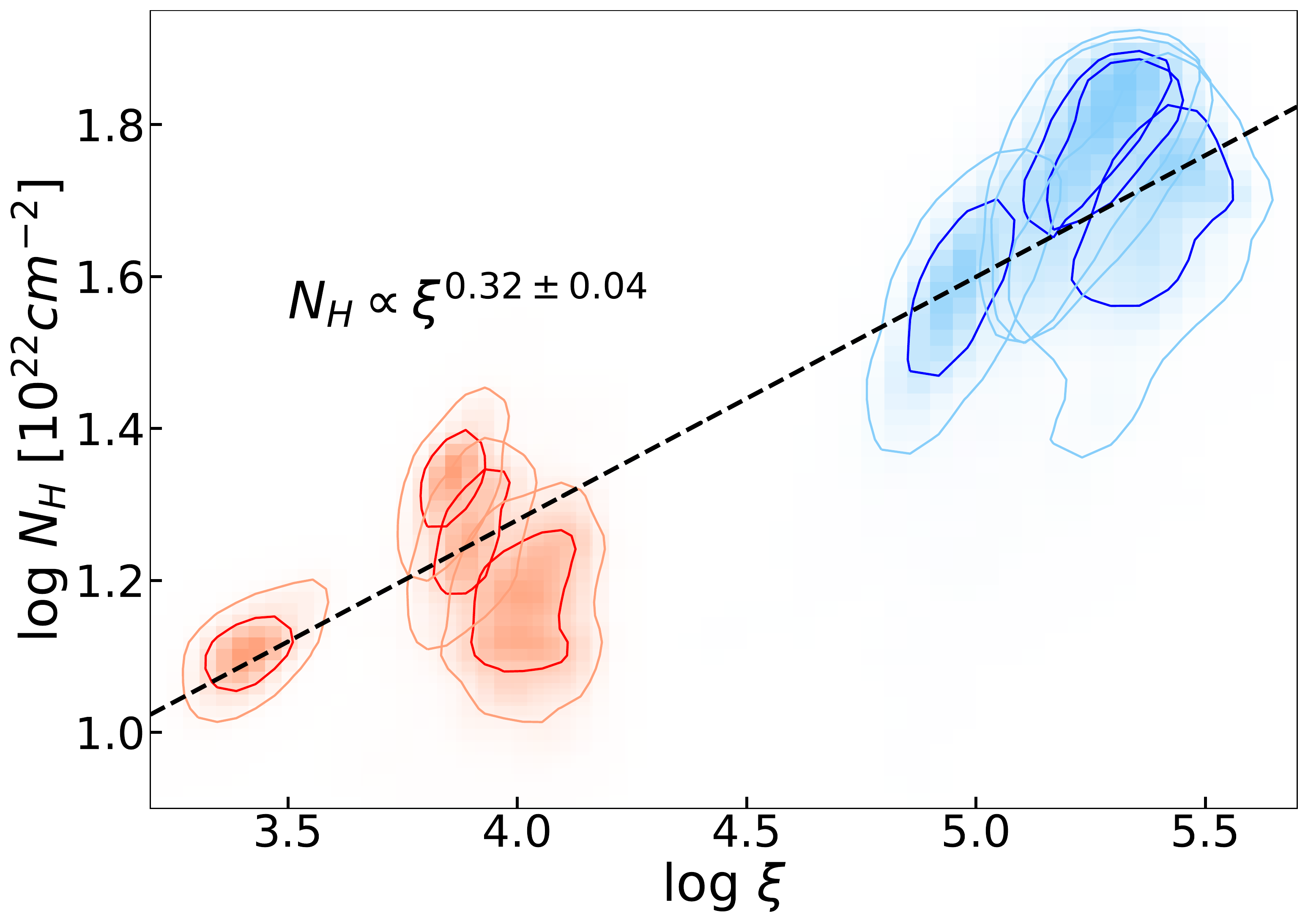}}
  \figcaption[t]{\footnotesize Average absorption measure distribution (AMD) for soft-state observations.
  Top panel: best-fit values and their $1\sigma$ errors as listed in Table \ref{tab:fit}. 
  A 2$-$D histogram of the MCMC chains are plotted as red and blue shaded regions.
  Bottom panel: same as the top panel except points are replaced by contours corresponding to the 68 and $90\%$ confidence intervals of the 2$-$D distribution. The resulting correlation appears to be largely unaffected by any degeneracy in some Zone 1 parameters\label{fig:py_AMD} 
   }
\end{figure}
%------------------------------------------------------------------------------------------------------------------------------------------------------------------------------------------------------------

%------------------------------------------------------------------------------------------------------------------------------------------------------------------------------------------------------------
%------------------------------------------------------------------------------------------------------------------------------------------------------------------------------------------------------------
\begin{figure*}
\centering
    \subfloat{\includegraphics[width=0.8\textwidth,angle=0]{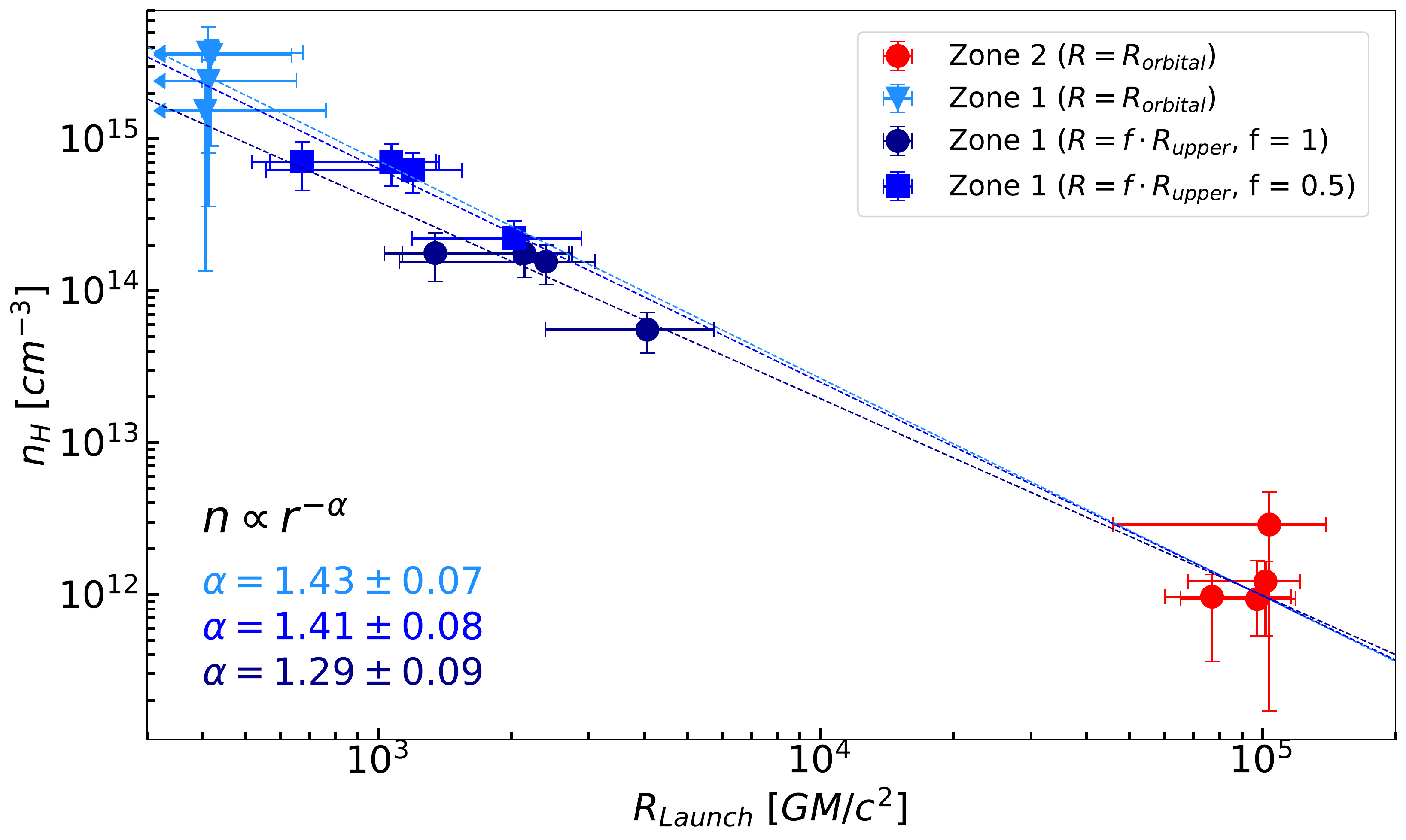}}

  \figcaption[t]{\footnotesize Equivalent hydrogen wind density values for soft-state observations plotted as a function of wind launching radius (${R}_{Launch}$).  
  An arrow at the end of an error bar indicates that a parameter is unconstrained in that direction.
  Zone 2 components were plotted as function of ${R}_{orbital}$, as their emission velocity broadening values are both well-constrained and yield ${R}_{orbital}$ values that are broadly consistent with the upper limit on the photoionization radius (or, ${R}_{upper}$).
  In the case of Zone 1 components, emission velocity-broadening values imply small orbital radii but are not well-constrained.  
  Wind density values for Zone 1 components are therefore plotted as a function of 3 different measurements of ${R}_{Launch}$:  
  (a) ${R}_{orbital}$, (b) ${R}_{upper}$ assuming a filling factor of 1, and (c) ${R}_{upper}$ assuming a uniform filling factor of 0.5 for all observations.
  Option (b) contains the combined largest radius and smallest density values for Zone 1 components.
  Separate radial wind density scalings were found using the combined Zone 2 ${R}_{orbital}$ values with each of the three ${R}_{Launch}$ (and corresponding density values)  for Zone 1.
  Dashed lines represent the best-fit linear scalings (in logarithmic space) of ${n}_{H} \propto {r}^{-\alpha}$ corresponding to each separate measurement of ${R}_{Launch}$ for Zone 1.
  The resulting radial density structure is consistent regardless which measurement of Zone 1 ${R}_{Launch}$ is used, as the best-fit values for scaling parameter $\alpha$ lie within 1$\sigma$ of each other. 
  The best-fit scaling found by Fukumura et al. (2017) for the MHD wind model of GRO J1655-40 (${n}_{H} \propto {r}^{-1.2}$), is also within 1$\sigma$ of option (b).
  Our data are consistent with an MHD outflow, as both the resulting density structure and specific density values are in agreement with numerical work on magnetic winds. 
  Please see the text for details.
     \label{fig:py_hden} }
\end{figure*}
%------------------------------------------------------------------------------------------------------------------------------------------------------------------------------------------------------------
%------------------------------------------------------------------------------------------------------------------------------------------------------------------------------------------------------------

Values of both ${R}_{upper}$ and ${R}_{\text{orbital}}$ lie comfortably below ${R}_{C}$ for both soft-state inner wind components (blue) 
and on average two orders of magnitude smaller than 0.1${R}_{C}$, the lowest estimate on the thermal driving limit.
From launching radii alone, these components are likely magnetic in origin.
Likewise, intermediate-state (cyan and magenta) values of ${R}_{\text{orbital}}$ are 1-2 orders of magnitude below 0.1${R}_{C}$, yet some of their corresponding ${R}_{upper}$ values lie around 0.2${R}_{C}$. 
Because ${R}_{upper}$ is simply an upper bound on the photoionization radius and, even when interpreted literally, these values just barely exceed the strictest limit on thermal driving, it is possible that these components are magnetic as well.

For Zone 2 soft-state wind components (red), both ${R}_{upper}$ and ${R}_{\text{orbital}}$ launching radii estimates lie above 0.1${R}_{C}$ ($\sim$ 0.25) and, in the case of S3, ${R}_{upper}$ extends up to $\sim 1 {R}_{C}$.
Again, ${R}_{upper}$ is only an upper limit and given the agreement in both ${R}_{\text{orbital}}$ and ${R}_{upper}$ among soft-state observations, it is likely that the launching radius of S3 is closer to its ${R}_{\text{orbital}}$ value. 
The large volume filling factors of these outer components approach unity and may be more consistent with thermal winds in this sense,
especially when compared to the small filling factors of the potentially magnetic components. 

The simultaneous detection of both a magnetic inner wind and an outer thermal wind would not be entirely unexpected, as Shakura-Sunyaev disks \citep{SS73} are predicted to have strong magnetic fields. Magnetic forces could then drive winds at the small radii where thermal driving becomes inefficient. 
This is perhaps the case during soft-state observations of 4U 1630-472, as the geometry of the wind suggests two distinct components of different origin.
A more complete picture, however, requires an examination of the physical properties and radial structure of these outflows.

Although analytical treatments suggest that Compton heating can drive winds at higher densities and outflow velocities than previously thought \citep{Done2018}, 
simulations have not been able to achieve outflow velocities larger than ${v}_{\text{out}} \sim 200$ km/s for wind densities above $n \sim {10}^{12} {\text{cm}}^{-3}$ \citep{HigginbottomProga,Higginbottom2017}. 
These values are similar to those we obtained for Zone 2 wind components in the soft-state (${v}_{\text{out}} \sim 200$ km/s and $n \sim {10}^{12-12.5} {\text{cm}}^{-3}$).
As with their launching radii, thermal driving cannot be ruled-out for these outer components based on their outflow velocities and densities.
Conversely, we find that the innermost wind components that we previously identified as magnetic (again, via launching radii estimates) also have considerably higher densities (${n} \sim {10}^{15-16} {\text{cm}}^{-3}$) and outflow velocities (${v}_{\text{out}} \sim 400-1300$ km/s) than the largest values predicted by these simulations.
This could be further indication that we may be simultaneously detecting both a magnetic wind component and a (separate) thermal wind component.

\subsubsection{Wind Structure}\label{sec:structure}

Our models were constructed as two separate wind zones, and our best-fit models suggested that the physical properties of these zones diverge significantly. 
However, at the resolution of the HEG, it is not clear whether these zones are truly separate wind components.
Although the data require two separate zones to model the absorbing wind, our models could simply be capturing two different portions of a continuous self-similar outflow, 
or perhaps something in-between.
%------------------------------------------------------------------------------------------------------------------------------------------------------------------------------------------------------------
%------------------------------------------------------------------------------------------------------------------------------------------------------------------------------------------------------------
\begin{figure*}

\centering

  \subfloat{\includegraphics[scale=0.285,angle=0]{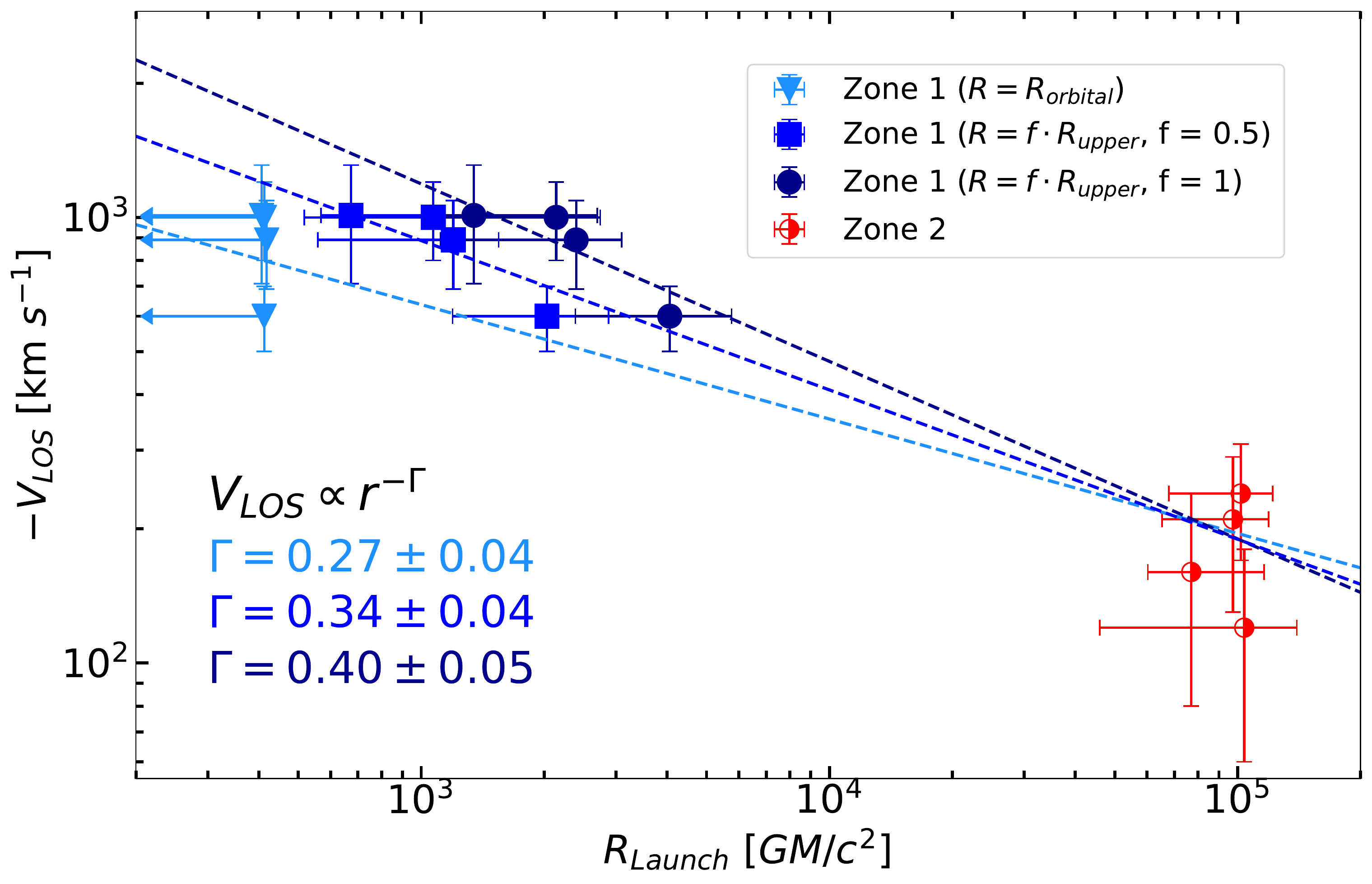}}
  \hspace{0.1 in}
   \subfloat{\includegraphics[scale=0.285,angle=0]{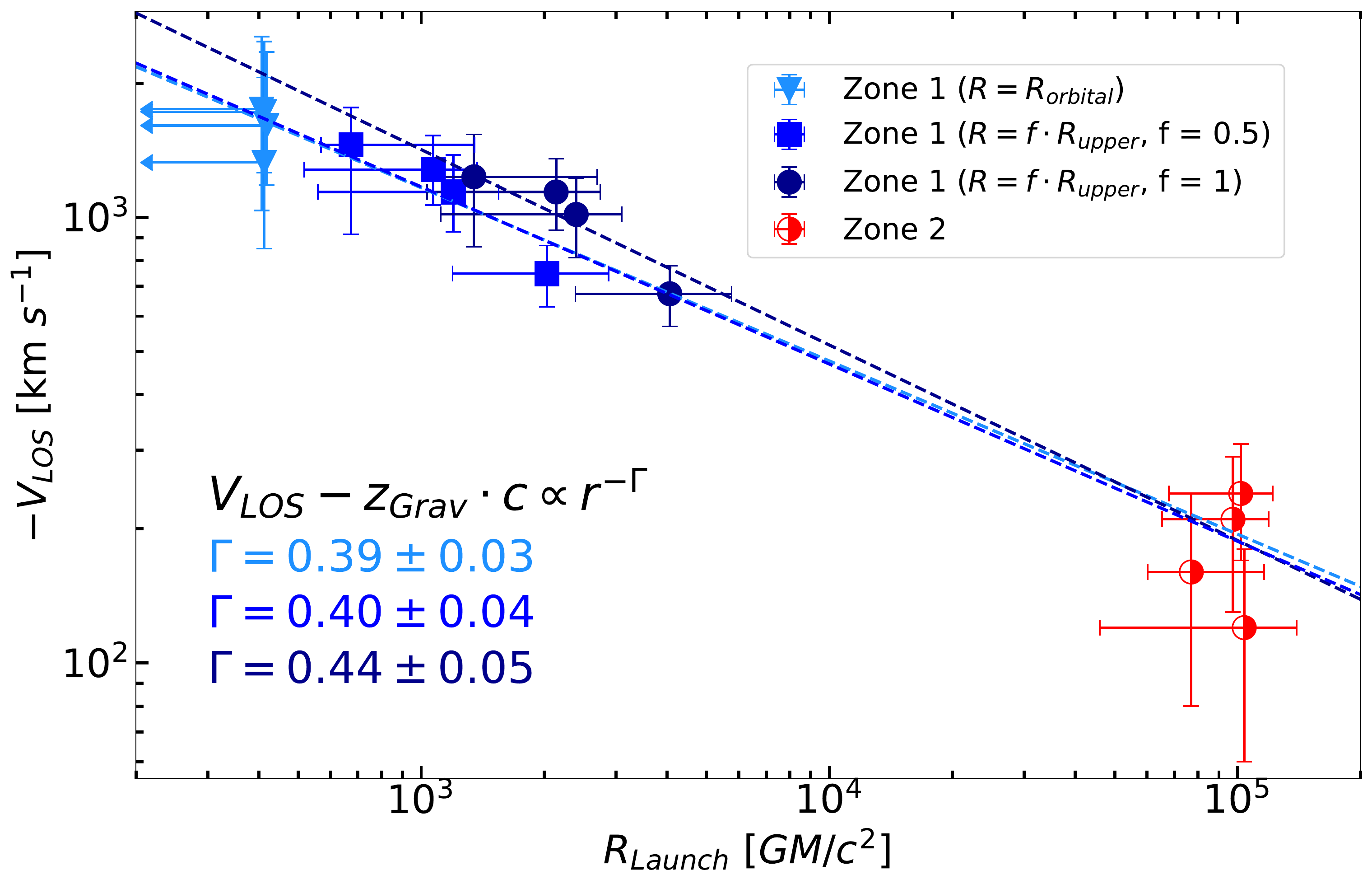}}
  \hspace{0.10 in}

  \caption{\footnotesize 
  Left: wind radial outflow-velocity structure for soft-state observations. 
  As in Figure \ref{fig:py_hden}, Zone 2 components are plotted as function of ${R}_{orbital}$, 
  while Zone 1 components are plotted using 3 separate measurements for ${R}_{Launch}$:  (a) ${R}_{orbital}$, (b) ${R}_{upper}$ assuming a filling factor of 1, and (c) ${R}_{upper}$ assuming a uniform filling factor of 0.5 for all observations. 
  Best-fit scalings (${v}_{LOS} \propto {r}^{-\Gamma}$, dashed lines) vary significantly depending on how ${R}_{Launch}$ of Zone 1 is measured.
  In addition, the observed line-of-sight velocity scalings are significantly flatter than the ${v} \propto {r}^{-0.5}$ scaling required for self-similarity. 
  Right: same as the left panel, except observed Zone 1 line-of-sight velocities are corrected for gravitational redshift (${z}_{Grav.} = \frac{GM}{{c}^{2}}/{R}_{Launch}$). 
  In Zone 1, the increase in ${v}_{LOS}$ ranges between 70 and 220 km/s if ${R}_{Launch}={R}_{upper}$, increasing up to $\sim$ 700 km/s if ${R}_{orbital}$ is used instead. This correction is negligible for Zone 2.
  Once corrected for ${z}_{Grav.}$, the resulting velocity structure is largely independent of how ${R}_{Launch}$ is measured in Zone 1, as the best-fit values for $\Gamma$ lie within 1$\sigma$ of each other.
  Compared to the left panel, these corrected scalings converge much closer to a ${v} \propto {r}^{-0.5}$ scaling, suggesting that this may be a self-similar wind   \label{fig:py_vel}}
\end{figure*}
%------------------------------------------------------------------------------------------------------------------------------------------------------------------------------------------------------------
%------------------------------------------------------------------------------------------------------------------------------------------------------------------------------------------------------------

Figure \ref{fig:py_AMD} shows the \textit{average} absorption measure distribution (or, AMD) for the four soft-state observations. 
The AMD relates two independently measured quantities (ionization and absorbing column), both of which depend on the density and geometry of the absorber.
The underlying radial density structure of the absorber can be revealed once constraints on the system can be obtained independently}.
Figure \ref{fig:py_hden} shows wind density values, ${n}_{H}$, plotted against launching radii, ${R}_{\text{Launch}}$ for the four soft-state observations. 
This plot presents the same information as Figure \ref{fig:py_AMD}, with the added constraint of including the distance of the absorber (and therefore the filling factor, f).
For Zone 2, ${R}_{\text{orbital}}$ provides a reliable estimate of ${R}_{\text{Launch}}$ as it is both well-constrained and largely agrees with ${R}_{\text{upper}}$,
the latter being easier to measure\footnote[2]{The degeneracy between ${\sigma}_{\text{emis}}$, 
the assumed turbulent velocity of the absorber (${v}_{turb}$),
and the emission covering factor ($\Omega$) is discussed in Section \ref{sec:soft}.
We found that lowering ${v}_{turb}$ and $\Omega$ has little effect on the resulting best-fit values and would likely only contribute some additional scatter in these plots. 
For simplicity, we only discuss the results of our original fits.}.
This is not the case for Zone 1- although the broadening values of the re-emission suggest ${R}_{\text{orbital}}$ is small, this value is not well constrained. 
Therefore we provide three separate estimates for ${R}_{\text{Launch}}$ (and corresponding density value) in Zone 1, each of which we compare against ${R}_{\text{orbital}}$-derived values of Zone 2:
(a) ${R}_{orbital}$, (b) ${R}_{upper}$ assuming a filling factor of 1, and (c) ${R}_{upper}$ assuming a uniform filling factor of 0.5 in Zone 1 across all observations.
For the latter two, ${R}_{upper}$ is relatively well-constrained and fairly uniform across all observations, 
with $f = 1$ resulting in the combination of the largest possible radii and lowest possible density values.

Strikingly, the underlying radial density structure of the wind is largely insensitive to which estimate of ${R}_{\text{Launch}}$ is adopted for Zone 1.
When fit separately, the resulting ${n} \propto {r}^{-\alpha}$ scalings (with $\alpha$ ranging from $1.29 \pm 0.09$ to $1.43 \pm 0.07$) are all within $1\sigma$ of each other.
Given the constancy of the wind during these four observations, we deemed using a uniform value of f as an acceptable assumption. 
However, the strong agreement between the scalings plotted in Figure \ref{fig:py_hden} suggests even if $f$ in Zone 1 varied significantly between observations, 
the resulting scaling would lie somewhere in this narrow range.
The specific values of $\alpha$ cluster around the $\alpha = 1.4$ scaling reported by \citet{Chakravorty2016} in their theoretical work on MHD winds in XRBs. 
This scaling, however, corresponds to their most extreme warm MHD solution and they were unable to produce outflows at the densities and small radii typical of XRB winds within the scope of their work.
%------------------------------------------------------------------------------------------------------------------------------------------------------------------------------------------------------------
%------------------------------------------------------------------------------------------------------------------------------------------------------------------------------------------------------------
\begin{figure*}

\centering
  \subfloat{\includegraphics[scale=0.25,angle=0]{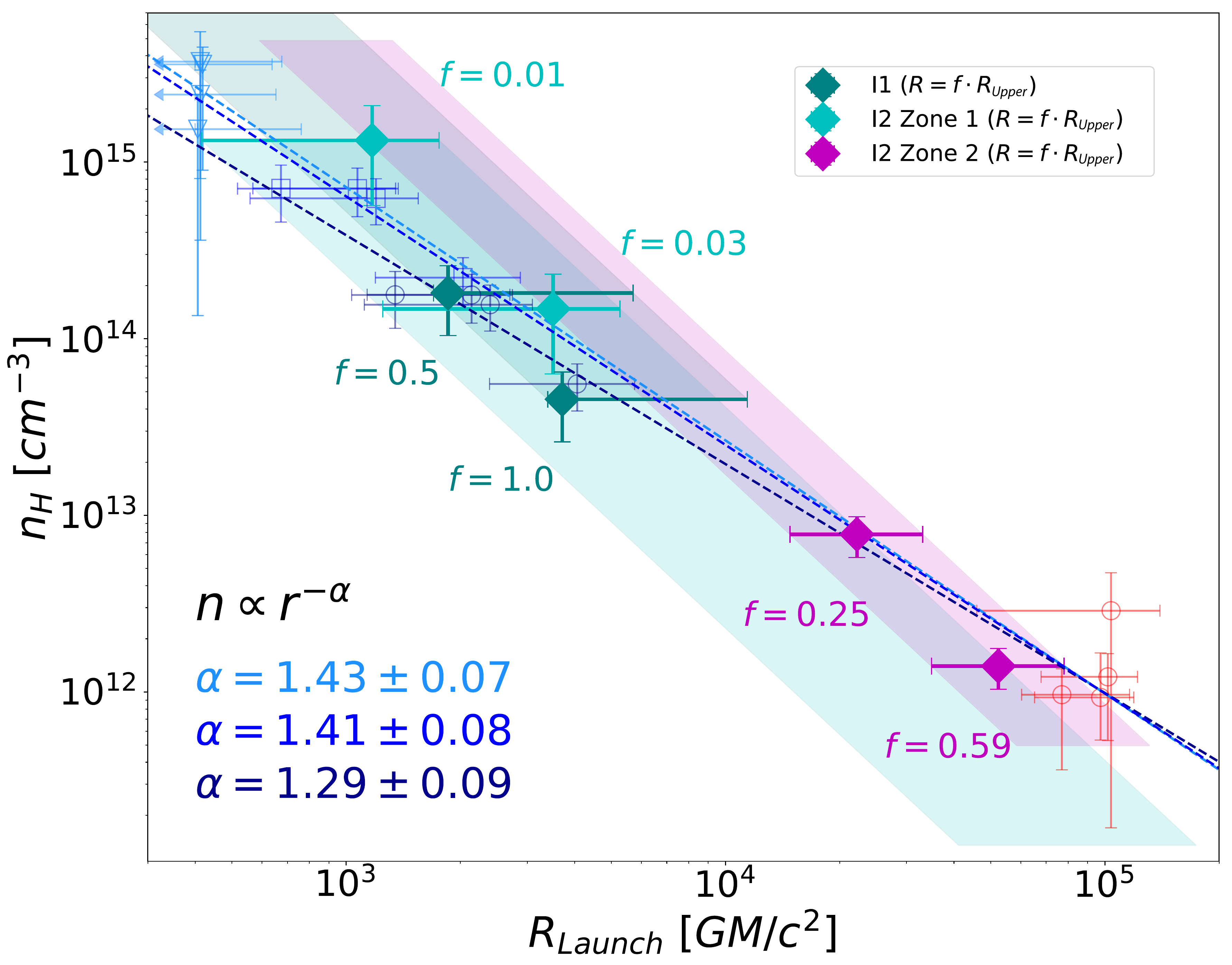}}
   \subfloat{\includegraphics[scale=0.25,angle=0]{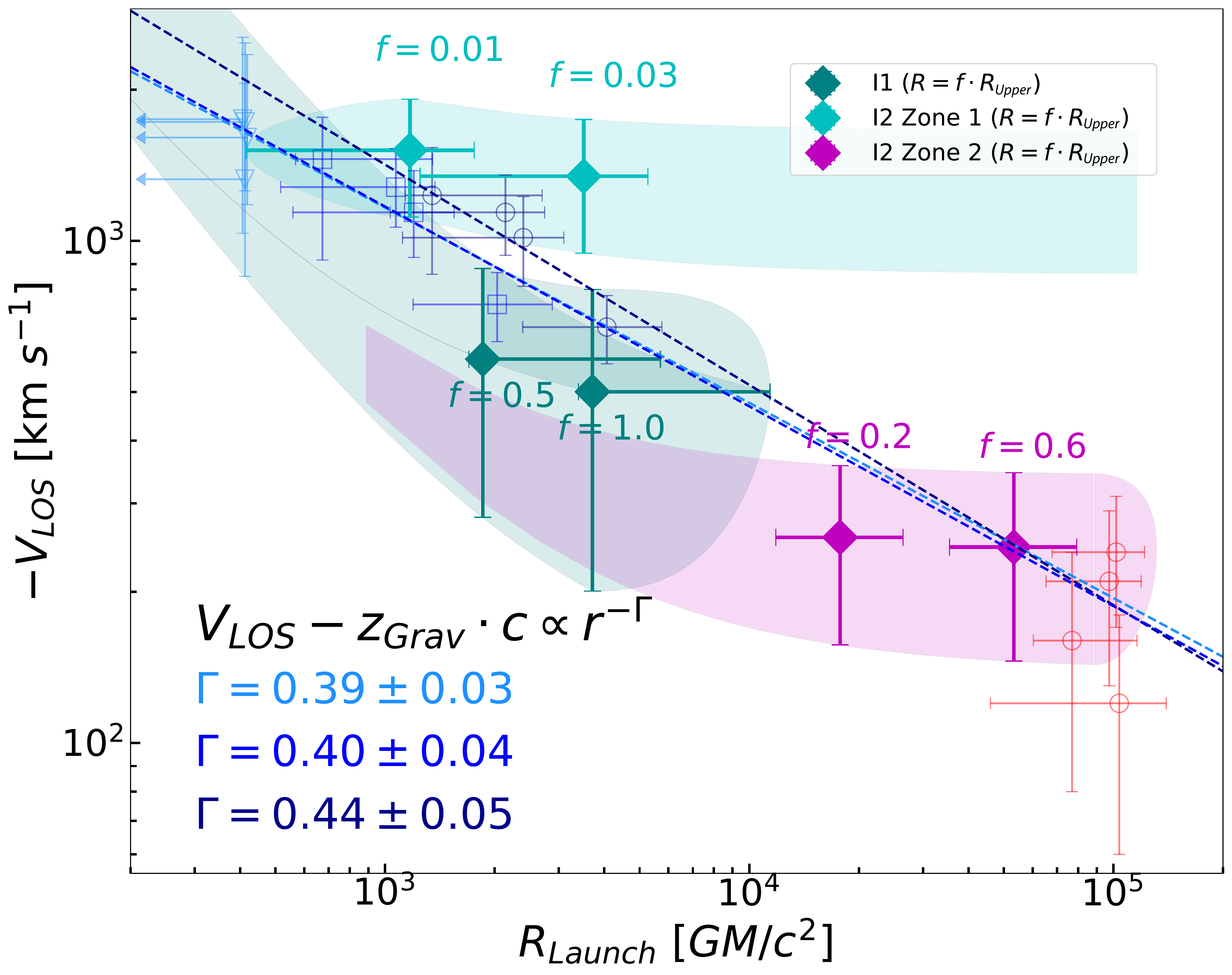}}
  \caption{\footnotesize 
Same as Figures \ref{fig:py_hden} (left pandel) and \ref{fig:py_vel} (right panel), except intermediate-state observations are included in both plots. 
These plots are meant simply to illustrate both how the observed winds in intermediate-state observations may fit a coherent picture along with soft-state observations,
while still highlighting the uncertainties involved when analyzing weakly absorbing and emitting winds.
For visual clarity, points from soft-state observations are faded. 
Due to the difficulties in constraining ${R}_{orbital}$, each intermediate-state wind component is plotted 
as function of ${R}_{Launch} = f \cdot {R}_{upper}$ as shaded regions. 
Each region spans a range of filling factor values (from 0.01 to 1) and its extent is determined by the $1\sigma$ error at each point.
For each wind component, the endpoints at which a region which is simultaneously $\sim1\sigma$ away from both the density and velocity scalings (obtained using soft-state observations) 
are plotted explicitly with error bars and labelled with the corresponding filling factor values. 
These points are meant to very roughly define the plausible range of filling factors and launching radii might be for these components if the wind density and velocity structure between soft and intermediate state observations is similar, and should not be interpreted literally. 
 \label{fig:py_inter} }
\end{figure*}
%------------------------------------------------------------------------------------------------------------------------------------------------------------------------------------------------------------
%------------------------------------------------------------------------------------------------------------------------------------------------------------------------------------------------------------

Unlike \citet{Chakravorty2016}, \citet{Fukumura2017} used their theoretical MHD wind framework in order to reproduce the absorption features in the Chandra/HETGS spectrum of GRO J1655-40.
They found that an MHD wind model with $\alpha = 1.2$ best describes the wind absorption present in that particular source, with outflow velocities up to ${v}_{\text{out}} \sim 4000$ km/s and very high absorbing columns (${N} \sim {10}^{24}$ ${\text{cm}}^{-2}$ for some ions). 
They also find wind density values of ${n} \sim 0.7 \times {10}^{15}$  and ${n} \sim 0.9 \times {10}^{11} {\text{cm}}^{-3}$  at
$r = 400$ and $r = {10}^{5} {GM/{c}^{2}}$ (characteristic radii of the inner and outer soft-state components in 4U 1630-472), respectively.
These values are close to what we found in 4U 1630-472 despite specifically being fit to the spectrum of GRO J1655-40 
and only require increasing the density normalization (${\tilde{n}}_{17}$, where $n = {\tilde{n}}_{17}  \times {(r/{r}_{g})}^{-1.2}$) 
by a factor of three in order to be \textit{broadly} consistent with our density structure.
Most notably, perhaps, the scaling of $\alpha = 1.2$ found by \citet{Fukumura2017} is $1\sigma$ away from what we obtained using $f=1$.
These similarities could perhaps mean that our outer soft-state components (which we previously identified as thermal) are a part of a broad MHD outflow.

The left panel in Figure \ref{fig:py_vel} shows how the line of sight outflow velocity scales with ${R}_{Launch}$. 
As with figure \ref{fig:py_hden}, we provide separate best-fit scalings of ${v}_{\text{LOS}} \propto {r}^{-\Gamma}$, 
for each of the three different estimates of  ${R}_{Launch}$ in Zone 1. 
The resulting scalings diverge from each other, and the underlying velocity structure appears to be highly dependent on how ${R}_{Launch}$ is estimated. 
In addition, self-similarity requires for $v \propto {r}^{-1/2}$ \citep{Fukumura2010, Fukumura2015,Fukumura2017,Chakravorty2016, Zanni2007, BlandfordPayne}, 
a scaling which is well outside the range the $\Gamma = 0.27 \pm 0.04$ to $\Gamma = 0.40 \pm 0.05$ we observed in this case.

The effect of gravitational redshift in the measured outflow velocities of black hole winds is rarely discussed.
This effect is negligible in many cases (e.g. in Zone 2, this redshift would likely not exceed 10 km/s based on plausible values of ${R}_{Launch}$),
yet ${R}_{upper}$ places a strict lower limit on what this correction should be. 
The right panel of Figure \ref{fig:py_vel} shows the resulting radial outflow velocity structures for the three different estimates of ${R}_{Launch}$ in Zone 1
after correcting velocity values by the gravitational redshift at that specific radius.
Although this correction can be as small as $100-200$ km/s given $f=1$, or as large as 700 km/s when using ${R}_{orbital}$,
the underlying velocity structure appears largely insensitive to the choice of ${R}_{Launch}$ estimate, 
with $\Gamma = 0.39 \pm 0.03$, $0.4 \pm 0.04$, and $0.44 \pm 0.05$. 
Besides consistency, these values also cluster much closer to a $v \propto {r}^{-1/2}$ scaling (just shy of $1\sigma$ away when using $f=1$).
Once outflow velocities are corrected for gravitational redshift, the velocity structure of the wind during soft-state observations 
closely resembles a self-similar wind. This would again be consistent with a single, continuous MHD outflow, rather than separate thermal and magnetic wind components.

The large discrepancy between ${R}_{orbital}$ and ${R}_{upper}$ in intermediate-state observations 
poses a challenge in trying to include them in this picture. 
Although the discrepancy between these two values is not inherently problematic (small filling factors are common in many astrophysical plasmas),
the discrepancy arises because re-emission in this wind is not very prominent, and therefore ${R}_{orbital}$ is hard to constrain. 
The left and right panels of Figure \ref{fig:py_inter} are the same plots as Figures \ref{fig:py_hden} and \ref{fig:py_vel}, respectively,
with intermediate-state observations included as shaded regions. 
Given the uncertainty with ${R}_{orbital}$, these regions are plotted entirely function of ${R}_{Launch}$ = $f \cdot {R}_{upper}$,
trancing its $1\sigma$ along a range of filling factor values.
Although our fits strayed away from the small re-emission velocity broadening values that would correspond to large filling factors,
the shaded regions span a range of $f = 0.01$ up to 1.0.

The left panel of Figure \ref{fig:py_inter} shows that, for a range of filling factors, these regions lie within $1\sigma$ of the plausible density structure obtained by fitting soft-state observations only. On the right panel, and especially for observation I2, this range is narrower by comparison.
We also explicitly plotted the points at which each region was $\sim 1\sigma$ away from the nearest scaling in \textit{both} panels simultaneously,
representing a very rough acceptable range of filling factors if the wind in intermediate-state observations had the same structure as those during soft-state observations.
It is important note that these points specifically, and Figure \ref{fig:py_inter}, are mainly included in this work for illustrative purposes. 
Even if the scalings found by fitting our results from soft-state observations are real, the winds found in intermediate-state observations do not necessary have to follow them, or even have the same normalizations.

%------------------------------------------------------------------------------------------------------------------------------------------------------------------------------------------------------------
%------------------------------------------------------------------------------------------------------------------------------------------------------------------------------------------------------------
\begin{table*}
\caption{Wind Outflow Parameters}
\begin{footnotesize}
\begin{center}
\begin{tabular*}{\textwidth}{l c @{\extracolsep{\fill}} ccc cccc }
\tableline
\\ [-3.0ex]
\tableline
Obs. Label & Zone &  $\dot{M}_{wind}$ & $\dot{M}_{wind}/\dot{M}_{edd}$ & $\dot{M}_{wind}/\dot{M}_{in}$ & ${L}_{rad}$ & ${L}_{rad}/{L}_{edd}$ & ${L}_{wind}$ & ${L}_{wind}/{L}_{rad}$\\
  	    & 	   & (${10}^{18}$ g/s)  &                                                    & 	                                                & (${10}^{38}$ erg/s) &			& (${10}^{32}$ erg/s) & (${10}^{-7}$)\\
\\ [-2.5ex]
\tableline 
\\ [-2.5ex]
S1 & 1 & $0.48^{+0.27}_{-0.16}$ & $0.034^{+0.020}_{-0.012}$ & $0.15^{+0.08}_{-0.05}$ & $2.93^{+0.30}_{-0.22}$ & $0.23^{+0.03}_{-0.02}$ & $17^{+24}_{-7}$ & $58^{+82}_{-24}$\\
	   & 2 &  $3.5^{+0.7}_{-2.7}$ & $0.25^{+0.05}_{-0.20}$ & $1.1^{+0.2}_{-0.8}$ & ... & .. & $3.7^{+3.3}_{-3.7^{\dagger}}$ & $20^{+17}_{-20^{\dagger}}$\\
\\ [-2.5ex]
\tableline 
\\ [-2.5ex]
S2 & 1 &$0.49^{+0.23}_{-0.30}$ & $0.035^{+0.017}_{-0.021}$ & $0.17^{+0.08}_{-0.10}$ & $2.68^{+0.29}_{-0.21}$ & $0.21(2)$ & $30^{+18}_{-27^{\dagger}}$ & $110^{+70}_{-100^{\dagger}}$\\
 		& 2 &$5.7 \pm 2.2$ & $0.41(16)$ & $1.9(8)$ & ... & .. & $12.5^{+7.9}_{-12.5^{\dagger}}$ & $65^{+41}_{-65^{\dagger}}$\\
\\ [-2.5ex]
\tableline 
\\ [-2.5ex]	    
S3 & 1 &  $0.20^{+0.12}_{-0.07}$ & $0.014^{+0.009}_{-0.005}$ & $0.08^{+0.05}_{-0.03}$ & $2.37^{+0.16}_{-0.19}$ & $0.18(1)$ & $2.9^{+3.8}_{-1.3}$ & $12^{+16}_{-6}$\\
& 2 & $3.2^{+1.2}_{-2.5^{\dagger}}$ & $0.23^{+0.08}_{-0.18^{\dagger}}$ & $1.2^{+1.4}_{-1.0^{\dagger}}$ & ... & .. & $2.4^{+2.6}_{-2.3^{\dagger}}$ & $14^{+15}_{-13^{\dagger}}$\\
\\ [-2.5ex]
\tableline 
\\ [-2.5ex]	    
S4 & 1 & $0.40^{+0.26}_{-0.15}$ & $0.029^{+0.019}_{-0.010}$ & $0.11^{+0.07}_{-0.04}$ & $3.19^{+0.39}_{-0.16}$ & $0.25^{+0.03}_{-0.01}$ & $14^{+17}_{-4}$ & $45^{+55}_{-14}$\\
	    & 2 &$5.9^{+1.7}_{-2.5}$ & $0.42^{+0.13}_{-0.18}$ & $1.7^{+0.45}_{-0.7}$ & ... & .. & $16^{+9}_{-15}$ & $86^{+45}_{-79}$\\
\\ [-2.5ex]
\tableline 
\\ [-3.0ex]
\tableline 
\\ [-2.5ex]
I1 & 1 &  $0.10^{+0.11}_{-0.06}$ & $0.007^{+0.008}_{-0.004}$ & $0.03^{+0.03}_{-0.02}$ & $3.11^{+0.25}_{-0.11}$ & $0.25^{+0.02}_{-0.01}$ & $0.49^{+2.51}_{-0.01^{\dagger}}$ & $1.60^{+8.10}_{-0.04^{\dagger}}$\\
\\ [-2.5ex]
\tableline 
\\ [-2.5ex]   
I2 & 1 &$0.04^{+0.04}_{-0.02^{\dagger}}$ & $0.003^{+0.003}_{-0.002^{\dagger}}$ & $0.013^{+0.012}_{-0.007^{\dagger}}$ & $2.92^{+0.05}_{-0.02}$ & $0.232^{+0.004}_{-0.002}$ & $2.2^{+4.8}_{-2.0^{\dagger}}$ & $7.7^{+16}_{-6.7^{\dagger}}$\\
	 & 2 &$0.05^{+0.03}_{-0.04}$ & $0.003^{+0.002}_{-0.003}$ & $0.01(1)$ & ... & .. & $0.08(8)$ & $0.28^{+0.26}_{-0.28^{\dagger}}$\\
\\ [-2.5ex]
\tableline 
\\ [-3.0ex]
\tableline

\end{tabular*}
\vspace*{-1.0\baselineskip}~\\ \end{center} 
\tablecomments{\footnotesize The table above lists estimates for various wind outflow parameters dervied from the best-fit models listed in Table  \ref{tab:fit}.
Quoted errors are at the $1\sigma$ level. Errors annotated with $\dagger$ indicate a parameter with unconstrained behavior outside the $1\sigma$ confidence interval.
Mass outflow rates were calculated by correcting for volume filling factor via $\dot{M}_{out} = 2\pi\mu{m}_{p}\cdot{v}_{abs} \cdot f \cdot L/\xi$, which is equivalent to 
$\dot{M}_{out} = 2\pi\mu{m}_{p}GM\cdot{v}_{abs}\cdot {N}_{H}/{\sigma}_{emis}^{2}$ 
(where L is the illuminating luminosity incident on a zone and not the intrinsic luminosity of the disk, $\mu$ is the mean molecular weight and fixed at 1.23, 
${m}_{p}$ is the proton mass, and M is the black hole mass).
Wind kinetic luminosity values were calculated using these filling factor corrected
outflow rates via $\dot{L}_{wind} = 0.5\dot{M}_{out}{v}_{abs}^{2}$. 
The intrinsic luminosity of the disk (${L}_{rad}$; equivalent to ${L}_{illum}$ for Zone 1) was used to estimate the mass accretion rate, $\dot{M}_{in}$,
using an assumed efficiency of $\eta=0.1$. 
Errors listed for ${L}_{rad}$ are based solely on the correlation with ${N}_{H}$ of the absorbing wind. 
See text for additional details.}
\vspace{1.0\baselineskip}
\end{footnotesize}
\label{tab:mdot}
\end{table*}

%------------------------------------------------------------------------------------------------------------------------------------------------------------------------------------------------------------

\subsubsection{Outflow Parameters}\label{sec:mdot}

Table \ref{tab:mdot} lists wind outflow properties derived from our best-fit models. 
The intrinsic source luminosity is given by the illuminating luminosity incident on Zone 1 (corrected for viewing angle), 
and a corresponding accretion rate of $\dot{M}_{in} = L/\eta{c}^{2}$, with $\eta = 0.1$.
The wind mass outflow rate was calculated using:
\begin{center}
$\dot{M}_{\text{out}} \simeq  2 \times (2\pi\mu{m}_{p}\cdot{v}_{\text{abs}}\cdot{r}^{2}n\cdot f)$, 
\end{center}
where $2 \times$ accounts for outflows on both sides of the disk. 
With ${r}^{2} n = L/\xi$ and $f = {R}_{\text{orbital}}/{R}_{upper} = ({N}_{H} \xi /L) \cdot GM/{\sigma}_{\text{emis}}^{2}$,
this becomes:
\begin{center}
$\dot{M}_{\text{out}} = 4\pi\mu{m}_{p}GM\cdot{v}_{\text{abs}}\cdot {N}_{H}/{\sigma}_{\text{emis}}^{2}$.
\end{center}
Wind kinetic luminosities were calculated using ${L}_{\text{wind}} = 0.5 \dot{M}_{\text{out}} {v}_{\text{abs}}^{2}$.

Although winds launched from Zone 1 only make up a small fraction of the total outflow rate, they constitute the majority of the total wind kinetic luminosity. 
Once corrected for $f$, $\dot{M}_{\text{out}}$ is no longer dependent on luminosity, making it possible to track how mass outflow rates evolve with each other without making circular arguments. Unsurprisingly, outflow rates trend roughly towards higher $\dot{M}_{\text{out}}$ with luminosity. 
However, all datasets can be considered flat within errors, and Zone 1 and total $\dot{M}_{\text{out}}$ not are not even rank-correlated with luminosity.

%----------------------------------------------------------------------

\section{Discussion and Conclusions}\label{sec:discussions}
%------------------------------------------------------------------------------------------------------------------------------------------------------------------------------------------------------------

We have re-analyzed all archival Chandra/HETG spectra of 4U 1630-472 that show definitive evidence of an absorbing disk wind.
The analysis was performed using PION \citep{Kaastra1996}, a SPEX absorption model that calculates the ionization of the absorbing plasma self-consistently with the unabsorbed source continuum.
Extended wind re-emission was implemented and dynamically broadened to the order of the local Keplerian velocity.
Fitting these self-consistent models using Markov chain Monte Carlo with physically motivated priors and a specialized fitting procedure resulted in: 
(1) Better statistical fits in Fe K band while simulateously capturing lines at lower energies (Ca XX and Ar XVIII) and the Fe K $\alpha$/$\beta$ line ratios (2), a better understanding of parameter errors and how they propagate when deriving outflow properties. 

With the exception of I1 (ObsID 4568), the spectra of 4U 1630-472 required two distinct photoionization zones to model the absorbing winds. 
For the four soft-state observations, we find that these photoionization zones follow the same pattern: 
A highly ionized and broadened inner wind component (Zone 1) launched at large outflow velocities, and an outer component (Zone 2), launched at a much lower velocity, ionization, and broadening.
This trend in ionization and velocities is consistent with that found by \citet{Miller2015} for S2, and with what is generally expected in these sources. 
Our results, however, indicate that the lower and higher ionization components are much lower and higher in ionization, respectively, than previously suggested.
We also find the absorbing column of the inner component to be much larger than previously reported,
with values for S1 and S4 being somewhat unconstrained (due to higher ionization in Zone 1) and approaching the regime of Compton-thick winds, a scenario that can be ruled-out by the lack of variability.
There are strong indications that these higher ionization parameter and equivalent hydrogen column density values are real: 
1) The absorbing columns in S2 and S3 are similarly large, yet well constrained and well below the Compton-thick regime; 2) The model luminosities for all soft-state observations strongly follow a ${T}^{4}$ trend despite very different absorbing columns, especially when compared to the models without wind absorption; and 3) the presence of lines at lower energies require a gas of lower ionization than what previously reported. 

Compared to the results by \citet{Gatuzz2019} using a single absorption zone and ``warmabs", we were able to achieve better fits for the broad absorption spectrum in the soft-state observations, including the strong Ca XX and Ar XVIII lines which their fits fail to capture, by using two separate absorbers.
Gatuzz et al. (2019) obtained adequate Fe $\alpha$/$\beta$ line ratios by fitting the turbulent velocity of the absorber down to values of 150-200 $\text{km/s}$, 
below what is typically observed in these sources.
The addition of wind re-emission in our model resulted in better fits to this Fe K$\beta$ region.

We report values for wind launching radii based on the velocity broadening of wind re-emission, and well as the upper limit on the photoionization radius. These estimates do not assume a fiducial density.
In both cases, estimates for wind launching radii rule out thermal driving for all Zone 1 components in soft-state observations ($r \sim {10}^{2-3}$ $GM/{c}^{2}$), 
as well as for all intermediate-state components. 
Launching radii estimates of the remaining soft-state Zone 2 winds lie between 0.1${R}_{C}$ and 1.0${R}_{C}$ ($r \sim {10}^{5}$ $GM/{c}^{2}$),
meaning that thermal driving cannot be ruled-out based on this criteria alone. 
The launching radii reported by \citet{Gatuzz2019} are broadly consistent with these Zone 2 winds, though their results assume a density. 

It has been suggested that massive thermal winds could be launched from radii below 0.1${R}_{C}$ as the source approaches ${L}_{Edd}$, 
once factors such as radiation pressure enhancement are implemented that more in realistic treatments of Compton heated winds \citep{Done2018},
with this limit extending down to 0.01${R}_{C}$ at $L \sim 0.67{L}_{Edd}$ \citep{ProgaKallman}.
This is likely not the case in 4U 1630-472, as the highest model luminosity we obtain is $0.25{L}_{Edd}$ for S4. 
Again, it is very likely that  $L < {L}_{Edd}$ given that continuum is disk-dominated and their luminosities follow a ${T}^{4}$ scaling.
Using our highest model luminosity of $0.25{L}_{Edd}$, \citet{Done2018} and the equations in \citet{ProgaKallman} would predict that thermal winds could be launched 
from radii as small as ${R}_{in} \sim {10}^{5} GM/{c}^{2}$, or $\sim 0.25{R}_{C}$.
This limit is still orders of magnitude larger than the launching radii of the innermost wind components,
while outermost wind components in the soft-state lie right at this limit.

\citet{Done2018} also propose that an additional, high-${N}_{H}$, 
completely ionized thermal wind could perhaps drive these outer components via pressure confinement. 
In this case, L would be approaching ${L}_{Edd}$ but appear less luminous due to electron scattering from 
this completely ionized component that is undetected due to the lack of absorption lines.
As mentioned earlier, there is significant evidence pointing to L being significantly lower than ${L}_{Edd}$.
In addition, our best-fit models already have relatively high absorbing columns, and therefore an additional, nearly co-spatial, high-${N}_{H}$ component would result in a Compton-thick photoionization zone.
These outflows would be clumpy \citep{King2015} and would result in high variability, something which we do not observe in our light curves.

Depending on the accepted theoretical model, the components lie at or below the lower limit for thermal driving ($\sim 0.25{R}_{C}$).
However, if the conditions in the disk are such that a thermal wind can be efficiently driven from 0.25${R}_{C}$,
then the lack of any wind components above $1{R}_{C}$ \citep[where most of the mass loss occurs for thermal winds;][]{Done2018} should raise some suspicions. Our best-fit models suggest that Compton heating is failing to drive disk winds at the very radii where it is expected to be the most efficient. One plausible explanation could be changes in the disk geometry that may obscure the central engine from the disk surface at large radii, but this is highly speculative. 
If instead these outer components are magnetically driven, then the lack of massive thermal winds at large radii may simply mean that, at the time of these observations, Compton heating is inefficient at all radii compared to magnetic driving.  

Some magnetic outflows may be the product of both magnetic and thermal driving, with varying degrees of contribution from each physical process 
\citep{WatersProga,Bai2016}.
These outer components may be one of these magneto-thermal hybrid winds \citep[such as those suggested via wind evolution in the 2005 outburst of GRO J1655-40;][]{NeilsenHoman}, with strong contributions from both magnetic and thermal driving. 
This could explain why these outer winds might be thermal in appearance (low density, low velocity, and high filling factor), 
yet are both the only thermal components detected in the source and are launched from the smallest radius possible for thermal driving.

A third plausible explanation could be that magnetic processes may, in some scenarios, suppress thermal winds. 
In their simulations of disk winds in LMXBs, \citet{WatersProga} found that by adding strong poloidal magnetic fields to systems with Compton-heated thermal winds,
the existing converging-diverging geometry which is conducive to wind acceleration can be disrupted at about $\sim {R}_{C}$.
This change in geometry may suppress thermal winds at their characteristic radii,
while still allowing for magnetic outflows and magnetically enhanced thermal winds. 
This may explain both why we see such high mass loss rates from these ``thermal" winds in our model despite being launched from the smallest radius possible for thermal driving, and why we do not see any thermal winds above 1${R}_{C}$. 
However, \citet{WatersProga} only perform their simulations down to 0.5${R}_{C}$, so we cannot make comparisons with the rest of our wind components.

Finally, our two wind zones may be different regions of a single continuous outflow, as we cannot tell from the data whether these are truly separate components. 
Our results are similar to those obtained by \citet{Fukumura2017}, where they modeled the wind absorption in GRO J1655-40 using their MHD wind models.
Although our radial density structure is slightly steeper (at $n\propto {r}^{-\alpha}$, $\alpha =1.29 \pm 0.09$ to 1.43 $\pm 0.07$), we obtain similar wind density values at similar radii and our data is broadly consistent with the their best fit $n \propto {r}^{-1.2}$ scaling (Figure \ref{fig:py_hden}). 
Our wind absorption measure distribution (${N}_{H} \propto {\xi}^{(\alpha-1/\alpha-2)}$) is also similar to their findings ($\alpha$ = 1.24 vs 1.2), 
although this depends on how this scaling is defined (see Section \ref{sec:rad_mdot}). 

One strong discrepancy with \citet{Fukumura2017} is their assumed ${v}^{-1/2}$ scaling, a product of self similarity. 
Although this scaling might indeed be a real description of the velocity structure of these outflows, this is not necessarily the case for the measured line-of-sight velocity, ${v}_{\text{LOS}}$. We find that our velocities have a significantly flatter scaling of  ${v}_{\text{LOS}} \propto {r}^{-\Gamma}$, $\Gamma = 0.27\pm 0.04$ to ${0.4\pm 0.05}$) and are highly dependent on how ${R}_{Launch}$ is measured in Zone 1.
Once outflow velocities are corrected for gravitational redshift, the resulting scalings are much closer to self-similarity (${v}_{\text{LOS}} \propto {r}^{-\Gamma}$, $\Gamma = 0.39\pm 0.03$ to $0.44\pm 0.05$), and are mostly insensitive the ${R}_{Launch}$ estimate used.
If these outflows are indeed self-similar, then a plausible explanation for the remaining discrepancy could be that the angle between the LOS and the outflow velocity vector decreases with radius. 
This would mean that the wind velocity vector would become increasingly orthogonal to the disk surface at smaller radii, 
a feature that is observed in simulations of MHD disk winds but rarely quantified \citep{WatersProga}.
Another factor is that faster wind components at smaller radii may be observed at lower elevations from the disk surface, and have not been fully accelerated 
along magnetic field lines \citep{Luketic2010}. 
Although this correction may perhaps be useful when connecting simulations of MHD winds to observation in the future, 
this will likely require the higher resolving power and sensitivity of the next generation of X-ray telescopes.

\acknowledgments
We would like to thank the anonymous referee for comments and suggestions that improved this manuscript. 
We acknowledge helpful discussions with Juliette Becker and Christopher Miller.

\appendix
\section{Markov chain Monte Carlo}\label{sec:MCMC}
MCMC analysis was implemented via \texttt{emcee} \citep{Foreman2013}, using the standard Metropolis-Hastings sampling algorithm.
At each step of a chain, SPEX is fed a set of parameter values: ${N}_{H}$, log $\xi$, ${v}_{\text{abs}}$, and ${\sigma}_{\text{emis}}$ for each wind zone. 
SPEX then performs a fit of the disk normalization, the only free parameter, using the continuum fitting range described in Section \ref{sec:cont}.
The more sophisticated atomic physics in SPEX, including wind re-emission, only affect the shape and depth of the absorption lines and are therefore turned off during this fit.
This process is substantially faster than including the normalization, which is degenerate with column density, as a degree of freedom in the MCMC analysis.
Once the continuum is fit, both re-emission and the full atomic physics in SPEX are turned back on, and a new $\chi^{2}$ is obtained using the line-focused fitting range (Section \ref{sec:cont}).

In addition to the parameter boundaries listed in Section \ref{sec:PIA}, we also set dynamic boundaries based on the geometry of the system. 
An upper limit on the photoionization radius, ${R}_{phot} = \sqrt{{L}/{{n}_{H}{\xi}}}$, can be set using $N = n \Delta r$.
By setting $f = {\Delta r}/{r} \leq 1$, the upper limit becomes ${R}_{phot} \leq {L}/{{N}_{H}{\xi}}$, where $f$ is the filling factor and $\Delta r$ is the thickness of the wind.
However, we relaxed this constraint to $f \sim {\Delta r}/{r} \leq 1.2$, in order to account for uncertainties in the distance and mass of the black hole.
We use this inequality as a prior to constrain our fitting parameters: 
The orbital radii implied by velocity broadening must be below the limit set by the wind's plasma properties, ${R}_{\text{orbital}}= GM/{{\sigma}^{2}_{\text{emis}}} \leq {L}/{ {\xi}{N}_{H}}$, 
where $ GM/{{c}^{2}} = {r}_{g}$. Although luminosity and ${r}_{g}$ depend on assumed values of either distance or black hole mass, the accepted ranges for these values
suggest that this uncertainty should have minimal impact on our results. 

We ran our chains with sixteen walkers for 4000 steps each. 
Several MCMC experiments were conducted in the process of finding a minimum for each, therefore the 64000 steps presented in this work only represent the final run in a series of experiments.
We checked for convergence using the Gelman-Rubin fitting statistic, where we only considered a parameter converged when this value reached below 1.2 while ensuring the variance on the posterior distribution was not a product of how walkers where initialized. 
Parameter errors were determined using the highest posterior density (or, HPD) interval method, in which the shortest interval that contains, for example, 
68\% of the posterior distribution corresponds to the 1$\sigma$ error interval.

\bibliographystyle{aasjournal}

%\clearpage

%----------------------------------------------------------------------

\end{document}